\documentclass[a4paper]{JHEP3}
\usepackage{cite}
\usepackage{epsfig}
\usepackage{amsmath}
\usepackage{xspace}
\usepackage{flafter}
\newcommand{\dd}{\ensuremath{\mathrm{d}}}
\newcommand{\ca}{\ensuremath{C_{\!A}}}

\newcommand{\Nc}{\ensuremath{N_{\!C}}}

\newcommand{\pHp}{\ensuremath{p_{\scriptscriptstyle H_\perp}}}
\newcommand{\as}{\ensuremath{\alpha_s}\xspace}
\newcommand{\gs}{\ensuremath{g_s}}
\title{Higgs Boson Production in Association with Multiple Hard Jets}
\author{Jeppe R.~Andersen\\
Theory Division, Physics Department, CERN, CH 1211 Geneva 23, Switzerland}
\author{Vittorio Del Duca\\
Istituto Nazionale di Fisica Nucleare,
Laboratori Nazionali di Frascati\\
Via E. Fermi 40 - 00044 Frascati (Roma), Italy}
\author{Chris D.~White\\
Nikhef, Kruislaan 409, 1098 SJ Amsterdam, The Netherlands}

\abstract{ We elucidate a new technique for estimating the production of
  multiple (at least two) hard jets in Higgs production via gluon-gluon
  fusion. The approach is based upon high energy factorisation, with the
  region of applicability extended by constraints on the analytic behaviour
  of the scattering amplitudes stemming from known all-order results. The
  method approximates both real and virtual corrections, and allows for the
  resummation in an $n$-parton inclusive event sample of the terms dominant
  in the high energy limit. The resulting approximation is matched to the
  known tree level matrix elements for the production of a Higgs boson in
  association with 2 and 3 jets, and implemented in a Monte Carlo
  generator. Example results are presented and characteristic radiation
  patterns discussed.}

\preprint{CERN-PH-TH/2008-178\\
NIKHEF-2008-010}
\begin{document}
\newpage
\section{Introduction}
\label{intro}

One of the main goals of the forthcoming Large Hadron Collider (LHC) will be
to study the nature of the electro-weak symmetry breaking (EWSB). If a
fundamental scalar is observed, one must determine whether or not it is the
Higgs boson, responsible for the EWSB in the Standard Model. Thus, it is
imperative to measure its couplings as accurately as possible, especially to
the gauge vector bosons. This is possible by observing its decay to vector
bosons, or by isolating the process of a Higgs boson production via vector-boson
fusion (VBF). The latter contributes to the general signal of a Higgs
boson in association with two jets, which also receives a significant
contribution from Higgs boson production via gluon-gluon fusion (GGF), where the
Higgs boson couples to gluons via a top quark loop. The large cross section
associated with GGF makes it a valuable discovery channel in its own right
\cite{Klamke:2007cu}. However, a precise determination of the coupling of the Higgs
boson to the weak gauge boson requires a suppression of the GGF contribution
to the production of a Higgs boson in association with two jets. This can be achieved with the {\it vector-boson fusion
  cuts} of Ref.~\cite{DelDuca:2001fn}. It is expected that the contribution
from the GGF process can be further diminished by vetoing jet activity in the
central rapidity region \cite{Dokshitzer:1987nc,Dokshitzer:1991he}, owing to the fact that GGF has
a $t$-channel colour octet exchange as opposed to the colour singlet of VBF.

The angular structure of the final state is very different in
the VBF and GGF processes at lowest order.
One expects a significant azimuthal correlation between the two jets in the
GGF case, which is not so for VBF~\cite{DelDuca:2001fn,Hankele:2006ja}.  The
nature of the correlation contains information about the CP properties of the
coupling of the Higgs boson to fermions or gauge bosons. The practical usefulness
of this correlation is potentially threatened, however, by multiple hard jet
emission which acts to decorrelate the jets. Thus it is clearly important to thoroughly
understand the multijet final states in GGF.

There are several established methods for estimating the production of a
Higgs with a number of accompanying partons. The first and best verified
approach is to calculate the processes in fixed-order perturbation theory, up
to as high an order in the strong coupling constant \as as is possible.  For
the VBF process, radiative corrections have been calculated both within
QCD~\cite{Han:1992hr,Djouadi:1991tk,Figy:2003nv,Berger:2004pc,Figy:2007kv,Harlander:2008xn}
and the electro-weak sector~\cite{Ciccolini:2007jr,Ciccolini:2007ec}. The radiative
corrections to the VBF channel are small, and there is even partial numerical
cancellation between the QCD and electro-weak contributions. Recently, the
quantum mechanical interference between QCD and electro-weak Higgs
boson production\cite{Andersen:2006ag} was also
investigated\cite{Andersen:2007mp,Bredenstein:2008tm}.

Higgs boson~+~2~jet production was computed at leading order in \as with full
top-mass dependence in Ref.~\cite{DelDuca:2001eu,DelDuca:2001fn}, and in the
large top mass limit in Ref.~\cite{Kauffman:1996ix}.  In the limit of
infinite top mass, the coupling of the Higgs boson to gluons through a top
quark loop can be described by a point
interaction\cite{Wilczek:1977zn,Dawson:1990zj,Djouadi:1991tk}. This
approximation has been applied in most studies of Higgs boson production in
association with jets, and will be applied also in the present one, although
this is not essential to the approach. The large top mass limit is valid as
long as the transverse energy of all the associated jets is smaller than the
Higgs and the top masses\cite{DelDuca:2003ba}\footnote{For jet transverse
  energies larger than $m_H$ or $m_t$, the full top-mass dependence must be
  taken into account.  Beyond leading order, that dependence is known only in
  the limit of high partonic centre-of-mass energy\cite{Marzani:2008az}.}.

In the large top mass limit, the radiative corrections to Higgs boson~+~two parton
final states were evaluated in
Ref.~\cite{DelDuca:2004wt,Dixon:2004za,Badger:2004ty,Ellis:2005qe} and Higgs
boson + 2 jet ($hjj$) production was calculated at full next-to-leading order (NLO)
in Ref.~\cite{Campbell:2006xx}. Higgs boson~+~3~jet ($hjjj$) production is known
only at leading order in $\as$~\cite{DelDuca:2004wt}; it is unlikely that the
process $hjjj$ will be available at full NLO in the near future. Likewise, no
full tree-level prediction for the production of a Higgs boson in association
with four jets has yet been presented.

One method for estimating the effects of perturbative corrections beyond
these lowest orders is to interface higher order 
tree level matrix elements with a parton
shower algorithm, which estimates the part of the radiative corrections
arising from soft and collinearly enhanced regions of the real emission phase
space. This entails a marked dependence of the jet cross section on the
parton-level generation cuts, and in particular
an unphysical behaviour of the real emission elements as the $p_t$
of a jet tends to zero. Collinear and soft singularities are generated, which
in a true NLO calculation would be cancelled by corresponding contributions
in the virtual corrections. Such a technique was applied to Higgs boson production  
in \cite{DelDuca:2006hk}, where the
discussion was tailored towards isolating the VBF signal. In that study, real
emission matrix elements for Higgs + $n$ parton production (where $n=2,3$)
were interfaced with a parton shower, and it was found that in a 
significant fraction of two-jet final states
one of the jets originated from the shower rather than the hard matrix element.

Thus, the question arises of whether it is possible to estimate multiparton
final states using a method which captures the hard-parton behaviour of the
multi-parton matrix elements to any order in $\alpha_s$. We present such an
algorithm in this paper. Our starting point is the Fadin-Kuraev-Lipatov
(FKL\cite{Fadin:1975cb,Kuraev:1976ge,Kuraev:1977fs}) factorised form of
multiparton amplitudes, an approximation to the full scattering amplitude
which becomes exact in the limit of multi-Regge
kinematics (MRK) with hard partons of infinite separation in rapidity, i.e.~in
the asymptotic limit when all interjet invariant masses $s_{ij}$
tend to infinity. These amplitudes allow one to define exclusive
multiparton final states, with the inclusion of virtual corrections whose
singularities cancel those associated with soft real emissions. The
conventional application of this framework, in the context of the
Balitsky-Fadin-Kuraev-Lipatov (BFKL\cite{Balitsky:1978ic}) equation, applies
the kinematical approximations everywhere in phase space, whilst integrating
the evolution fully inclusively. By comparing the result order by order with
the corresponding full fixed order results, we will demonstrate that this
approach does not lead to a good approximation. However, we show that it is
possible to modify the FKL framework to take into account the known 
structure of singularities of the full scattering amplitude to any order in the coupling. We thereby construct
approximate scattering amplitudes, which have the same MRK limit as the FKL amplitudes, but
maintain the correct position of singularities away from the high energy limit and thus
better approximate the true amplitudes over all of phase space.
One can
view our approach as systematically building up approximate matrix elements,
using FKL factorisation as a (hard) starting point, in contrast with
applying a soft shower algorithm to a hard matrix element. Thus, our technique should be
better suited at describing results sensitive to the \emph{jet multiplicity} rather
than the \emph{internal structure of each jet}, which is best described by 
soft and collinear resummation.

We will validate our technique by comparing to known tree level matrix elements
at low orders in \as. We then implement our prescription for calculating amplitudes
in a Monte Carlo generator for Higgs boson production via GGF, where the tree level results
for 2 and 3 parton final states are included via a matching procedure which avoids any
double counting of radiation. This matching procedure could in principle be implemented at higher orders in \as. However,
the evaluation of full tree level matrix elements becomes computationally
punitive for more than 3 jets.

The structure of the paper is as follows. In section~\ref{sec:higgs-+-n}, we
introduce the GGF process using LO results. In section~\ref{sec:high-energy-fact}, we introduce the
factorisation properties of amplitudes in the \emph{high energy limit}
(HEL), before discussing the traditional BFKL implementation of this
framework, applied to 
Higgs boson production via GGF in association with at least two jets. We then present our alternative implementation, based
on the imposition of known analytic constraints, and validate the approach by comparing
order by order in \as with fixed order results. In section~\ref{matching}, we discuss
how the resummed amplitudes can be matched to the fixed order results at low orders in \as, 
and implemented in a Monte Carlo event generator. Some results from this generator are shown 
in section~\ref{results}. Finally, in section~\ref{discussion}, we summarise and discuss the possibilities for
further systematic improvements.
\section{Higgs Boson Production with Multiple Jets at Fixed Order}
\label{sec:higgs-+-n}
\subsection{Production of Higgs + 2 Jets}
\label{sec:central-higgs-+}
The fully differential cross-section for the production of a Higgs boson with
two accompanying partons $c,d$ may be written as follows:
\begin{align}
    &\frac{\dd\sigma}{\dd^2p_{c_\perp}\ \dd^2p_{d_\perp}\ \dd^2\pHp\
      \dd y_{c}\ \dd y_{d}\ \dd y_H} \nonumber\\
    & = \sum_{a,b} x_{a} f_{a/A}(x_{a},\mu_F^2)\ x_{b} f_{b/B}(x_{b},\mu_F^2)\
    \frac{|\mathcal{M}_{ab\rightarrow cdh}|^2} {256\pi^5\ \hat s^2}\ \delta^{(2)}(p_{c_\perp}+p_{d_\perp}+\pHp),
	\label{xsec}
\end{align}
where:
\begin{equation}
    x_a=\sum_{f\in\{c,d,h\}}\frac{|m_{f_\perp}|}{\sqrt s} \exp(-y_f), \qquad 
    x_b=\sum_{f\in\{c,d,h\}}\frac{|m_{f_\perp}|}{\sqrt s} \exp(y_f)
\end{equation}
are the momentum fractions of the incoming partons; $m_{(c,d)_\perp}$ the
transverse mass of the final state partons and Higgs boson;
$y_{(c,d,h)}$ the corresponding rapidities. Here $|\mathcal{M}_{ab\rightarrow cdh}|^2$ is
the matrix element for production of a Higgs boson in association with partons
$c$ and $d$ from partons $a$ and $b$ after averaging / summing over colour
and helicity, $\hat s$ is the squared partonic centre of mass energy, and the
remaining numerical factors arise from the parameterisation of phase space.

Unless otherwise stated, 
we will concentrate on the contribution from QCD generated
Higgs boson production within cuts optimised for the selection of events
originating from the weak boson fusion process, as they were used in the NLO
calculation of the QCD contribution to the $hjj$-channel~\cite{Campbell:2006xx}. These are listed in
Table~\ref{tab:cuts}.
\begin{table}[tbp]
  \centering
  \begin{tabular}{|rl||rl|}
    \hline
    $p_{j_\perp}$ & $> 40$ GeV & $y_c\cdot y_d$& $<0$ \\
    $|y_{j,h}|$ & $<$ 4.5 & $\vert y_c-y_d \vert$ & $> 4.2$ \\\hline
  \end{tabular}
  \caption{The cuts used in the following analysis which bias the Higgs boson
    plus jet sample towards VBF.  The suffices $c,d$ label cuts that must be 
   satisfied by at least two jets, whereas $j$ labels conditions that must be 
   satisfied by all jets; $h$ labels the Higgs boson.}
  \label{tab:cuts}
\end{table}
As we will demonstrate, the good performance of the approximations we will
make later does not rely crucially on these cuts - specifically we will also
study events selected with a rapidity cut of just 2 units of rapidity.
\begin{figure}[tbp]
  \centering
  \epsfig{width=9cm,file=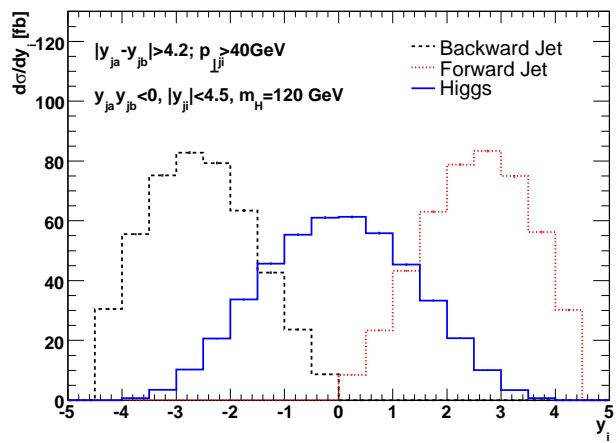}
  \caption{The rapidity distributions of the final state particles in $hjj$ production via GGF at LO.}
  \label{fig:h2p}
\end{figure}
In Figure~\ref{fig:h2p} we have plotted the rapidity distribution of the
forward and backward jet, and the Higgs boson for the sum of all channels
contributing to Higgs boson plus dijet production through gluon fusion at leading
order. We use the following values for the Higgs
boson mass, vacuum expectation value of the Higgs boson field and top quark
mass respectively:
\begin{equation}
m_H=120\text{GeV},\qquad \langle \phi\rangle_0=\frac v {\sqrt 2}, v=246\text{GeV},\qquad m_t=174\text{GeV}.
\label{params}
\end{equation}
We also include a factor multiplying the effective Higgs boson vertices,
accounting for finite top-mass effects~\cite{Dawson:1993qf}:
\begin{equation}
K(\tau)=1+\frac{7\tau}{30}+\frac{2\tau^2}{21}+\frac{26\tau^3}{525},\quad \tau=\frac{m_H^2}{4m_t^2},\nonumber
\label{K}
\end{equation}
which increase the cross section by 5.9\% for the parameter values in
Eq.~(\ref{params}). Furthermore, we choose the NLO set of parton density functions from
Ref.\cite{Martin:2004ir} (for all the studies presented here), and 
choose factorisation and renormalisation scales equal to $m_H$. 
However, the resummation and approximation presented later is based explicitly on
the kinematical part of the amplitude, and \emph{not} on the running
coupling terms. Thus, it does not relate to any specific scale choice,
at least to the discussed accuracy. Tree level matrix elements have been obtained
using MadGraph\cite{Alwall:2007st}. We report results for a 14~TeV $pp$-collider.

We find the tree-level cross-section from the QCD generated $Hjj$-channel to
be $230^{+167}_{-90}$fb, where the uncertainty is obtained by varying the
common factorisation and renormalisation scale, as given above,
by a factor of two. The variation is far less if
the renormalisation and factorisation scales are varied simultaneously in
opposite directions, where we find $230^{+88}_{-55}$fb. 

\subsection{Observables for Distinguishing VBF and GGF}
\label{sec:discr-observ}

For a large rapidity-separation of the two jets, Higgs boson production via
gluon fusion is dominated by a $t$-channel exchange. The different $C\!P$
structures of the $ggH$ and $VVH$ vertices (where $V$ is a $W$ or $Z$ boson) 
give rise to different azimuthal
correlations of the two leading jets. While VBF has a very mild dependence on
the azimuthal angle $\phi_{j_a j_b}$ between the two jets, for the GGF
process the azimuthal correlation between the two jets exhibits a
characteristic dip at $\phi_{j_aj_b}=\pi/2$~\cite{DelDuca:2001fn}.  However,
it is expected that higher order corrections will somewhat fill the dip,
thereby decreasing the discriminating power. The effect of such higher order
corrections were estimated in Ref.\cite{DelDuca:2006hk} using a parton shower
approach to resum the soft and collinear radiation. It is the purpose of this
current study to calculate the effect caused by the emission of several
\emph{hard} gluons, which obviously can result in more decorrelation. In the
following two subsections we will introduce some of the suggested variables
for discriminating the QCD and VBF contributions to $H+(n\ge2)j$.

\boldmath
\subsubsection{Azimuthal angle distribution}
\label{sec:discr-observ-1}
\begin{figure}[tb]
  \centering
    \epsfig{width=9cm,file=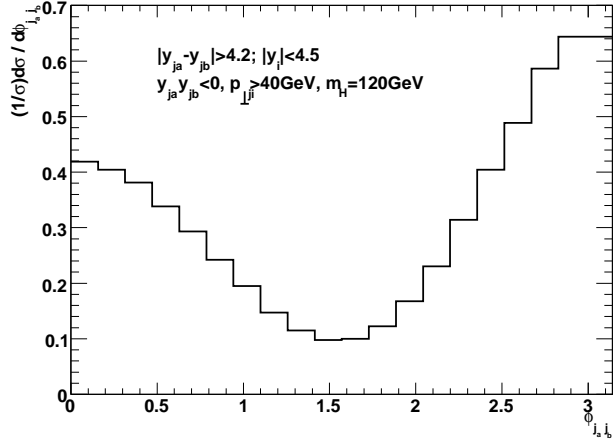}
  \caption{$d\sigma/d\phi_{j_aj_b}$ for Higgs boson production in association with two jets. $\phi_{j_aj_b}$ is the azimuthal angle between the two jets.}
  \label{fig:2jLOdsdphi}
\end{figure}

\unboldmath The LO result for the azimuthal angle distribution 
$d\sigma/d\phi_{j_a j_b}$ in $hjj$ production via GGF 
is shown in Fig.~\ref{fig:2jLOdsdphi}, for the Standard Model
Higgs boson with $C\!P$-even couplings to two gluons, and shows the 
characteristic dip mentioned above.
By contrast, the contribution from the VBF channel is much
flatter (see e.g.~Ref~\cite{Hankele:2006ja}). 

\boldmath
\subsubsection{$A_{\phi}$}
\label{sec:discr-observ-2}
\unboldmath As suggested in Ref.\cite{Hankele:2006ja} the structure of the
distribution $d\sigma/d\phi_{j_aj_b}$ can be distilled into a single number
$A_\phi$ given by:
\begin{align}
  \label{eq:Aphi}
  A_\phi=\frac{\sigma(\phi_{j_aj_b}<\pi/4)-\sigma(\pi/4<\phi_{j_aj_b}<3\pi/4)+\sigma(\phi_{j_aj_b}>3\pi/4)}{\sigma(\phi_{j_aj_b}<\pi/4)+\sigma(\pi/4<\phi_{j_aj_b}<3\pi/4)+\sigma(\phi_{j_aj_b}>3\pi/4)}
\end{align}
According to this definition, one has $-1\leq A_\phi\leq 1$, with $A_\phi=0$
representing no azimuthal correlation between the tagging
jets. A $C\!P$-even coupling for the vector boson fusion into a Higgs boson
leads to a positive value for $A_\phi$, whereas a $C\!P$-odd coupling results
in a negative value. The Standard Model coupling for the weak gauge boson
leads to $A_\phi\approx0$ in the VBF-sample. Using our standard cuts and scale choice
in the GGF sample of a Higgs boson plus two partons, we obtain
$A_\phi=0.456^{+0.003}_{-0.003}$, in agreement with the CP-even nature of the effective
coupling between the Higgs boson and two gluons, and the dominance of the
$t$-channel gluon fusion processes within the QCD contribution.

\subsection{Production of Higgs + 3 Jets}
\label{sec:production-higgs-+}
By analogy with equation (\ref{xsec}), the fully differential cross-section for the production of a Higgs boson with
three accompanying partons $c,d,e$ may be written as follows:
\begin{align}
    &\frac{\dd\sigma}{\dd^2p_{c_\perp}\ \dd^2p_{d_\perp}\ \dd^2p_{e_\perp}\ \dd^2\pHp\
      \dd y_{c}\ \dd y_{d}\ \dd y_{e}\  \dd y_H}\notag\\
    & = \sum_{a,b} x_{a} f_{a/A}(x_{a},\mu_F^2)\ x_{b}
    f_{b/B}(x_{b},\mu_F^2)\ \frac 1 {4\pi (2\pi)^2}\ 
    \frac{|\mathcal{M}_{ab\rightarrow cdeh}|^2} {256\pi^5\ \hat s^2}\ \delta^{(2)}(p_{c_\perp}+p_{d_\perp}+p_{e_\perp}+\pHp),
	\label{xsec3jh}
\end{align}
where $p_{(c,d,e)_\perp}$ are the transverse momenta of the final state
partons, and $y_{(c,d,e)}$ their rapidities. 

\begin{figure}[tbp]
  \centering
    \epsfig{width=.49\textwidth,file=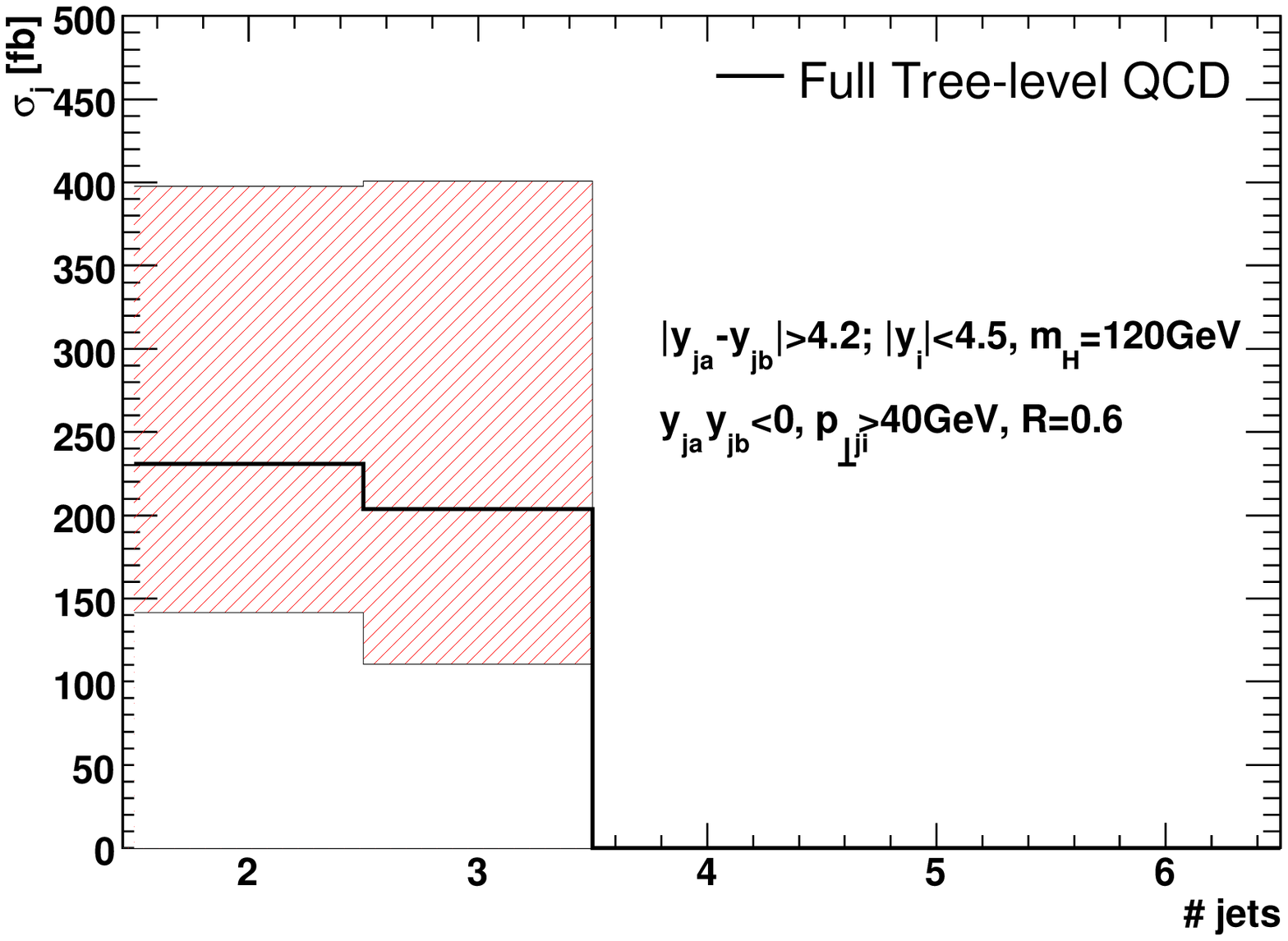}
    \epsfig{width=.49\textwidth,file=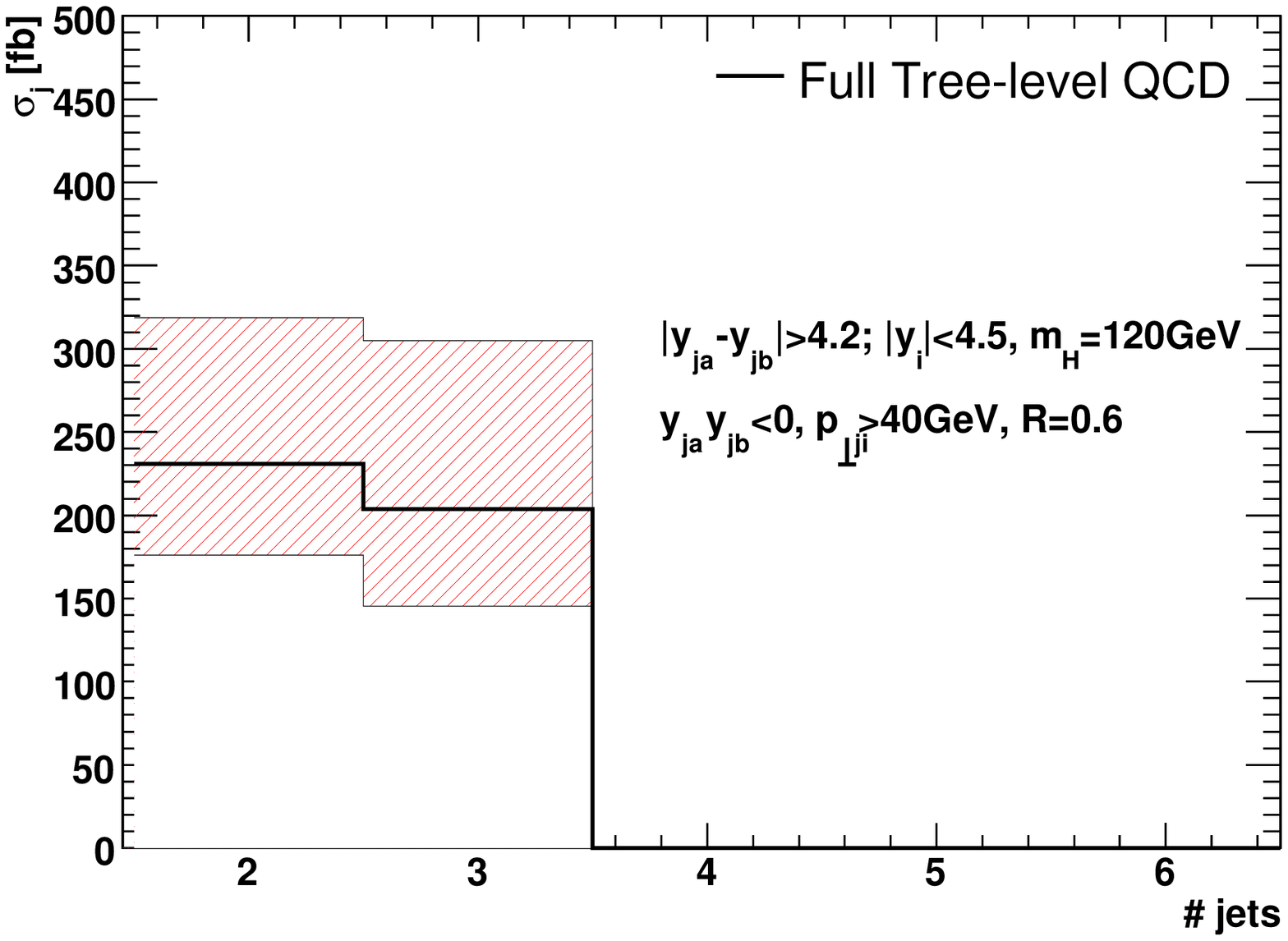}
    \caption{The LO $hjj$ and $hjjj$ jet cross-sections obtained using the tree level
      matrix elements for $h$+4 partons and $h$+5 partons respectively. The
      uncertainty band of the tree level results are obtained by varying the
      factorisation and renormalisation scale by a factor 2, either as a
      common scale (left) or in opposite directions (right).}
  \label{h+jetsLO1}
\end{figure}

We start by considering the fully inclusive 3-hard-jet radiative correction to the
two jet sample. By this we mean that an event with 3 hard jets will be accepted, if 
two of the jets fulfill all the requirements of
Table~\ref{tab:cuts}. Contrary to the case for the 2-jet calculation with
these cuts, the jet
algorithm is relevant in defining the phase space for the 3-jet
configuration. We choose the $k_t$-algorithm as implemented in
Ref.\cite{Cacciari:2005hq}, with $R=0.6$ and using the energy recombination
scheme. With the relevant cuts, and by evaluating the extra coupling compared
to the two-jet case at scale $m_H$, we find the
leading order cross section for the production of Higgs Boson plus three jets
to be $203^{+170}_{-94}$fb. The uncertainty obtained by varying the
factorisation and renormalisation scales in opposite directions is far smaller,
resulting in an estimate for the cross section of $203^{+74}_{-86}$fb. The requirement of an extra jet with the
accompanying \as-suppression leads to a change in the tree-level cross
section of less than 12\%! In
Figure~\ref{h+jetsLO1} we have plotted the tree-level results for the $hjj$ and $hjjj$
processes. The shaded bands indicate the size of the
renormalisation and factorisation scale uncertainty obtained by varying the
factorisation and renormalisation scales by a factor of two, either as a
common scale (left) or in opposite directions (right). The full NLO K-factor for
$\sigma_{hjj}$ with this set of cuts is found to be 1.7-1.8\footnote{We thank
  John Campbell for this information.}. The value of $A_\phi$ obtained with this 3-jet
sample is $A_\phi=0.203^{+0.002}_{-0.002}$, significantly lower than the leading order
two-jet value.

The apparent lack of suppression of the three parton final state using the cuts
of Table~\ref{tab:cuts} has led to the alternative suggestion that when events containing
three jets or more are considered, one should require that the two hardest jets in the event
also satisfy the rapidity separation constraint~\cite{Figy:2003nv,DelDuca:2004wt,Campbell:2006xx,Figy:2007kv}.
This significantly reduces the accepted three-jet phase
space: A central jet will have a slightly harder transverse momentum spectrum
than any forward jet, simply because of the smaller impact on the parton
momentum fraction of a hard central jet, and the resulting lack of suppression 
from the PDFs. We will return to this point in Section~\ref{sec:cross-sections-jet}. Therefore, if considering only the two hardest jets in the
event, one is less likely to satisfy the rapidity separation constraint. For the 
above parameters we find $\sigma_{hjjj}^{LO}=76^{+74}_{-35}$fb
($\sigma_{hjjj}^{LO}=76^{+37}_{-22}$fb when varying the scales in opposite directions).
The NLO K-factor obtained in the calculation of
Ref.\cite{Campbell:2006xx} with this set of cuts is 1.3-1.4, the difference
with respect to the previous case obviously arising from the reduced three-jet phase
space. The value of $A_\phi$ for this set of three-jet cuts is
$0.374^{+0.004}_{-0.004}$, i.e.~much closer to the value found for the 2-jet sample.
\begin{figure}[tbp]
  \centering
    \epsfig{width=.49\textwidth,file=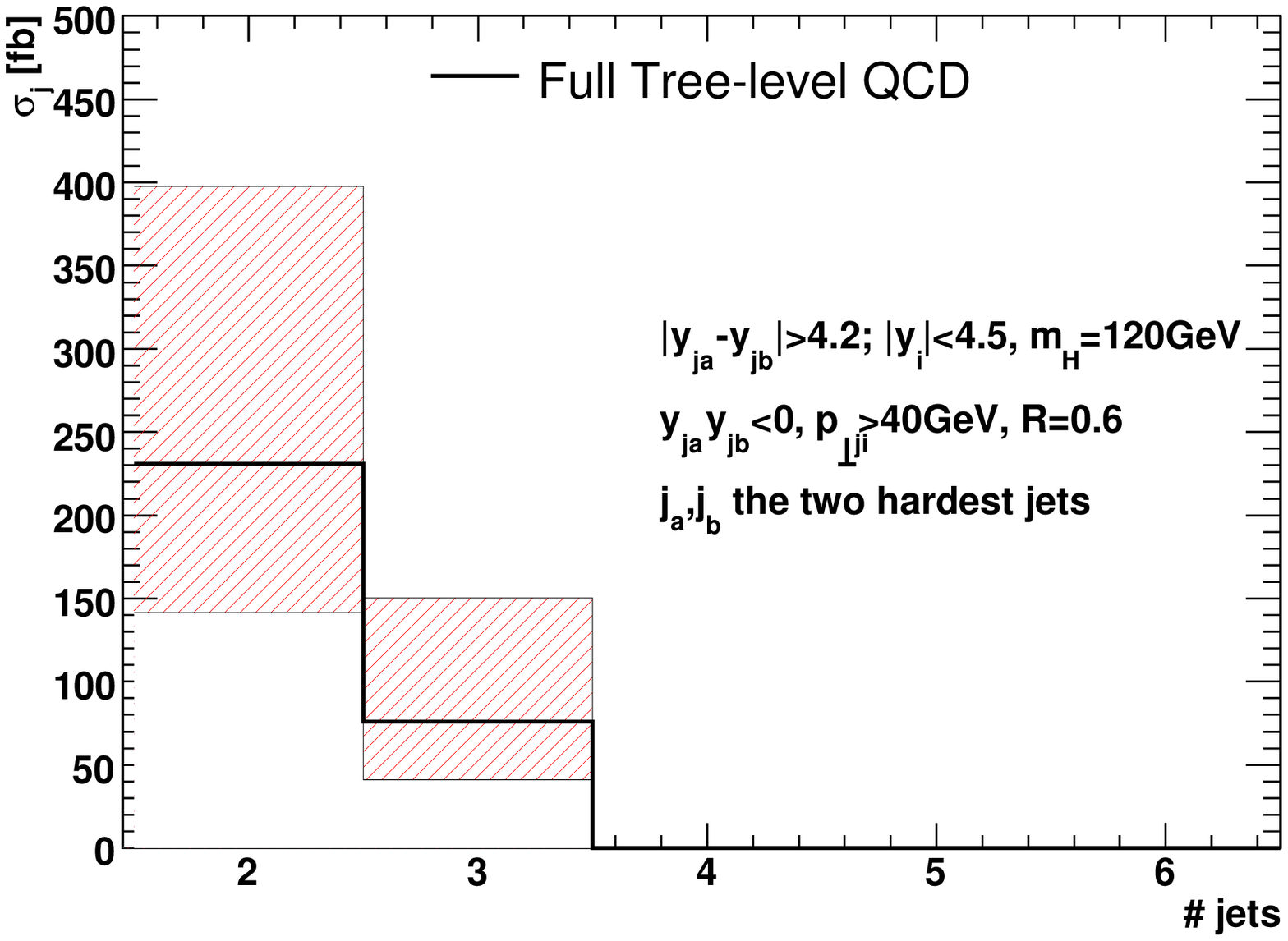}
    \epsfig{width=.49\textwidth,file=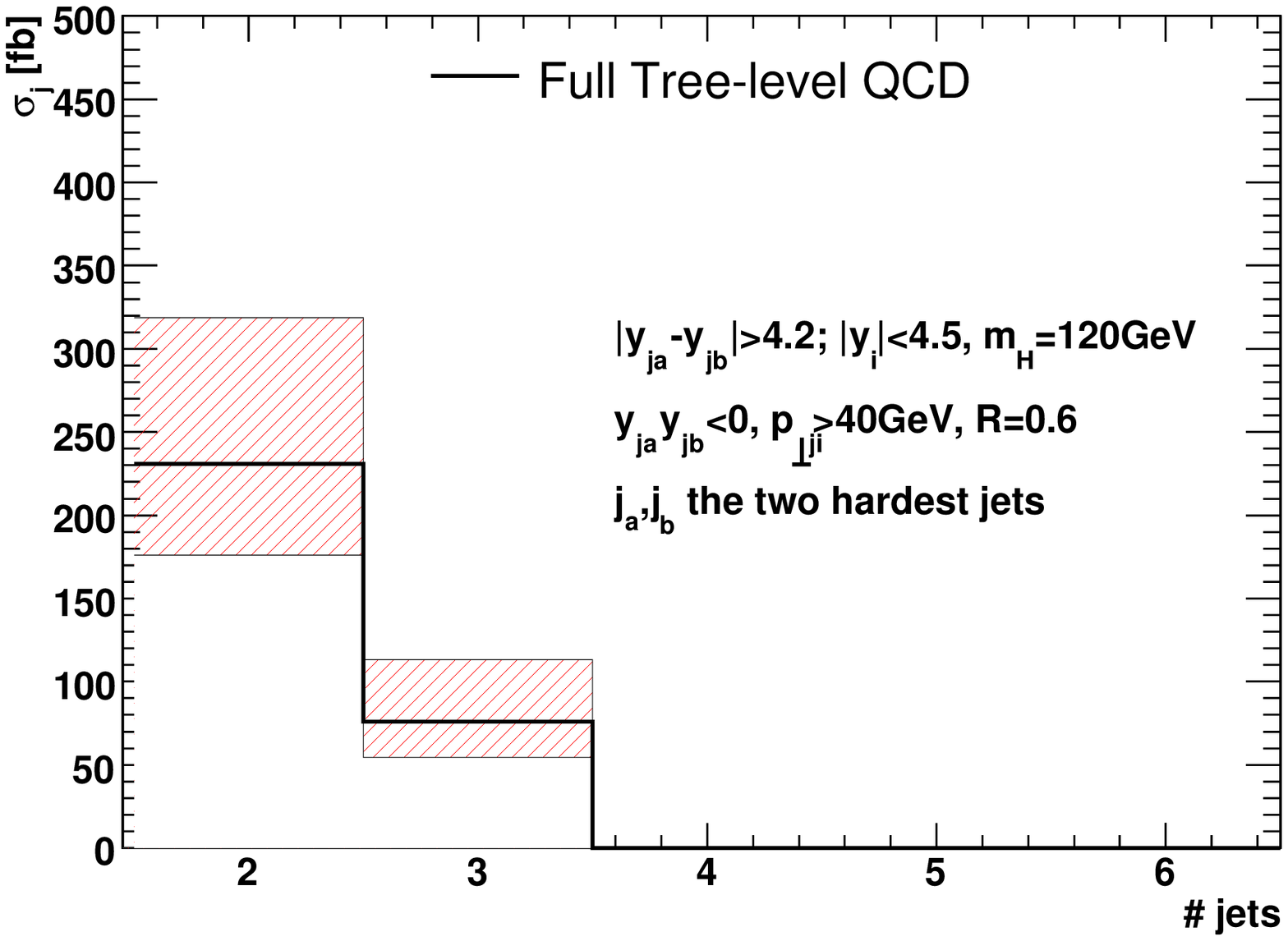}
    \caption{The LO $hjj$ and $hjjj$ cross-sections, obtained using similar cuts as Figure 3, but where
   the two hardest jets must satisfy the rapidity separation constraint. The uncertainty band arises from
   varying the renormalisation and factorisation scale by a factor or two,
   either as a common scale (left) or in opposite directions (right).}
  \label{h+jetsLO2}
\end{figure}

The reason for the lack of a suppression in the 3-jet rate compared to the
2-jet rate of order \as is simply due to the large size of the $(n+1)$-body
phase space compared to the $(n)$-body equivalent. At the LHC, and when a
large rapidity span is already required, the balance between the impact of
additional central jets on the parton momentum fractions followed by a PDF
(and \as) suppression and the increase in $(n+1)$-body phase space can be
such that the additional emission is not suppressed. We have here
demonstrated this at the lowest orders in perturbation theory. Notice that
the large size of the 3-jet rate is not due to a divergent matrix element
(the divergent region is explicitly cut out by the jet algorithm and the
requirement of all three jets having a significant transverse momentum), but is simply a
well-behaved matrix element integrated over a large phase space. In order to
stabilise the perturbative series by resumming the effects of a large
$n$-body phase space to all orders, we would need to construct an
approximation (since we cannot calculate exactly to all orders) to the
$n$-leg, $l$-loop amplitude, including the hard-parton region. This is the
aim of the next section.

\section{High Energy Factorised Matrix Elements and Inclusive Jet Samples}
\label{sec:high-energy-fact}
In this section we will outline how we can arrive at a useful approximation
for the $n$-leg, $l$-loop amplitude relevant for the calculation of Higgs
boson production in association with jets. We will start by general
observations based on the analyticity of the
S-matrix\cite{Regge:1959mz,Regge:1960zc,Brower:1974yv} for $n$-particle
scattering in a certain kinematical limit. We will extend the region of
applicability for the results obtained in this limit to all of phase space,
by further restricting the analytic behaviour of the amplitudes away from the
specific limit. The amplitudes thus obtained will allow for a resummation of
the leading behaviour as the Mandelstam variable $\hat
s\to\infty$, with fixed, perturbative transverse scattering momenta. We will demonstrate
explicitly that the dominant contribution to the multi-jet cross section is
captured within this framework, by comparing order-by-order with the results
obtained in the previous section.

We will discuss the traditional implementation of the framework through the
BFKL equation\cite{Balitsky:1978ic}, and explain why this fails to reproduce
the results obtained in the fixed order approach, even if it is built upon
the same asymptotic limit as the framework presented here.

\subsection{FKL Factorisation}
\label{sec:fkl-factorisation}
It has long been known (see e.g.~\cite{Collins:1977jy,Landshoff:1966} and references therein)
that in any Lorentz-invariant quantum field theory, the
scattering amplitude for $2\rightarrow 2$ scattering in the {\it Regge limit}
of large centre of mass energy $\hat{s}$ and fixed momentum transfer $t$ assumes
the form
\begin{align}
  \label{eq:regge2to2}
  \mathcal{M}^{p_a p_b\to p_{1} p_{2}}\stackrel{\mathrm{Regge\
      limit}}\longrightarrow\ \hat s ^{\hat\alpha(\hat t)}\ \gamma(\hat t),
\end{align}
where $\gamma(\hat t)$ and $\hat\alpha(\hat t)$ depend on the dynamics of the underlying
theory. The generalisation to $2\to3$-scattering $p_a p_b\to p_1 p_2 p_3$ in
the double-Regge limit of
$s_{12}=(p_1+p_2)^2,s_{23}=(p_2+p_3)^2,s_{ab}=(p_a+p_b)^2\to \infty$,
$t_1=(p_a-p_1)^2,t_2=(p_a-p_1-p_2)^2=(p_3-p_b)^2$ fixed was expected to be
\begin{align}
  \label{eq:regge2to3}
  \mathcal{M}^{p_a p_b\to p_{1} p_{2}p_3}\stackrel{\mathrm{double-Regge\
      limit}}\longrightarrow\ \hat s_{12}^{\hat\alpha(\hat t_1)}\
  s_{23}^{\hat\alpha(\hat t_2)}\ \gamma(\hat t_1,\hat t_2, s_{ab}/(s_{12} s_{23})).
\end{align}
For the specific case of QCD, this factorisation was shown explicitly to hold 
\cite{Balitsky:1979ap} for
general QCD multi-gluon amplitudes in a multi-Regge limit (i.e.~all $s_{ij}$
tending to infinity) to all orders in \as. It was later proved to hold also
when one invariant mass $s_{ij}$ is allowed not to tend to
infinity\cite{Fadin:2006bj}. The factorised amplitudes allows one to
calculate the behaviour of the scattering as $s_{ab}\to\infty$ for fixed
scattering momenta (of typical size $\sqrt{|t|}$) to leading and
next-to-leading logarithmic accuracy in $\ln(s_{ab}/|t|)$. To leading
logarithmic accuracy, one needs only to take into account the dominant
contribution as all $s_{ij}\to\infty$, which arises from processes with only
gluon quantum numbers exchanged in the $t$-channel(s). In the present study
we will work only within this approximation.

We will denote the QCD amplitudes derived from the multi-Regge limit the
\emph{FKL-amplitudes} (Fadin-Kuraev-Lipatov), after the people who 
proved\cite{Fadin:1975cb} the factorisation property and derived the form of
$\hat \gamma$ and $\hat \alpha$ in QCD.  It is found in this case that the
function $\hat \gamma$ factorises further into functions of non-overlapping
momenta. We will denote by \emph{FKL factorisation} the factorisation of QCD
amplitudes into a product of building blocks of the form $s_{ij}^{\hat\alpha
(\hat t_i)} \gamma(q_i,q_{j})$ (at LL, and similarly at NLL) in the limit of
all $s_{ij}\to \infty$, $t_i$ fixed.

It is worth noting that the VBF cuts of Table~\ref{tab:cuts} approach at
least the limit of large invariant mass between two particles, since a
large rapidity span of the event is required. However, in the calculation of
cross sections, there will be no cut on the invariant mass between all pairs
of particles, and so the multi-Regge limit is not necessarily
approached. Our starting point for obtaining an approximation to the
matrix elements will be the FKL amplitudes obtained in the multi-Regge limit,
and we will later discuss how to extend their region of applicability outside
the ultimate MRK limit, so that inclusive cross-sections can be
calculated reliably.

The FKL factorised $(2\to n+2)$-gluon amplitudes, derived for the MRK
limit, and adapted to include also the production of a Higgs boson with a
rapidity between the produced jets, are given by
\begin{eqnarray}
i{\cal M}_{\mathrm{HE}}^{ab\rightarrow p_0\ldots p_jhp_{j+1}\ldots p_{n+1}}&=&2i\hat s
\left(i g_s f^{ad_0c_1} g_{\mu_a\mu_0}\right)\nonumber\\
&\cdot&
\prod_{i=1}^j \left(\frac{1}{q_i^2}\exp[\hat\alpha(q_i^2)(y_{i-1}-y_i)]\left(i g_s f^{c_id_ic_{i+1}}\right)C_{\mu_i}(q_i,q_{i+1})\right)\nonumber\\
&\cdot&\left(\frac 1 {q_h^2}\exp[\hat\alpha(q_i^2)(y_{j}-y_h)]C_{H}(q_{j+1},q_{h})\right)\label{FKL}
\\
&\cdot&\prod_{i=j+1}^n
\left(\frac{1}{q_i^2}\exp[\hat\alpha(q_i^2)(y'_{i-1}-y'_i)]\left(i g_s f^{c_id_ic_{i+1}}\right)C_{\mu_i}(q_i,q_{i+1})\right)\nonumber\\
&\cdot&\frac 1 {q_{n+1}^2}\exp[\hat\alpha(q_{n+1}^2)(y'_{n}-y'_{n+1})]\left(i g_s f^{bd_{n+1}c_{n+1}} g_{\mu_b\mu_{n+1}}\right)\nonumber
\end{eqnarray}
where $g_s$ is the strong coupling constant, and $q_i, q_h$ are the
4-momentum of gluon propagators (e.g.~$q_i=p_a-\sum_{k=0}^{i-1}p_k$ for
$i<j$), $C_{\mu_i}$ is the {\it Lipatov effective vertex}, which at the
leading logarithmic accuracy applied here results in gluon emission only (no
quark-anti-quark pairs produced), and $C_H$ is the effective vertex for the
production of a Higgs boson, as calculated in Ref.\cite{DelDuca:2003ba}. In
Eq.~(\ref{FKL}), the rapidities $y_i$ of gluons are denoted with primes, if the
rapidities are larger than that of the Higgs boson, such that we can set
$y'_{j}=y_h$ and $y'_{i}=y_i$ for $i>j$. The quantities $\hat{\alpha}(q_i^2)$
occur from the Reggeisation of the gluon propagator, and encode virtual
corrections (see e.g.~Ref.~\cite{DelDuca:1995hf}). We have suppressed the
obvious Lorentz indices on the LHS of Eq.~\eqref{FKL}. This amplitude is
pictorially represented in Fig.~\ref{FKLfig} - each zigzag $t$-channel
Reggeised gluon corresponds to a factor $\frac 1 {q^2}
\exp[\hat\alpha(q^2)(y_{i-1}-y_i)]$, and each real emission vertex
corresponds to a factor $(ig_s f^{abc}C_\mu(q_i,q_{i+1}))$. For the
production of a Higgs boson outside (in rapidity) of the partons, we will use
the equivalent factorised amplitudes with an impact factor for the production
of a parton in association with a Higgs boson (see
Ref.\cite{DelDuca:2003ba}), and gluon production vertices (no $C_H$)
connecting to the far end of the ladder.

\FIGURE{
\epsfig{file=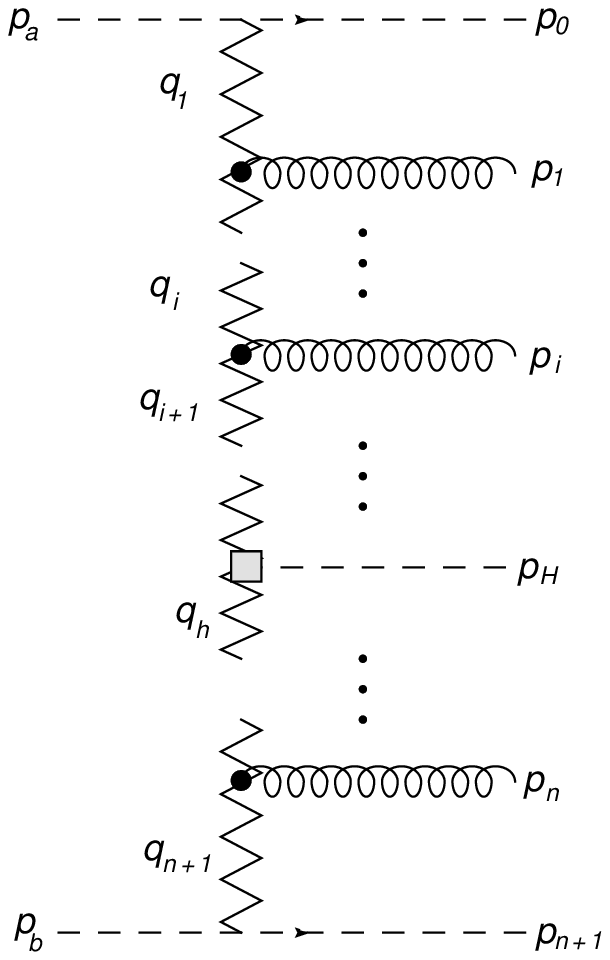}
\caption{Pictorial representation of the factorised amplitudes of
Eq.~(\ref{FKL}). The dashed lines represent incoming quarks or gluons, and
  the zigzag line represents Reggeised gluons. The gluons emitted from the
  $t$-channel exchange are coupled using Lipatov vertices. Final state
  particles as shown are ordered in rapidity as in Eq.~(\ref{MRK}).\label{FKLfig}}}
The MRK limit can be expressed in terms of the rapidities $\{y_i\}$ of the
outgoing partons and their transverse momenta $\{p_{i\perp}\}$, so the MRK
limit where the amplitudes of Eq.~(\ref{FKL}) become relevant is
\begin{equation}
y_0\gg y_1\gg \ldots\gg y_{n+1};\quad p_{i\perp}\simeq p_{i+1\perp};\quad q_i^2 \simeq q_j^2.
\label{MRK}
\end{equation}
Each emitted gluon is coupled to the $t$-channel gluon exchange by a {\it
  Lipatov effective vertex}, given by (see e.g.~Ref.~\cite{Fadin:1998sh}):
\begin{align}
C^{\mu_i}(p_a,p_b,q_i,q_{i+1})&=\left[-(q_i+q_{i+1})^{\mu_i}-2\left(\frac{{\hat{s}}_{ai}}{\hat{s}_{ab}}
+\frac{\hat{t}_{i+1}}{\hat{s}_{bi}}\right)p_b^{\mu_i}\right.
\left.+2\left(\frac{\hat{s}_{bi}}{\hat{s}_{ab}}
+\frac{\hat{t}_i}{\hat{s}_{ai}}\right)p_a^{\mu_i}\right],
\label{lip1}
\end{align}
where $\hat{s}_{ai}=2p_a\cdot p_i$ etc., $\hat{t}_i=q_i^2$ is the propagator
denominator associated with the $i^{\text{th}}$ Reggeised gluon. The set of
Feynman diagrams entering the calculation of the Lipatov vertex is gauge
invariant up to subleading terms in the MRK limit. The MRK limit determines
only the asymptotic form of the Lipatov vertex, and forms differing
only by sub-asymptotic terms are often used, see
e.g.~Ref.\cite{DelDuca:1995hf,Andersen:2008ue}. However, we will choose the
form in Eq.~\eqref{lip1}, since it is manifestly Lorentz and gauge invariant
in all of phase space, as discussed in Sec.~\ref{sec:modified-high-energy-1}.
The exponential factors in Eq.~(\ref{FKL}) result from the \emph{Lipatov
  Ansatz} for the Reggeisation of the $t$-channel gluon propagator, which
encodes the leading virtual corrections and contains the function,
\begin{equation}
\hat{\alpha}(t_i)=\alpha_sN_ct_i\int\frac{d^2k_\perp}{(2\pi)^2}\frac{1}{k_\perp^2(q_i-k)_\perp^2},
\label{alpha}
\end{equation}
with $N_c$ the number of colours. Note that the colour factors in equation
(\ref{FKL}) are derived for incoming gluons. The form of the amplitude is
unchanged for incoming quarks apart from colour factors, such 
that the overall normalisation
receives a factor $C_F/C_A$ if an incoming gluon is replaced by a 
quark~\cite{Combridge:1984jn}. This arises from
the fact that in the MRK limit the coupling of the $t$-channel gluons to
external particles is insensitive to their spin.

In equation (\ref{FKL}), the effective vertex $C_H$ for Higgs boson production
is given by:
\begin{align}
    C^H(q_a,q_b)&=\lim_{\mathrm{MRK}}\frac{\mathcal{M}_{gg\to ghg}}{2i\gs^2\ \hat{s}
      \hat{t}_1 \hat{t}_2}    \label{CH}
\\
    &=2\frac{\gs^2m_t^2}{v}\left[m_{H\perp}^2A_1(q_a,q_b)-2A_2(q_a,q_b)\right]\nonumber,
\end{align}
where the coefficients $A_i$ are given in \cite{DelDuca:2003ba}, and we have
not denoted colour factors. Here $q_a$
and $q_b$ are the $t$-channel momenta entering the vertex, and $m_{H\perp}$
the transverse mass of the Higgs. This result is derived in
Ref.\cite{DelDuca:2003ba}, and is valid for all values of the top
mass. However, Eq.~(\ref{CH}) simplifies in the large top mass limit
$m_t\rightarrow\infty$, when the top quark loop coupling the $t$-channel
gluons to the Higgs boson can be replaced by a contact interaction. In this
limit one finds:
\begin{align}
    \lim_{m_t\rightarrow\infty}C^H(q_a,q_b)&=i\frac{A}{2}\left(\left|p_{H\perp}\right|^2-|q_{a\perp}|^2-|q_{b\perp}|^2\right),\nonumber\\
    A&=\frac{\alpha_s}{3\pi v}.    \label{CHlim}
\end{align}
As one would expect, this is the same result one would obtain by starting
directly from the effective Lagrangian in this limit, which was used to
obtain the fixed-order results of the previous section. We will adopt the
large top-mass limit from now on, in order to make direct comparison with the
fixed order results. However, it is clear that the limit of large top-mass is
not a necessary ingredient of our approach. We will also multiply the
effective vertex for Higgs boson production with the same factor of
Eq.~(\ref{K}) used in the fixed-order analyses.

As stated above, Eq.~(\ref{FKL}) applies to the case when the Higgs boson is
produced with a rapidity inbetween partons 0 and $n+1$. When this is not the
case, a similar factorised form still applies, but with one of the jet impact
factors (that of the jet closest in rapidity to the Higgs boson) replaced by
the impact factor for combined Higgs boson plus one jet
production\cite{DelDuca:2003ba}.
\subsubsection{Contribution from Processes with a suppressed High Energy Limit}
\label{sec:contr-from-proc}
The above amplitudes dominate the cross-section for Higgs boson production
with multiple partons in the MRK limit, which gives the leading logarithmic
contribution to the cross section as $s_{ab}\to\infty$ with fixed scattering
momentum. It is clear that only some of the possible partonic configurations
are present i.e. those of the form:
\begin{equation}
\alpha(p_a) + \beta(p_b) \to \alpha(p_0) + \sum_{i}^{n-1} g(p_i) + \beta(p_n)+h(p_h)\label{parts}
\end{equation}
where $\alpha,\beta\in\{q,\bar q,g\}$, and the partons are ordered 
according to increasing rapidity in both the initial and final states (but
the Higgs boson rapidity is unconstrained). 
Thus, there are two types of 
contribution to the scattering amplitude which are absent. Firstly, 
those containing final state partons other
than the incoming species and multiple gluons. Secondly, diagrams with the external parton content
of Eq.~(\ref{parts}), but where the final state partons are not ordered in increasing rapidity.
These possibilities are shown schematically in Figure~\ref{partsfig}.
\FIGURE[tb]{
\epsfig{file=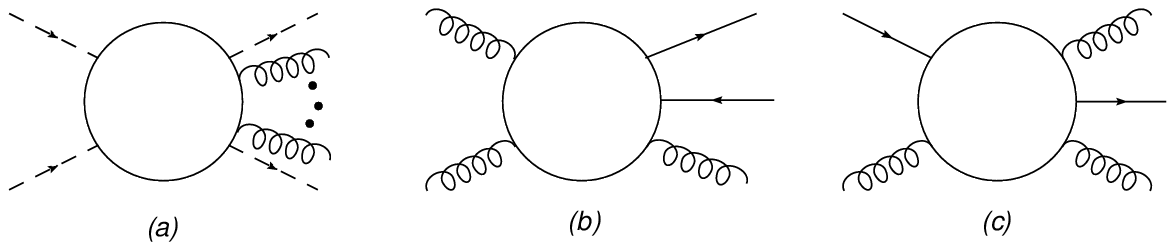}
\caption{Schematic representation of the $h$ $+$ multiparton scattering amplitude, where 
partons emerging from the amplitude are ordered in increasing rapidity, showing:
(a) a non-suppressed configuration in the MRK limit, which contributes to the
LL FKL amplitude; 
(b) a partonic configuration not included in the LL FKL factorisation; (c) a partonic
configuration with the right parton content, but with incorrect rapidity ordering. 
Dashed lines represent (anti-)quarks or gluons, and the Higgs boson 
is not shown.\label{partsfig}}}
In order to be able to base approximate scattering amplitudes on the FKL description, one must first
check that the missing partonic subprocesses do not give a significant contribution to the
cross-section in the fixed order results. In
Section~\ref{sec:central-higgs-+} we evaluated $hjj$-production at leading
order using the cuts of Table~\ref{tab:cuts} and the standard choice of
scales. The result for the total cross section was 230fb. We have explicitly checked
the contribution to this arising from non-FKL like parton configurations, and the result
is 0.7fb, i.e.~less than 0.3\%.

The non-FKL contributions are so heavily suppressed in the two-jet channel
because of the requirement of a large rapidity separation between the two
jets. However, there is no requirement of a large separation between all
three jets for the production of a Higgs boson in association with three
jets. Here, there is a leading logarithmic FKL contribution to e.g.~$qQ\to hqgQ$ but not $qQ\to
hqQg$ (with ordering of rapidity as indicated), and there is no contribution
at all to channels such as $q\bar q\to hggg$. We will denote the contributing
channels as \emph{FKL configurations}. In Section
\ref{sec:production-higgs-+} we evaluated the tree-level cross section for
$hjjj$-production within the ``inclusive'' cuts to be 203fb. The contribution
from non-FKL configurations is 19fb, i.e.~less than 10\%, even with no
requirement of a large invariant mass between all three jets. When requiring
the two hardest jets to satisfy the jet cuts, we found the cross section for
$hjjj$-production to be 76fb. Within these cuts, the contribution from
processes with a suppressed MRK limit is 6.0fb, less than 8\%. One
could worry about the growing trend of missing contributions. This arises as
a result of calculating the contribution from an increasing number of jets
within a fixed rapidity interval (e.g.~the detector coverage), which
automatically decreases the maximum invariant mass between each jet. Within
the description of factorised amplitudes, many subleading channels could be
included by directly applying the Feynman rules for Reggeised particle
exchanges as derived in Ref.\cite{Bogdan:2006af} and references
therein. However, in the present study we will consider only the same underlying
processes which enter the leading logarithmic BFKL resummation scheme, as a
first test of the importance of the improvements we can make to the
analytic behaviour of each amplitude.

We have established that the partonic channels included in the approximation
scheme indeed do dominate the cross section within the cuts of
Table~\ref{tab:cuts} (and in Section~\ref{sec:jet-activity} we will see that
the description is equally good for less stringent cuts). It is perhaps
surprising that the approximation works so well also in the three-jet case,
but the requirement of individual jets could act as a sufficient requirement
of a minimum invariant mass between partons to ensure the dominance of the
terms taken into account by the FKL approximation.

\subsection{Connection with the BFKL equation}
\label{sec:connection-with-bfkl}
The FKL result and the factorisation of amplitudes is proven in the MRK
limit. The question remains of how to apply this result to the calculation of
radiative corrections outside this limit i.e. with no restriction 
on inter-parton invariant masses.

In this section we will describe how the FKL framework is traditionally
implemented through the use of the BFKL equation. We will compare the
description so obtained order by order with the results obtained from a full
tree level calculation of the production of a Higgs boson in association with
two or three jets. We will implement the relevant BFKL description of this
process using the formalism developed in Ref.\cite{Andersen:2006sp},
respecting energy and momentum conservation.

In the MRK limit, the virtual momenta $q_i$ are dominated by their
transverse components such that $t_i\simeq q_{i\perp}^2$. Thus, it is
conventional to neglect the longitudinal components of the $q_i$'s when
evaluating the propagators. Starting from a lowest order $n$-parton amplitude,
obtained by setting all exponential factors to unity in
Eq.~(\ref{FKL}), the emission of an additional gluon $i$ between gluon $j$ and
$j+1$ leads to a change in all the momenta, but also to the emergence
of an extra factor in the squared amplitude:
\begin{align}
  \label{eq:FKLextraemissionfactor}
  \frac{-C^{\mu_i}\cdot C_{\mu_i}}{t_i\ t_{i+1}}
\end{align}
In the MRK limit, with all longitudinal degrees of freedom suppressed,
this factor becomes
\begin{equation}
\frac{-C^{\mu_i}\cdot C_{\mu_i}}{t_i\ t_{i+1}}\to \frac 4 {\left|p_{i\perp}\right|^2}
\label{liplimit}
\end{equation}
Taking into account the colour factors and couplings, the effect of one
emission (apart from a change in momenta to account for overall
energy and momentum conservation) then reduces to an extra factor in the
squared matrix element (summed and averaged over colours and spins) of:
\begin{align}
  \label{eq:BFKLemissionfactor}
  \frac{4\ \gs^2\ \ca}{\left|p_{i\perp}\right|^2}
\end{align}
The \emph{BFKL} approximation for the $\alpha_s^4$ term of the colour and
spin summed and averaged matrix element squared for $gg\to hgg$ with the
Higgs boson produced with a rapidity between that of the jets is (see also
Ref.\cite{DelDuca:2003ba}):
\begin{align}
  \label{eq:BFKLgghgg}
  \left|\mathcal{M}^{gg\to hgg}\right|^2=\frac {4\hat s^2} {\Nc^2-1}\frac{\ca
    \gs^2}{|p_{0\perp}|^2}\left| C^H\left(-p_{0\perp},p_{1,\perp}\right)\ \right|^2 \frac{\ca \gs^2}{|p_{1\perp}|^2},
\end{align}
and the approximation for the $\alpha_s^5$ term of the colour and spin summed
and averaged  matrix element square for $gg\to hggg$ is:
\begin{align}
  \label{eq:BFKLgghggg}
  \left|\mathcal{M}^{gg\to hggg}\right|^2=\frac {4\hat s^2} {\Nc^2-1}\frac{\ca
    \gs^2}{|p_{0\perp}|^2}\ \left|
    C^H\left(q_{a\perp},q_{b,\perp}\right)\ \right|^2 \frac{4\ \ca \gs^2}{|p_{1\perp}|^2} \frac{\ca \gs^2}{|p_{2\perp}|^2}.
\end{align}
In Eqs.~\eqref{eq:BFKLgghgg}-\eqref{eq:BFKLgghggg},
$q_{a\perp}=-\sum_{i=0}^jp_{i\perp}$ where $j$ counts the number of partons
with a rapidity smaller than that of the Higgs, and $q_{b\perp}=q_{a\perp}-p_{H\perp}$.

The soft divergence for $p_{i\perp}\to 0$ in the amplitudes of
Eqs.~\eqref{eq:BFKLgghgg}-~\eqref{eq:BFKLgghggg} and their obvious
multi-gluon generalisations is regulated by the soft-gluon divergence in the
Reggeised propagators, as discussed in Sec.~\ref{sec:regul-ampl}.

While the original factorisation of the amplitudes, and the form of
the invariants used in deriving the results in
Eqs.~(\ref{eq:BFKLgghgg})-(\ref{eq:BFKLgghggg}) are valid only in the MRK
phase space region of
\begin{equation}
y_0\gg y_1\gg \ldots\gg y_{n+1};\quad p_{i\perp}\simeq p_{i+1\perp};\quad q_{\perp i}^2 \simeq q_{\perp j}^2\, ,
\label{BFKL-MRK}
\end{equation}
in the BFKL equation they are applied to the fully inclusive phase space,
where the constraint on large rapidity separations between all partons is
dropped.  The simple form of Eqs.~(\ref{eq:BFKLgghgg})-(\ref{eq:BFKLgghggg})
(generalised to all orders),
allows for the calculation of the approximate sum over $j$ and $n$ and the
infinite phase-space integral of the emitted gluons in the squared amplitude of
Eq.~(\ref{FKL}). The partonic cross section for e.g.~$gg\to g\cdots
h\cdots g$ as a function of the momenta of the Higgs boson and the extremal
partons only then takes the following form:
\begin{align}
  \begin{split}
    \label{eq:BFKLapproximation}
    &\frac{d\hat\sigma_{gg\to g\cdots h\cdots g}}{dp_{a\perp}^2dy_a\ dp_{b\perp}^2dy_b\ dp_{H\perp}^2dy_H}\\
    &=\int{d^2 q_{a\perp}d^2 q_{b\perp}} \left(\frac{\alpha_s\ N_c}{p_{a\perp}^2}\right)
    f(-p_{a\perp},q_{a,\perp},\Delta y_{aH}) \left|\
      C_{HEL}^H\left(q_{a,\perp},q_{b,\perp}\right)\ \right|^2
    f(q_{b\perp},p_{b,\perp},\Delta y_{Hb}) \left(\frac{\alpha_s\
        N_c}{p_{b\perp}^2}\right),
  \end{split}
\end{align}
where $f(q_{b\perp},k_{b,\perp},\Delta y_{Hb})$ is the solution of the BFKL
equation, which has the form:
\begin{align}
  \label{eq:BFKLeqn}
\omega \ f_\omega\! \left({\bf k}_a,{\bf k}_b\right) = \delta^{(2+2\epsilon)} 
\left({\bf k}_a-{\bf k}_b\right) + \int \mathrm{d}^{2+2\epsilon}{\bf k} ~
\mathcal{K}_\epsilon\!\left({\bf k}_a,{\bf k}+{\bf k}_a\right) \ f_\omega\!\left({\bf k}+{\bf k}_a,{\bf k}_b \right).
\end{align}
Here $\mathcal{K}_\epsilon$ is the BFKL kernel, and ${\bf k}$ the
two-dimensional transverse part of 4-momentum $k$ (i.e. $k_\perp^2=-{\bf
  k}^2$). The implicit integrations over all of (rapidity ordered) momenta
for the emitted gluons performed when solving the BFKL equation to find
$f(q_{b\perp},k_{b,\perp},\Delta y)$ can create a problem though. In
Eq.~(\ref{eq:BFKLapproximation}), a factor of $\hat s$ has been cancelled
between the approximate squared matrix element and the flux factor in the
partonic cross section. This might leave the impression that the resulting
(partly integrated) partonic cross sections of
Eq.~(\ref{eq:BFKLapproximation}) do not depend on the momenta of the incoming
particles. For a given final state configuration, the incoming momenta are of
course given by momentum conservation. However, the final states which are
integrated and summed over to arrive at Eq.~(\ref{eq:BFKLapproximation})
arise from different initial state momenta, even if these cannot be
reconstructed after the solution to the BFKL equation has been
substituted. This is the problem of energy and momentum conservation in the
BFKL formalism.

In order to obey factorisation when calculating hadronic cross
sections, the parton distribution functions (PDF) must be evaluated at the
light-cone momentum fractions of the incoming partons, which are relevant for
the given final state momentum configuration. This requires the PDFs to be
convoluted with the solution of the coupled BFKL equations in
Eq.~(\ref{eq:BFKLapproximation}) (thus altering the BFKL evolution), which
can be achieved by solving the BFKL equation iteratively, through the method
outlined in
Ref.\cite{Orr:1997im,Schmidt:1997fg,Andersen:2001kt}. Specifically, we choose
to follow the implementation advocated in Ref.\cite{Andersen:2006sp}. These
methods allow for a straightforward expansion of the solution in powers of
$\alpha_s$. It is thereby easy to extract the contribution proportional to
$\alpha_s^4$ for 2-parton final states, and $\alpha_s^5$ for 3-parton final
states (these obviously agree with Eq.~\eqref{eq:BFKLgghgg} and
Eq.~\eqref{eq:BFKLgghggg} respectively), and compare the results obtained for
the 2 and 3-jet final states with those obtained using the full matrix
element.

On Fig.~\ref{xsecfig_BFKL} we have compared the results for the cross section
for the production of a Higgs boson in association with two and three jets
obtained within the BFKL approach with the results obtained for the full
matrix elements in the FKL configurations (to be compared with
Fig.~\ref{h+jetsLO1}). We find that the results based on the BFKL
approximations for the $hjj$ ($458$fb) and inclusive $hjjj$ ($660$fb) cross sections
differ from their full leading order counterparts by $99\%$ and $225\%$
respectively. The kinematic approximations are clearly inadequate in
describing amplitudes in general at the LHC. The matching corrections to a
resummation based on these results would be uncomfortably large, and
encourage very little trust in the predictions based on this approximation.
\FIGURE[tb]{ \centering
  \epsfig{width=.49\textwidth,file=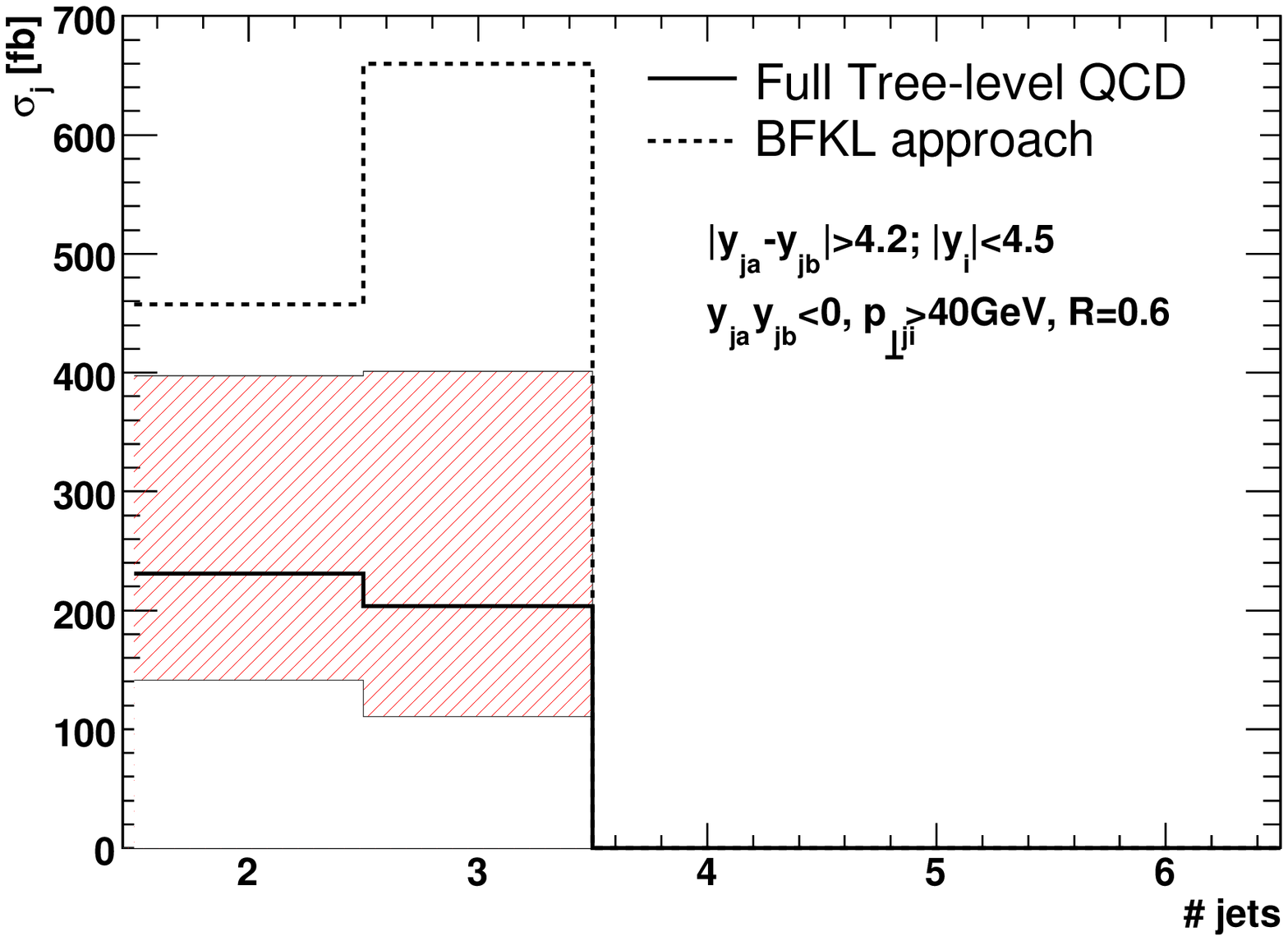}
  \epsfig{width=.49\textwidth,file=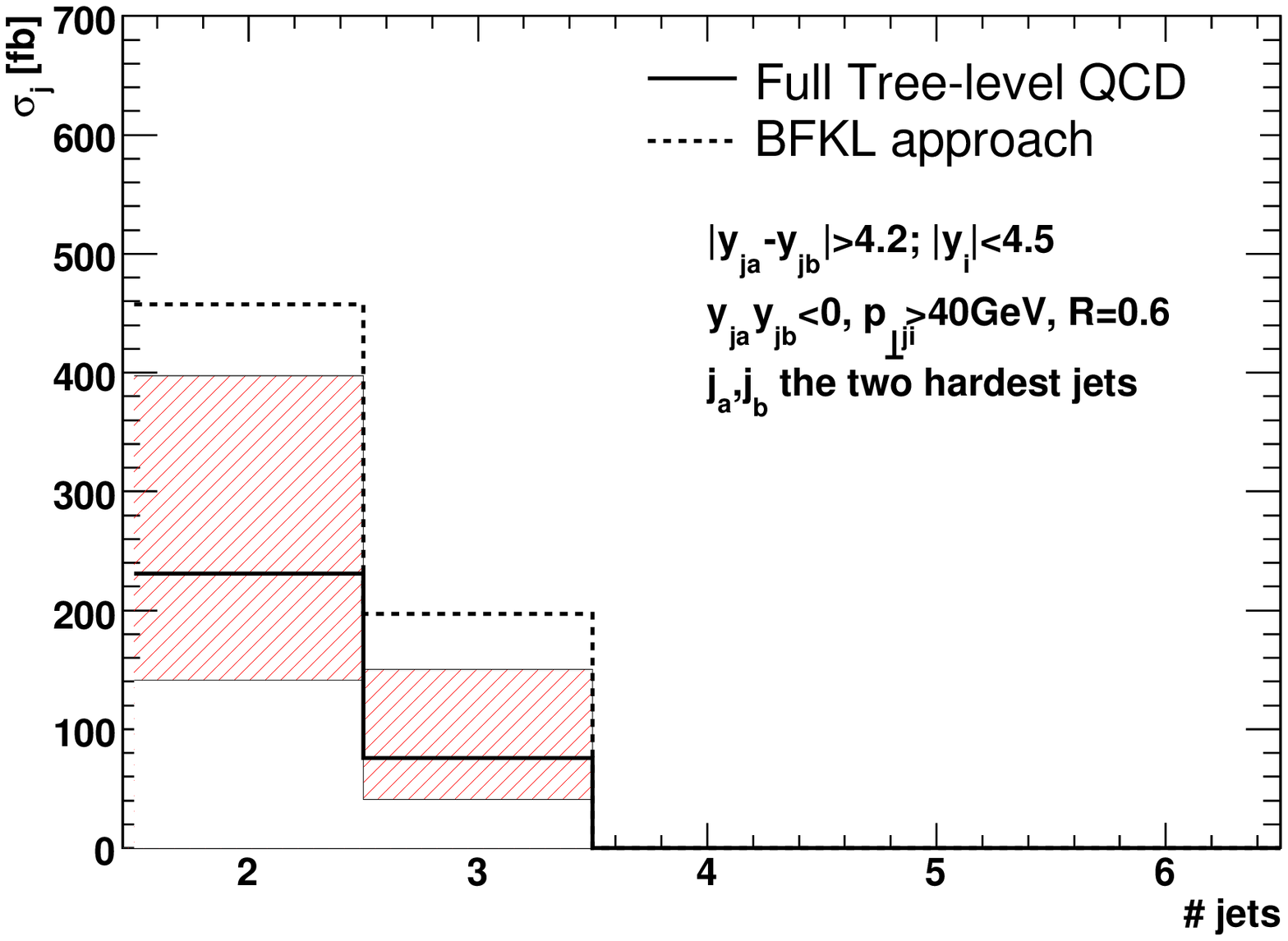}
    \caption{The 2 and 3 jet cross-sections calculated using the known LO
      matrix elements (solid) as in Fig.~\ref{h+jetsLO1}, and the BFKL
      approximation with energy and momentum conservation (dotted). The
      results are shown for the two sets of 3-jet cuts defined in
      Section~\ref{sec:production-higgs-+}. \label{xsecfig_BFKL}}
} 
We note again that there is no restriction on the rapidity separation between
the Higgs boson and any jet, nor between the rapidity of the middle (in
rapidity) jet and any other object. So the BFKL amplitudes are here, just as
in the BFKL equation, applied far outside of the MRK region.

The results of Figure~\ref{xsecfig_BFKL} were found imposing 4-momentum
conservation on the BFKL equation. However, this is traditionally neglected
in the BFKL formalism since it is formally subleading in $\ln\hat
s/\hat t$, and also because the BFKL equation explicitly requires the
longitudinal momentum dependence to be dropped, which leads to violation of
momentum conservation in all but the transverse components.
However, such results are clearly not sensible, due to the unbounded
phase space integration of the emitted gluons.
It is easy to see that capturing just the dominating behaviour of the
partonic cross section as $\hat s\to\infty$, $t$ fixed will not describe
correctly the large-invariant mass limit at a fixed energy collider: Consider
the limit $\hat{s}\to s$ (i.e. the hadronic centre of mass energy), obtained
when the final state partons of extremal rapidity become widely separated.
This limit occurs before the strict Regge limit of $\hat{s}/t\to\infty$ is
reached, and there is then no phase space left for the emission of additional
gluons, and the kinematics return to that of the lowest order. This is not
respected by any description based upon an analytic solution of the BFKL equation.

\subsection{Modified High Energy Factorised Amplitudes}
\label{sec:modified-high-energy-1}
In the previous section, we saw that the BFKL implementation of the FKL 
factorisation formula does not accurately approximate fixed order matrix elements
when applied over the phase space relevant to the LHC, even when energy and 
momentum conservation are implemented. In this section, we show that it is possible 
to modify the FKL description outside of the MRK limit, in such a way that its 
applicability can be extended. These modifications go beyond any logarithmic 
order in $\hat{s}/|t|$ in the traditional BFKL expansion. The resulting amplitudes can be used
as an approximation to scattering amplitudes with many final state partons, i.e.
at orders in \as where fixed order perturbation theory is infeasible. This is analogous to
the use of the soft- and collinear factorisation, which forms the basis of
any parton shower Monte Carlo and most resummation formulae, outside of the
strict soft and collinear region. Only outside the strict soft and
collinear regions is there any observable effect of the radiation, and one 
then hopes that the results obtained in this limit are also relevant for a
larger region of phase space. This is also the case in the present framework. However, 
because it does not rely on soft and collinear approximations, it should be better
at describing the hard jet topology in an event, as opposed to the jet substructure. This 
section elaborates on the presentation in Ref.\cite{Andersen:2008ue},
although with some modifications to the formalism that we discuss in what follows.

We take as our definition of our approximate scattering amplitudes for $h$ + 
multiparton production the FKL factorisation formula (Eq.~(\ref{FKL})), 
which is then applied subject to the following guidelines:
\begin{enumerate}
  \item Use of full virtual 4-momenta: Rather than substituting
  $q_i^2\to-q_{i\perp}^2$ as in the BFKL equation and the original work of
  Fadin and collaborators, we keep the dependence on
  the full 4-momenta of all particles. This ensures that outside of the MRK
  limit, the singularity structure of the approximate amplitudes coincides
  with known singularities of the full fixed order scattering amplitude.

\item Use of the Lipatov vertex as defined in Eq.~(\ref{lip1}). The results
  of the MRK limit constrains only the asymptotic form of the Lipatov
  vertex. Our choice for the sub-asymptotic behaviour is enforced by an added
  requirement of Lorentz invariance and fulfilment of the Ward identity
  throughout \emph{all} of phase space. The
  latter condition is expressed by $-C.C>0$, where the minus sign arises from
  the gluon polarisation tensor.
\end{enumerate}
These guidelines distinguish our approach from previous applications of the
FKL factorisation formula, and thus we refer to our amplitudes from now on as
{\it modified FKL amplitudes}. The above requirements do not impact on the
logarithmic accuracy (in $\hat{s}/t$) of the amplitudes, but enforce
constraints on the sub-asymptotic behaviour stemming from known features of
all-order perturbation theory.

\subsubsection{Results for the Modified FKL Amplitudes}
\label{sec:results-modified-fkl}
In Fig.~\ref{xsecfig} we have plotted the results obtained using the
expansions to $\mathcal{O}(\as^4)$ for $hjj$ and
$\mathcal{O}(\alpha_s^5)$ for $hjjj$ of the matrix elements in
Eq.~(\ref{FKL}), supplemented with the guidelines outlined above. 
We compare these with results obtained for the FKL configurations of
full tree level QCD (dotted line), as discussed in 
subsection~\ref{sec:contr-from-proc} and Figure~\ref{partsfig}, and with the
results of all tree-level QCD configurations (full line). The same choice of
renormalisation and factorisation scale ($m_H$) has been applied in both the
modified FKL approximation and the full QCD results. To give a sense of the
level of agreement between the FKL approximation and the full results, we
indicate with the hashed band the scale uncertainty from varying the common
renormalisation and factorisation scale by a factor of two in the full QCD
result. 
\FIGURE[tb]{
    \epsfig{width=.49\textwidth,file=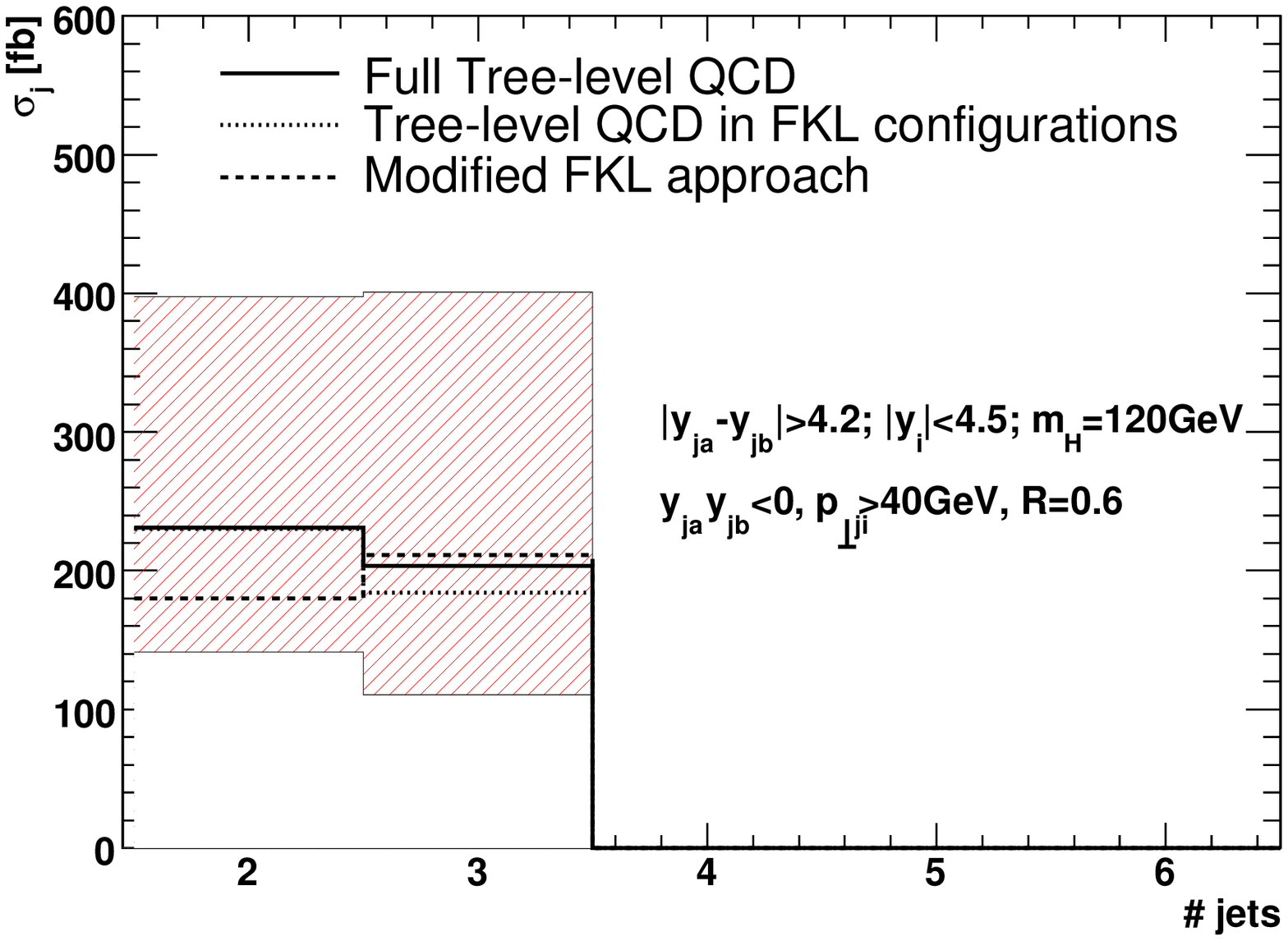}
    \epsfig{width=.49\textwidth,file=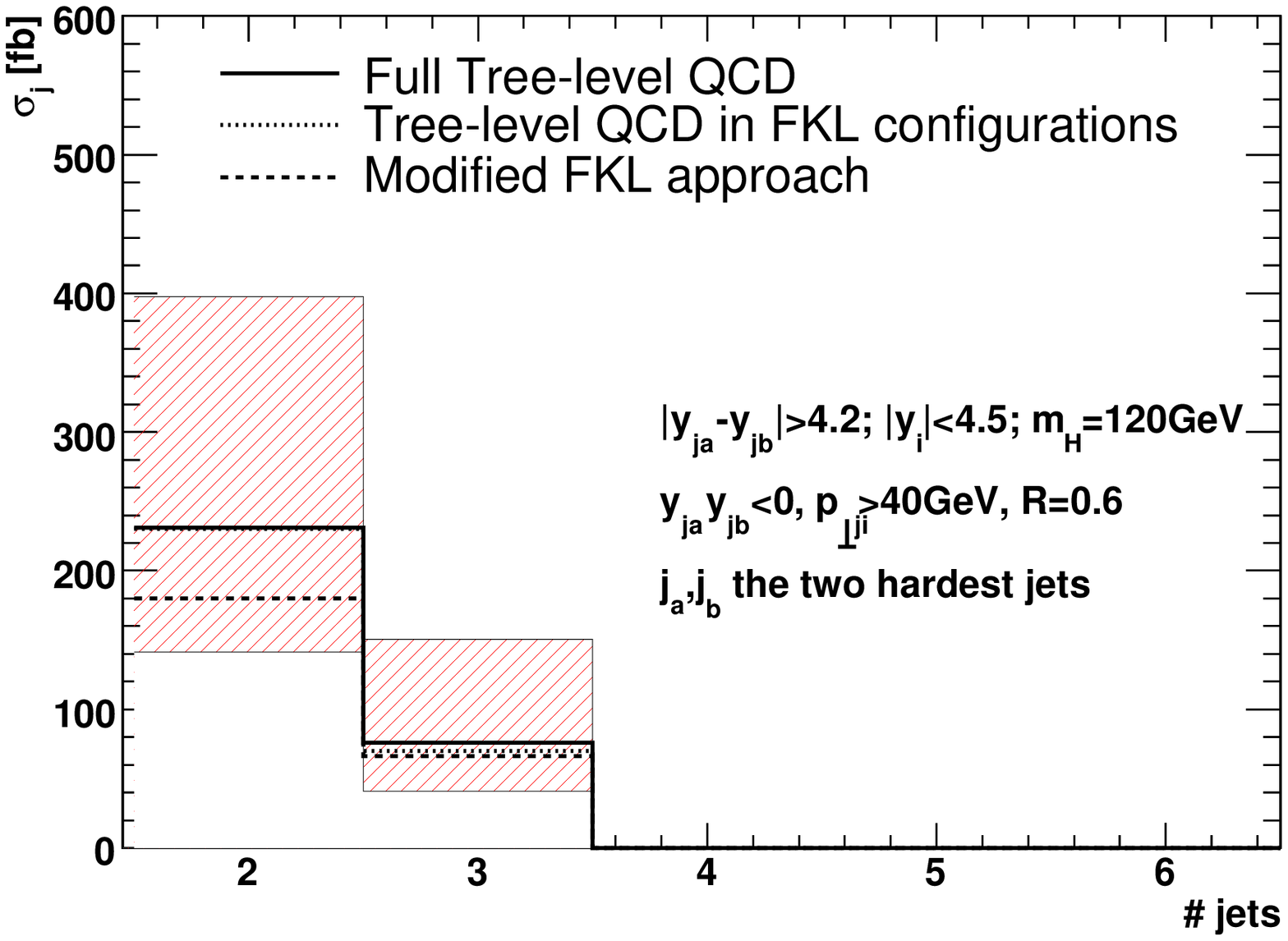}
    \caption{The LO $hjj$ and $hjjj$ cross-sections for full tree-level QCD
      (full line), and those obtained from tree-level QCD in FKL
      momentum-configurations (dotted line), compared with the results
      obtained by expanding the FKL amplitudes with the modifications of
      Sec.~\ref{sec:modified-high-energy-1} (dashed line). For two jets, the difference
      between the solid and dotted lines is not visible on this scale, and neither is,
      for the hardest jet cuts, the
      difference between the results of tree-level QCD in FKL-configuration (dotted) and the
      modified FKL amplitudes (dashed). The scale
      uncertainty relates to the tree-level QCD results.\label{xsecfig}}
}
It is clear that the modified FKL amplitudes result in a much better
approximation than those obtained using the BFKL description. The matching
corrections in the BFKL approach of the previous section would be of the
order of 100\%, whereas the use of the modified FKL amplitudes results in
matching corrections of less than 25\% compared to both the result for FKL
configurations only, and the sum over all subprocesses and rapidity
configurations (i.e.~the full result at LO). One sees that the LO $hjj$ and $hjjj$
cross-sections (in FKL configurations) are produced to within 22\% and 15\%
respectively in the case of the inclusive cuts, and within 22\% and 6\%
respectively in the case of the hard cuts.
We stress that the good agreement obtained is not specific to the particular choices
made for either the cuts, or the renormalisation and factorisation scales. In
Sec.~\ref{sec:jet-activity} we will discuss results obtained with the
requirement of a rapidity span reduced to 2 units of rapidity.
The very good level of approximation obtained using the modified FKL
prescription is important for several reasons: It gives a viable platform for
building a resummation scheme, and it demonstrates that at least for the hard
parton configurations considered here, there is no other source of large
systematic effects not taken into account by the current prescription.

In the case of the Higgs boson produced in-between the jets (in rapidity),
the three jet rates are in fact produced to a slightly better accuracy than
the two-jet rates. This might seem strange, since the factorised three-jet
formula builds on the factorised two-jet formula, and so one could expect the
approximations to worsen order by order. However, in the $hjj$-sample, the
jets can have very different transverse scales, leading to a large variation
of the $t_i$'s. In the three-jet sample, however, large transverse momenta
are increasingly suppressed by constraints from the PDFs; the requirement of
an extra hard jet limits the available transverse phase space. Therefore, the
results for three-jet configurations approximate the full QCD better than the
results obtained with the two-jet sample. A much better agreement for the two
jet case arises when the Higgs boson lies outside the two jets in rapidity.
The resulting FKL amplitude contains an impact factor for jet + Higgs boson
production separated from another jet by the exchange of a single $t$-channel
Reggeised gluon with the transverse momenta on either side of the $t$-channel
propagator balancing, and the two-jet approximation is then found to be
better than 1\%! It is no surprise that the approximation works better in
such cases; the factorisation assumes an infinite invariant mass between the
system of one jet and a jet with the Higgs boson, rather than a infinite
invariant mass between all particles. We note in passing that the mass of the
Higgs boson introduces an extra scale to the problem, which is actually
expected to worsen the quality of the approximations over the situation in a
pure jet study.

One must also check that kinematic distributions are well approximated by the formalism,
and we show here some sample results for the LO $hjj$ and $hjjj$ channels using the
inclusive cuts of  table \ref{tab:cuts}. In figure \ref{yjetsfixed} we show the rapidity
distributions of the extremal partons, for the LO $hjj$ and $hjjj$.
One sees that the modified FKL formalism agrees well with the full tree level results. 
\FIGURE[tb]{
    \epsfig{width=.49\textwidth,file=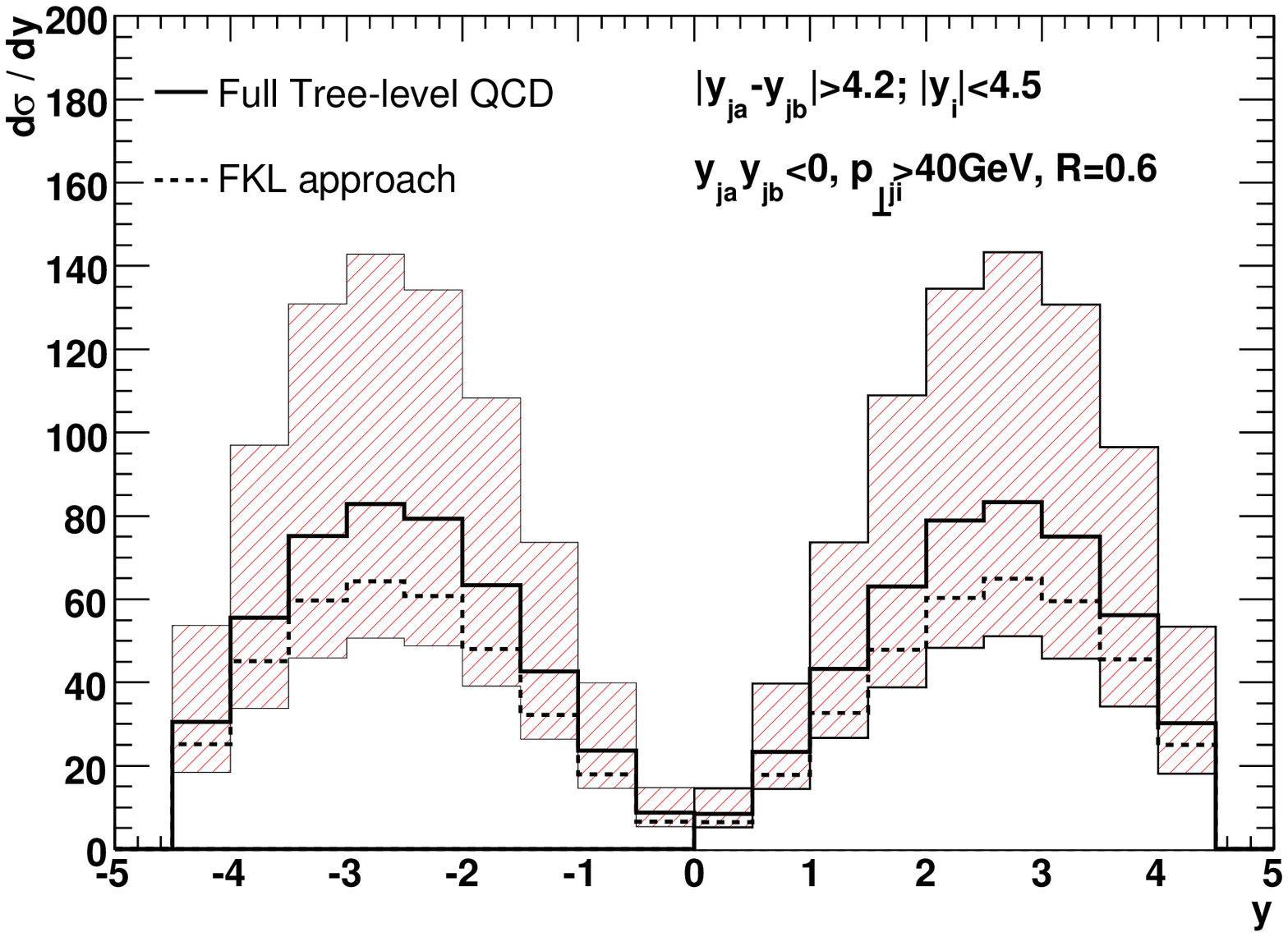}
    \epsfig{width=.49\textwidth,file=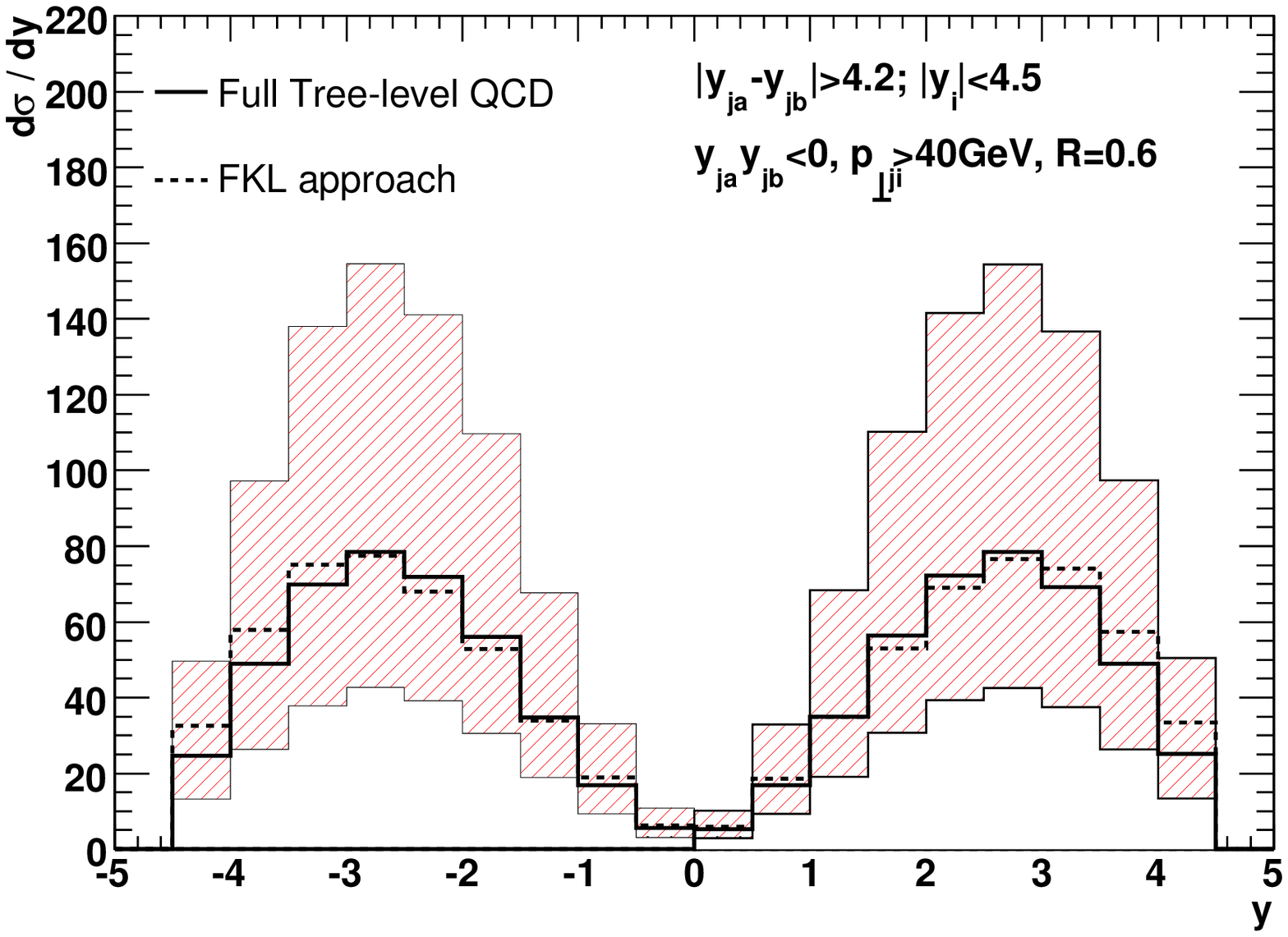}
    \caption{The rapidity distributions of the extremal partons, for the LO
  $hjj$ (left) and $hjjj$ (right) channels. The scale variation corresponds to the full tree
   level results.\label{yjetsfixed}}
}
Similarly, the rapidity distribution of the central parton in $hjjj$ is shown in figure \ref{y3},
and that of the Higgs boson (for both $hjj$ and $hjjj$) in figure \ref{yHfixed}.

\FIGURE[tb]{
  \scalebox{0.5}{\includegraphics{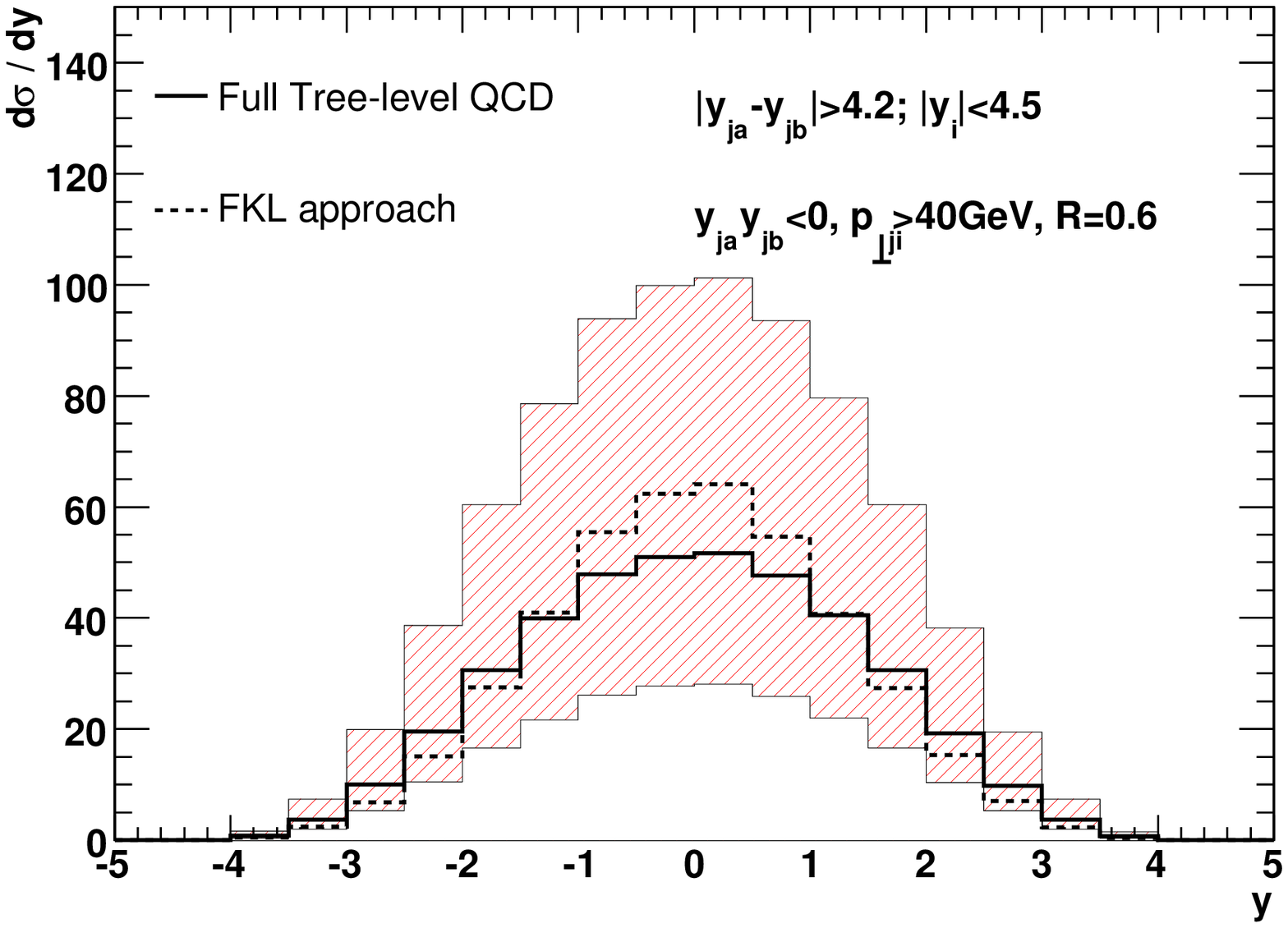}}
    \caption{The rapidity distribution of the central parton in the $hjjj$ channel at LO. 
The scale variation corresponds to the full tree level result.\label{y3}}
}

\FIGURE[tb]{
    \epsfig{width=.49\textwidth,file=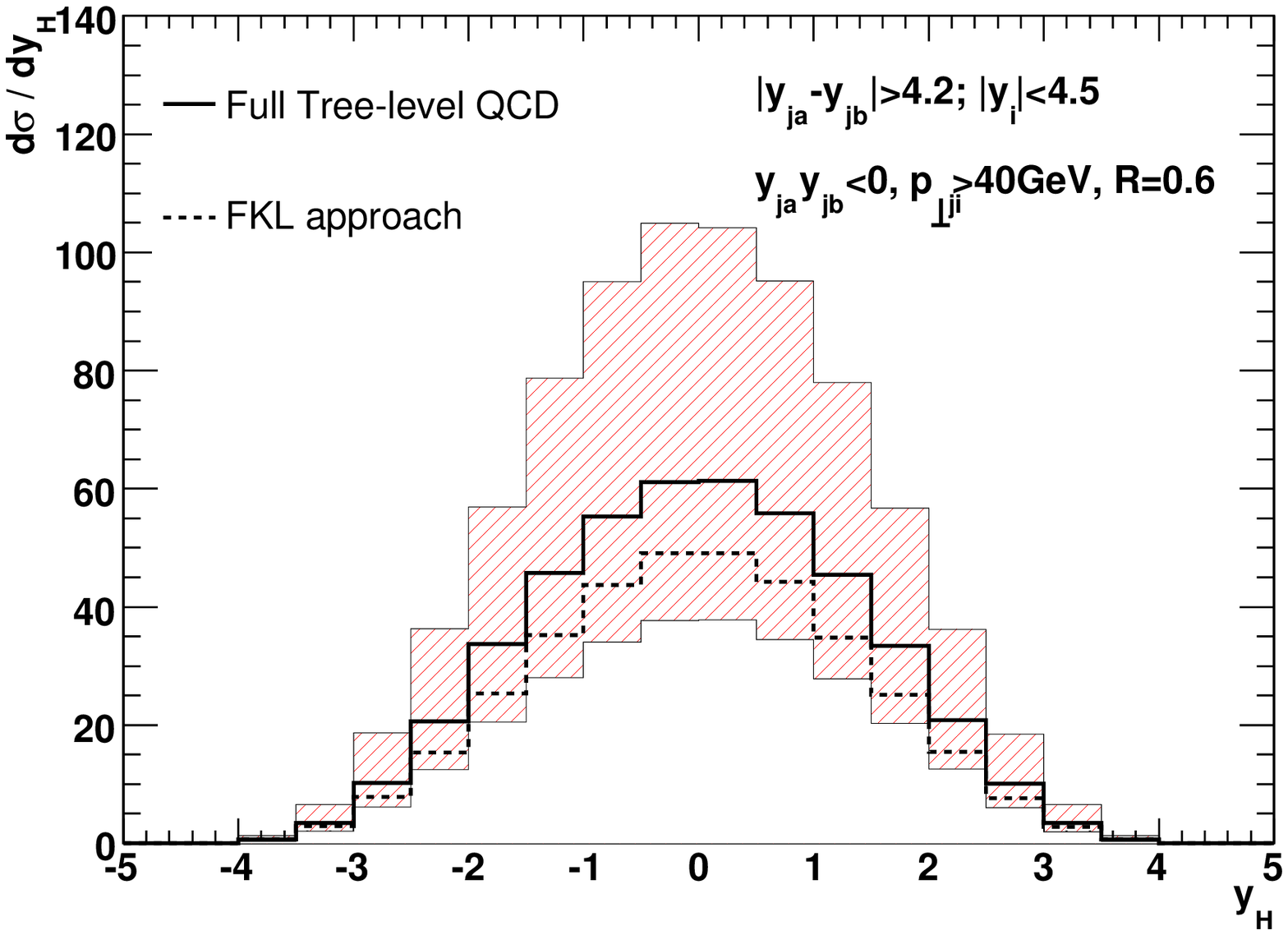}
    \epsfig{width=.49\textwidth,file=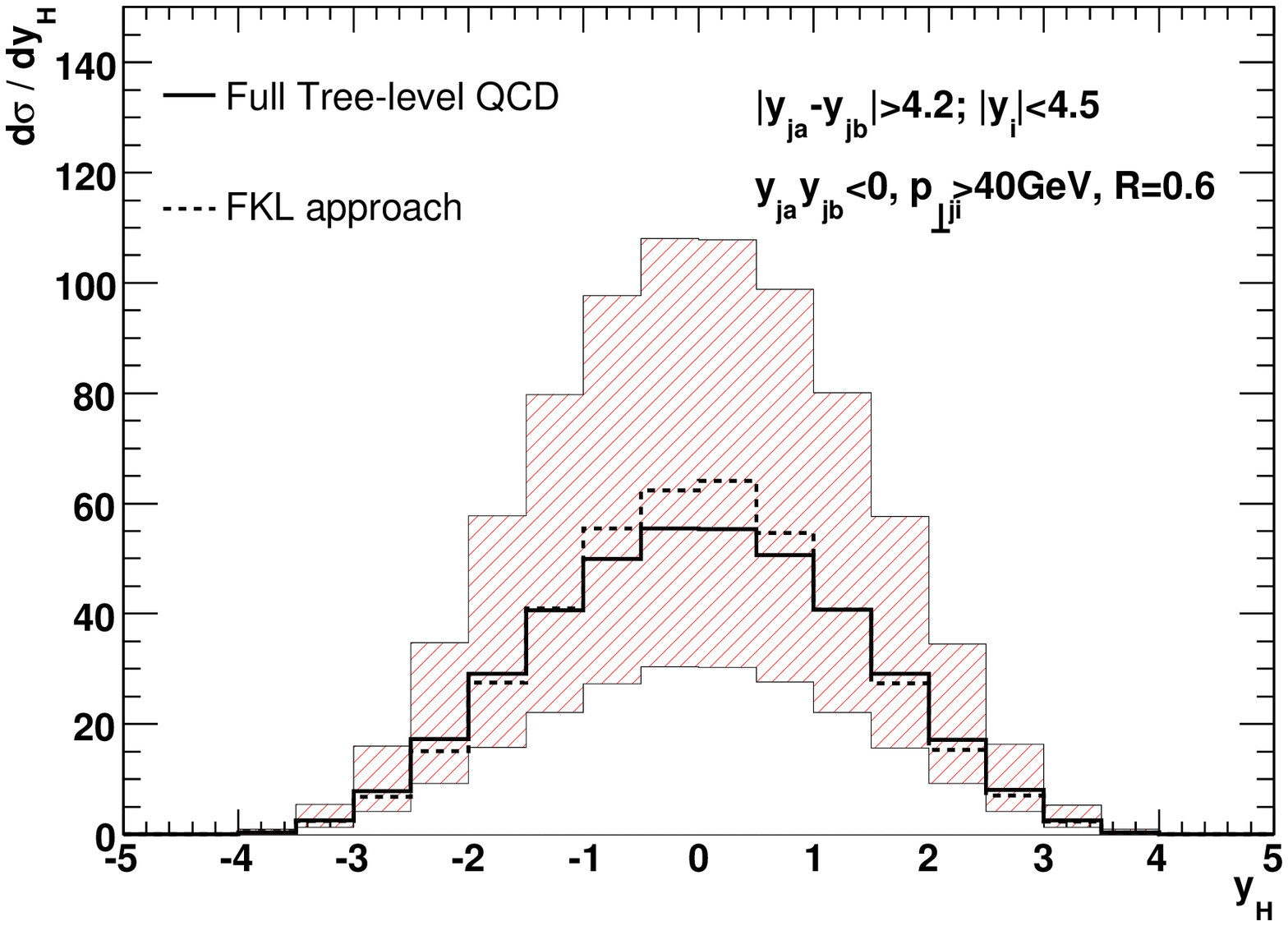}
    \caption{The rapidity distributions of the Higgs boson, for the LO
  $hjj$ (left) and $hjjj$ (right) channels. The scale variation corresponds to the full tree
   level results.\label{yHfixed}}
}
In figures \ref{ptafixed}, \ref{pt3fixed} and \ref{ptH} we show the transverse momentum
distributions for the extremal partons, central parton (in $hjjj$) and Higgs
boson respectively. The shape of the Higgs boson $p_t$ spectrum is discussed further in section \ref{sec:transv-moment-spectr-1}.
For now we note that there are large qualitative differences between the result for
$hjj$ and $hjjj$. This is shown in figure \ref{phifixed}.
\FIGURE[tb]{
    \epsfig{width=.49\textwidth,file=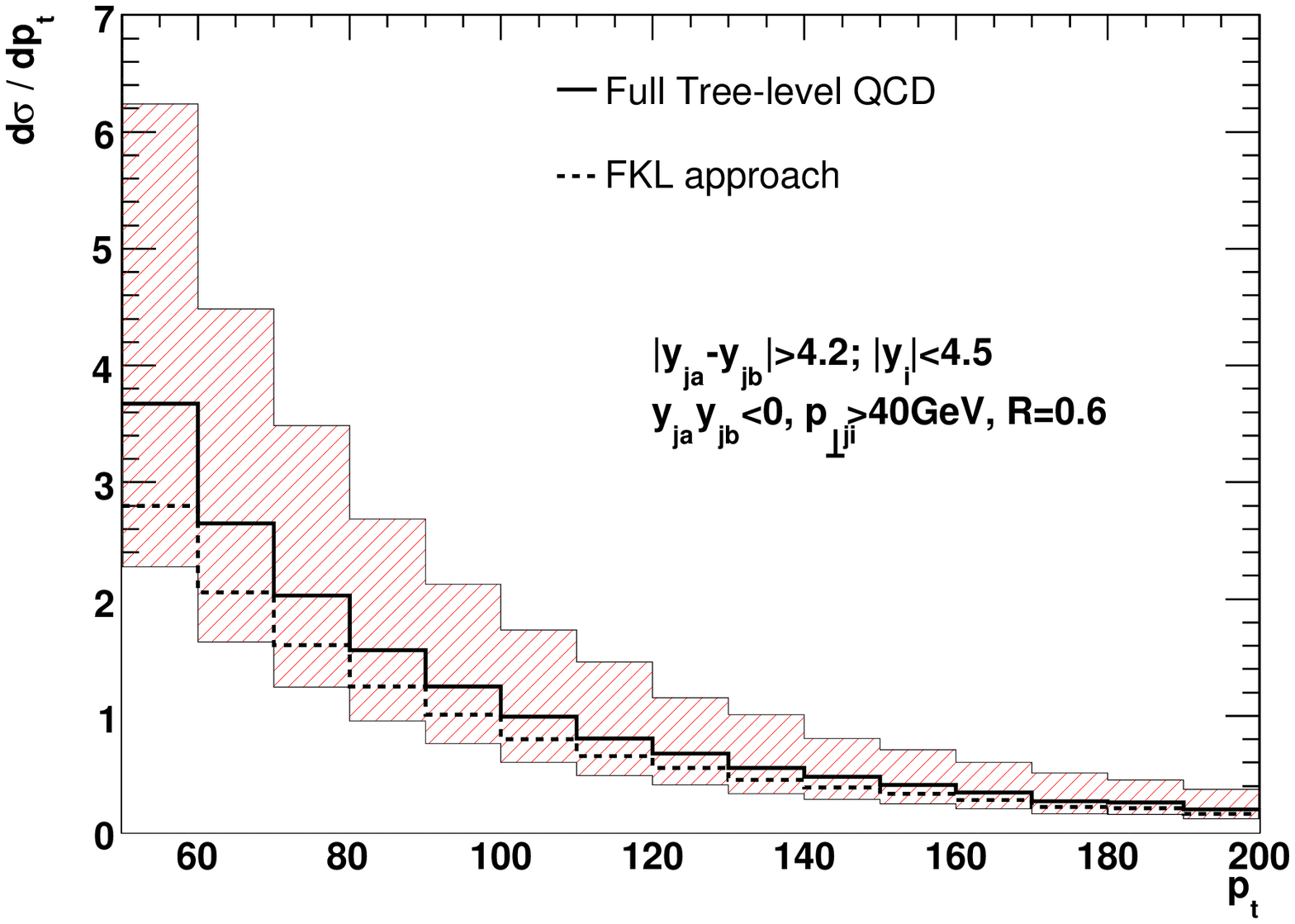}
    \epsfig{width=.49\textwidth,file=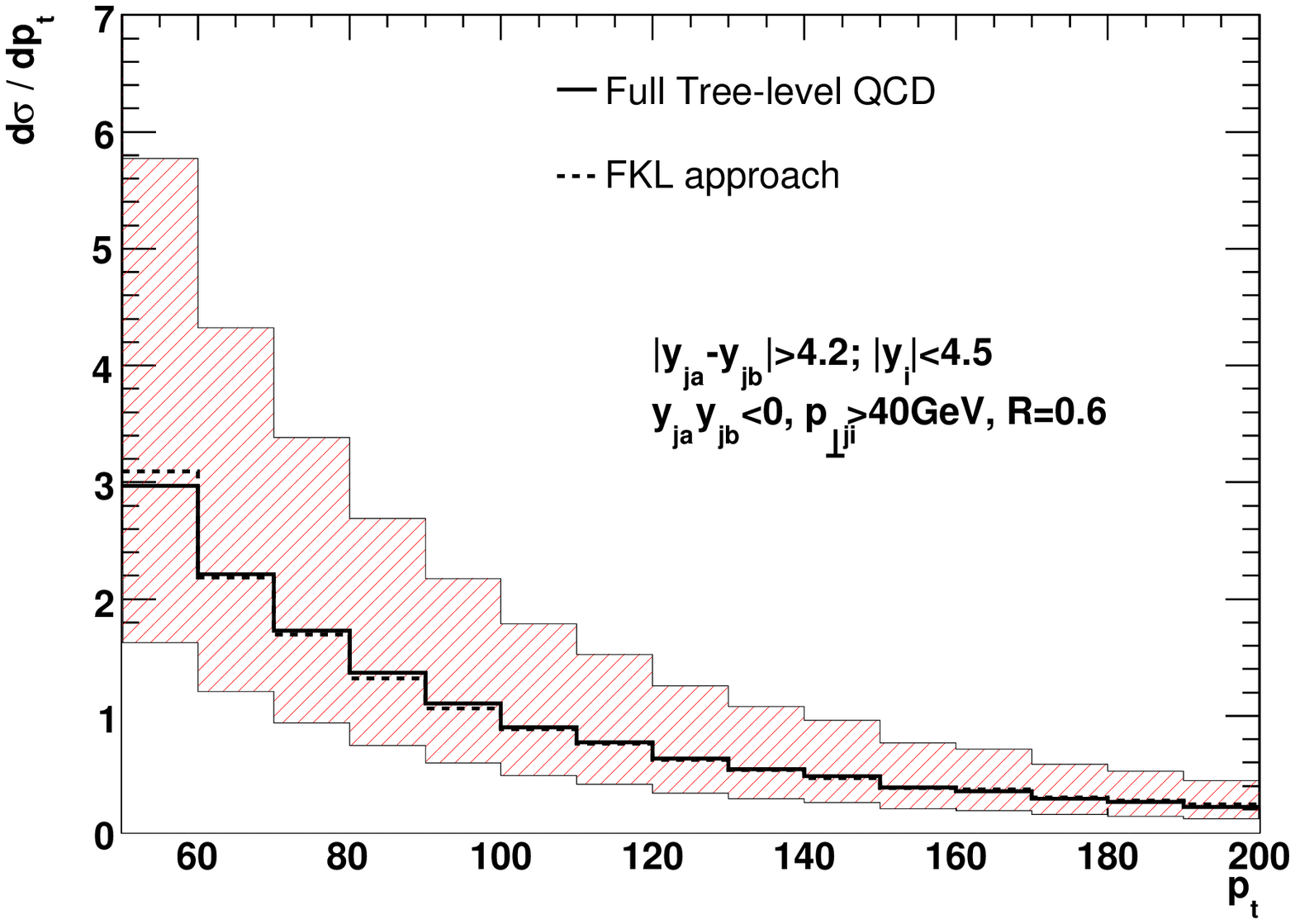}
    \caption{The transverse momentum distributions of the extremal partons, for the LO
  $hjj$ (left) and $hjjj$ (right) channels. The scale variation corresponds to the full tree
   level results.\label{ptafixed}}
}
\FIGURE[tb]{
    \epsfig{width=.49\textwidth,file=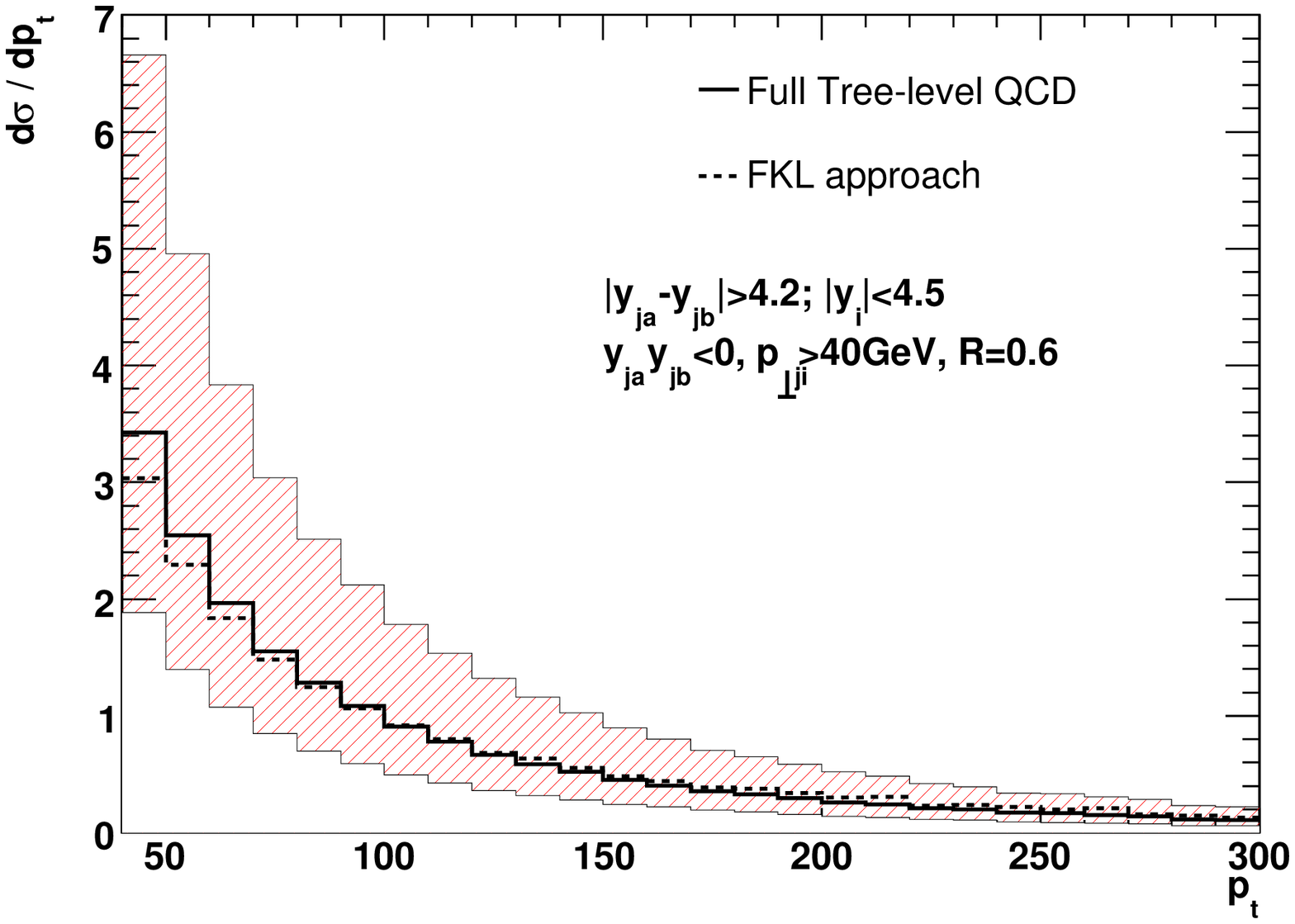}
    \caption{The transverse momentum distribution of the central parton in the $hjjj$ channel at LO. 
The scale variation corresponds to the full tree level result.\label{pt3fixed}}
}
\FIGURE[tb]{
    \epsfig{width=.49\textwidth,file=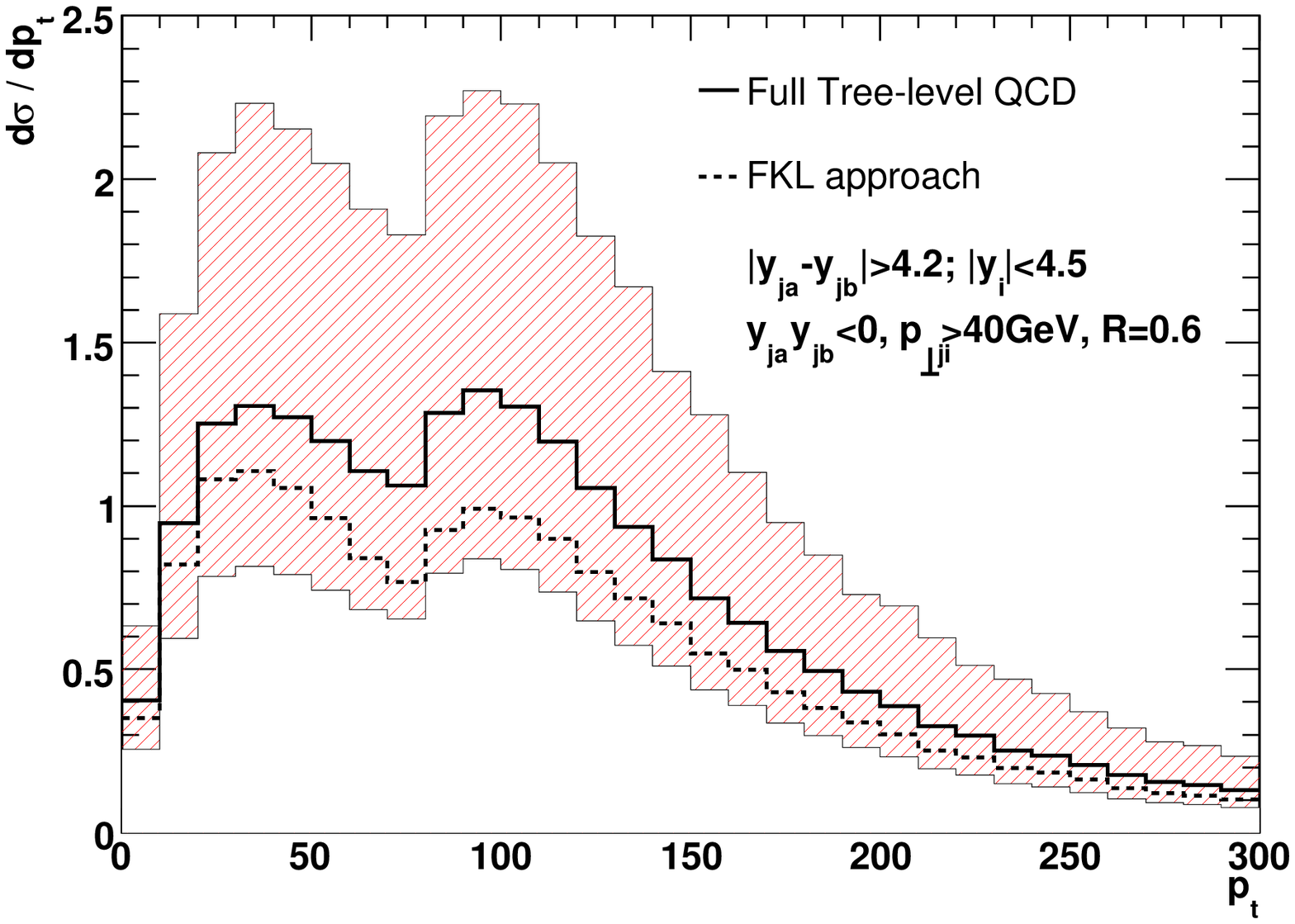}
    \epsfig{width=.49\textwidth,file=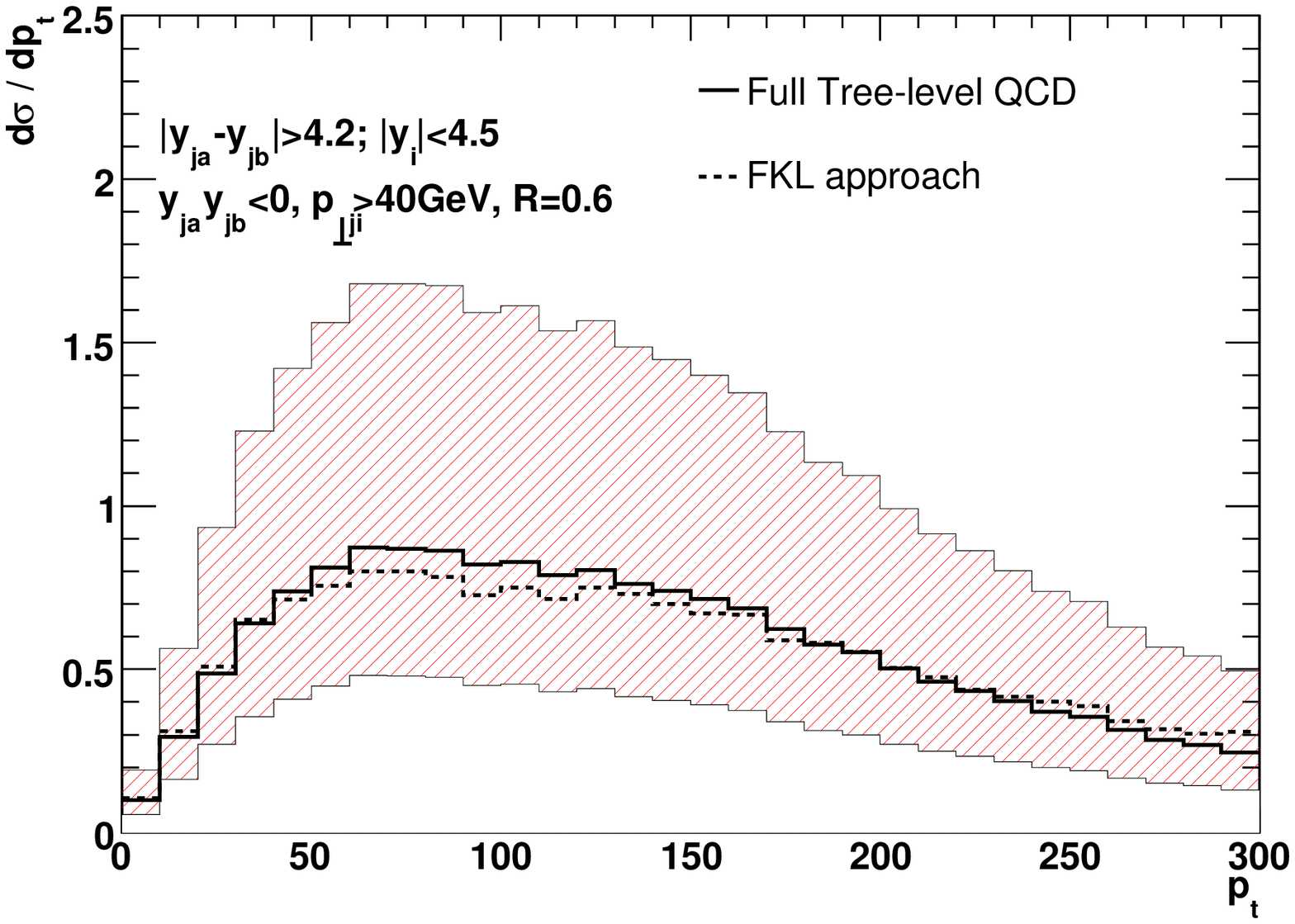}
    \caption{The transverse momentum distributions of the extremal partons, for the LO
  $hjj$ (left) and $hjjj$ (right) channels. The scale variation corresponds to the full tree
   level results.\label{ptH}}
}
\FIGURE[tb]{
    \epsfig{width=.49\textwidth,file=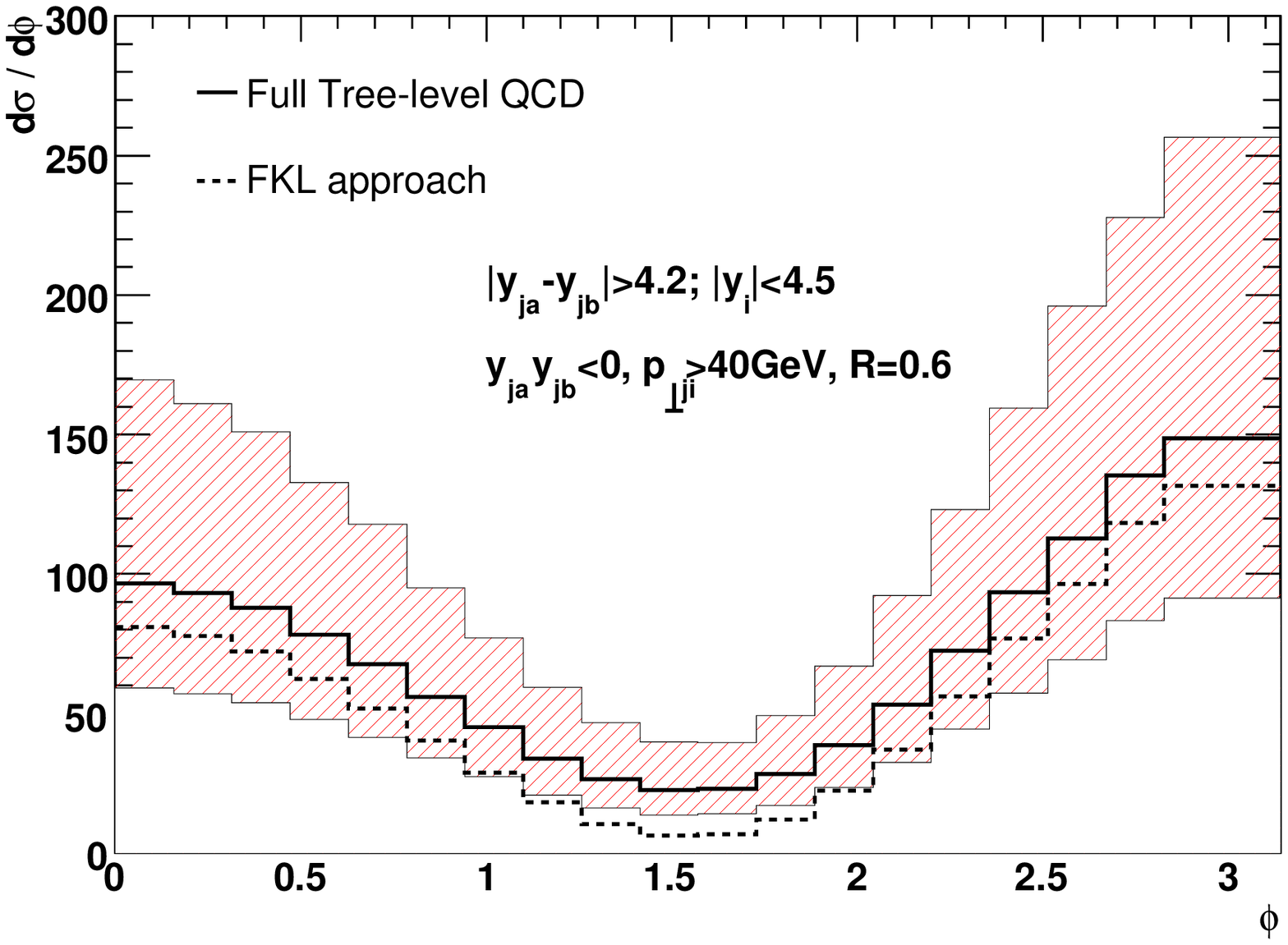}
    \epsfig{width=.49\textwidth,file=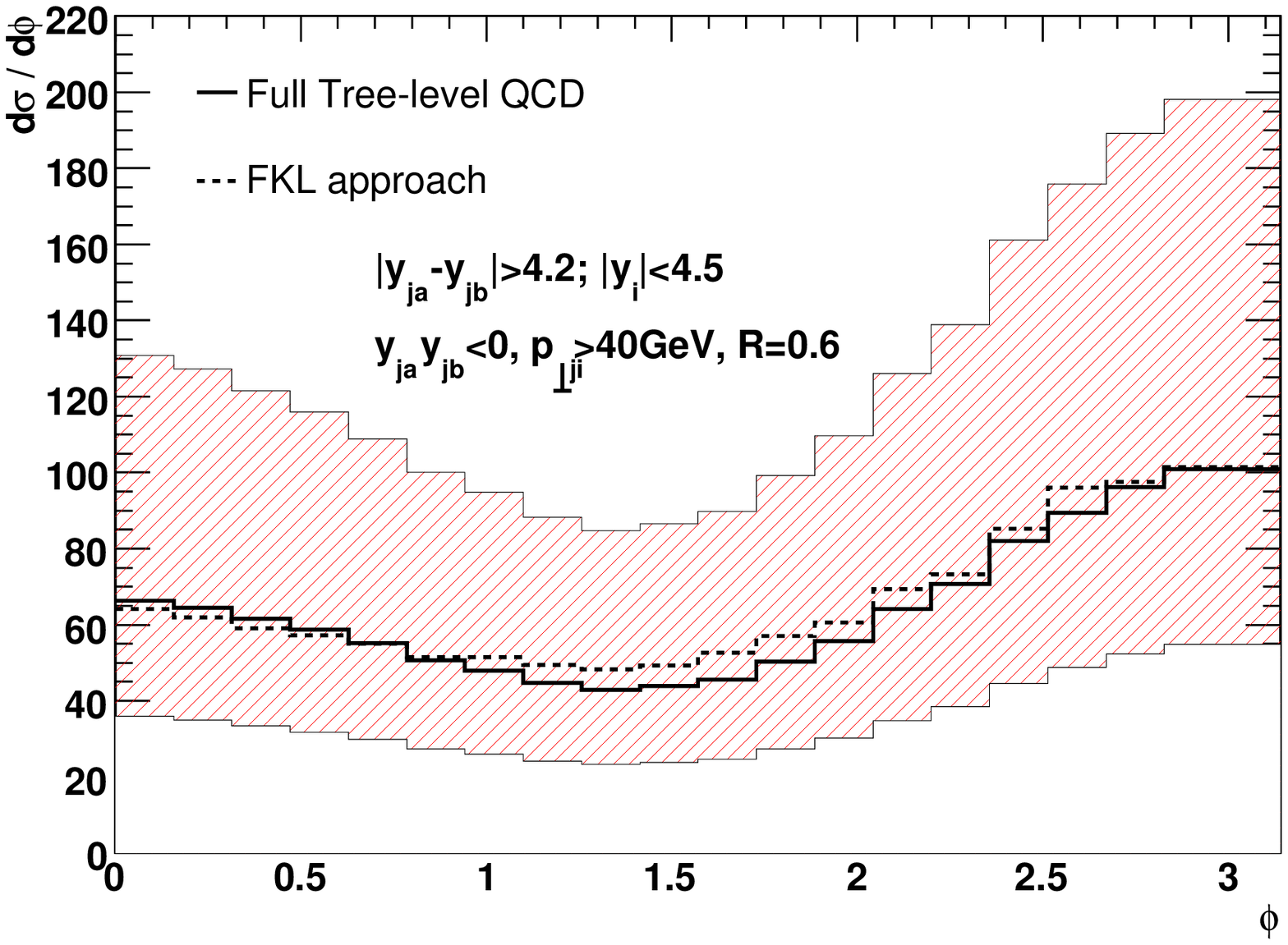}
    \caption{The azimuthal angle between the extremal partons, for the LO
  $hjj$ (left) and $hjjj$ (right) channels. The scale variation corresponds to the full tree
   level results.\label{phifixed}}
}
For each of the observables above, we find that shapes of distributions are generally well-estimated
by the modified FKL formalism. Note that although the results in figures 
\ref{yjetsfixed}-\ref{phifixed} have been presented for one choice of cuts (the inclusive cuts of table \ref{tab:cuts}),
we checked that a similar level of agreement was obtained for the other cut choices used in this paper.

Having validated our framework for estimating Higgs boson + multiparton scattering amplitudes, 
we now consider the physics motivations underlying the modifications to the FKL amplitudes 
presented in section~\ref{sec:modified-high-energy-1}.
\subsubsection{Restoration of Analytic Properties and the Relation to
  Next-to-Leading Logarithmic Corrections}
The argument for using the longitudinal (as well as transverse) components in
evaluating the Reggeised gluon propagators (contrary to what is advocated in
e.g.~Ref.\cite{Fadin:2006bj}) can be rationalised as follows. Firstly, as
already hinted above, this restores the correct position of the divergences
for the amplitude outside of the true MRK limit. This is obviously important
when the amplitude is evaluated without any explicit requirement of 
infinite invariant mass between each and every parton.  Using full
propagators also removes the problem of diffusion~\cite{Bartels:1993du} encountered in the BFKL
formalism when $q_{i\perp}\to0$, since this is not a special point for the
FKL amplitudes. In fact, within the relevant physical region of phase space
encountered at the LHC, the transverse momentum of the imagined $t$-channel
gluons $|q_{i\perp}|^2$ vary to a much larger extent than the square of the
full $t$-channel 4-momentum $q_{i}^2$. This immediately seems to endorse a
setup tailored to the constraint of Eq.~\eqref{MRK} rather than one using the
constraints in Eq.~\eqref{BFKL-MRK}.

Secondly, the next-to-leading logarithmic corrections to the Lipatov vertex
begin to address the same issue, by reinstating the dependence on the
longitudinal momenta - albeit only that part relating to the emission of two particles 
from the same Lipatov vertex at NLL. However, as long as the BFKL equation is
kept (at either LL, NLL accuracy or beyond), a replacement of $q^2_i
\rightarrow -q^2_{i\perp}$ between each Lipatov vertex is performed, which
ignores the longitudinal components of propagators connecting
$N^n\!LL$-vertices (this ensures that the BFKL equation depends only on the
transverse momenta), and there is no information on longitudinal momentum
flow between each Lipatov vertex. Therefore, the
restoration of the full propagator is beyond any fixed logarithmic
accuracy. For example, the full propagator for the ${\cal O}(\alpha_s^4)$
matrix element for Higgs~$+ 2$~partons would never be restored in a
description based on the BFKL equation. By completely avoiding the framework
of the BFKL equation, we can implement these important corrections to any
logarithmic accuracy (and beyond).

\subsubsection{Connection to the \emph{Kinematic Constraint}}
\label{sec:modified-high-energy}
The results of FKL constrain only the form of the square of the Lipatov
vertex in the strict MRK limit, as given in Eq.~(\ref{liplimit}). The use of
any form differing only by sub-asymptotic terms (i.e.~leading to the same
asymptotic limit) is obviously allowed, if one is trying to maintain only a
certain logarithmic (in $\hat s/|\hat t|$) accuracy. However, we would like to
ensure a physical behaviour of the amplitudes in all of phase space (not just
in the MRK limit), in order to construct an approximation which can be
trusted to deliver reliable results for processes relevant for collider
phenomenology, specifically with no requirement of infinite invariant mass
between all partons. We choose to require gauge invariance (or fulfilment of
Ward's identity) and positive definiteness of the squared emission vertex,
which severely constrains the sub-asymptotic terms. The former corresponds to
the requirement $k.C=0$, where $k$ is the emitted gluon momentum. It is
easily verified from Eq.~(\ref{lip1}) that this is the case over all of phase
space (i.e. not just in the MRK limit).  Positive definiteness $-C.C>0$
arises from the fact that $C^\mu$ as defined in Eq.~(\ref{lip1}) is a sum of
light-like and space-like 4-vectors, and thus is itself space-like. The fact
that $C^\mu$ is a 4-vector also implies that the squared emission vertex is
Lorentz invariant. Other choices of vertex do not satisfy this property in
general (although all are longitudinally boost invariant); the terms breaking
Lorentz invariance and the gauge dependent terms are suppressed in the MRK
limit. The extra constraints go beyond any fixed logarithmic accuracy, but
are needed in order to enforce a correct physical behaviour of the scattering
amplitudes, when applied to the calculation of scattering processes of
relevance to collider phenomenology.

Other procedures for modifying the sub-asymptotic behaviour in the FKL 
formalism have been studied before. In Ref.~\cite{Andersen:2008ue}, a
different choice of the Lipatov vertex was used (that of 
Ref.~\cite{DelDuca:1995hf}). This was not positive definite over all of phase
space, and thus the constraint $-C.C>0$ was additionally imposed. This
 removed a part of the sub-MRK
phase space, and thus was related at least partially to the so-called
\emph{kinematic constraint} \cite{Ciafaloni:1987ur,Andersson:1995ju}
implemented in the CCFM
equation~\cite{Catani:1989sg,Catani:1989yc,Kwiecinski:1996td}. Both
approaches limit the region of phase space in which the factorised amplitudes
are applied by requiring a varying degree of dominance of the transverse
momenta over the longitudinal momenta of the $t$-channel propagator momenta,
and specifically apply to the sub-MRK behaviour of the result.

The kinematic constraint is most often discussed in connection with the
small-$x$ evolution of parton distribution functions, and restricts the region
where the BFKL evolution is applied on the basis of the $t$-channel
momenta $q_i$ and the transverse momenta of the gluons
$k_{i\perp}$. Introducing the light-cone coordinates:
\begin{equation}
q_i^\pm=q_i^0\pm q_i^3,
\label{lightcone}
\end{equation}
we consider an incoming parton with light-cone momentum
$p_a^+$. Then the kinematic constraint for the $i$'th $t$-channel gluon can be given in
terms of the light-cone momentum fractions:
\begin{equation}
z_i^+=\frac{q_i^+}{p_a^+}.
\label{zi}
\end{equation}
In our study, we have incoming partons with both negative and positive
light-cone momenta, so we will also need the negative light-cone momentum fraction
\begin{equation}
z_i^-=\frac{q_i^-}{p_b^-}.
\label{zim}
\end{equation}
We note that the MRK limit corresponds to $z_i^-,z_i^+\to 0$ for all $i$.
We can now study three incarnations of the kinematic constraint, which
require the momenta of the emitted partons $k_i$ and the $t$-channel momenta
$t_i$ to fulfill the following conditions ($z_i$ can be either $z_i^+$,
$z_i^{-}$ or both, depending on the direction of the evolution)
\begin{enumerate}
\item The form of Ref.~\cite{Kwiecinski:1996td}:
\begin{equation}
|k_{i\perp}|^2<\frac{(1-z_i)}{z_i}|q_{i\perp}|^2,
\label{kin1}
\end{equation}
\item The form of equation (\ref{kin1}) arising from considering the limit
  $z_i\rightarrow0$ \cite{Ciafaloni:1987ur,Catani:1989sg,Catani:1989yc}:
\begin{equation}
|k_{i\perp}|^2<\frac{1}{z_i}|q_{i\perp}|^2.
\label{kin2}
\end{equation}
This is a weaker constraint than (\ref{kin1}), given that $0<z_i<1$.
\item The condition that the transverse components of the Reggeised momentum
   is larger than the longitudinal parts:
\begin{equation}
|q_{i\perp}^2|>|q_i^+q_i^-|.
\label{kin3}
\end{equation}
\end{enumerate}
Radiation not fulfilling these inequalities is vetoed. All three constraints
are trivially satisfied in the MRK limit.

Starting from the FKL-based approximation to the cross section, we collect in
Table~\ref{kinresults} the result of imposing the kinematic constraints
mentioned above in the Higgs boson plus three-jet rate.
\begin{table}
\begin{center}
\begin{tabular}{c|c}
Constraint&$\sigma_{hjjj}$\\
\hline
None&211 fb \\
$|k_{i\perp}|^2<(1-z_i^+)/z_i^+\,|q_{i\perp}|^2$&57 fb \\
$|k_{i\perp}|^2<(1-z_i^\pm)/z_i^\pm\,|q_{i\perp}|^2$ &33 fb\\
$|k_{i\perp}|^2<1/z_i^+|\,q_{i\perp}|^2$ &140 fb\\
$|k_{i\perp}|^2<1/z_i^\pm|\,q_{i\perp}|^2$&97 fb\\
$|q_{i\perp}|^2>|q_i^+q_i^-|$&178 fb
\end{tabular}
\caption{Values of the three-parton rate for the various kinematic
  constraints discussed in the text. The notation $z_i^\pm$ implies
that the constraint is applied to both plus and minus momenta.}
\label{kinresults}
\end{center}
\end{table}
Without any constraint, the
integrated cross section is 211fb (to be compared with the
contribution from full LO QCD of 203fb). 
Imposing an additional kinematic constraint (absent in our current
implementation of the FKL formula) can be seen to remove significant
fractions of the (non-MRK) phase space, which is relevant to LHC physics,
reducing the cross section by up to a factor of 6!

As expected, the weaker forms of the constraint cut out less of the phase
space, but would still seem to be removing too many events to then be able to
describe the cross-section well. 

\subsubsection{Regularisation of the Amplitudes}
\label{sec:regul-ampl}
As shown in the previous sections, the modified FKL results for the $hjj$ and $hjjj$ 
cross-sections agree well with the full fixed order matrix elements at low orders
in \as. However, the aim of the framework
is not just to reproduce known results, but to allow for an approximation of
higher order amplitudes which are not calculable using present day full fixed
order perturbative 
techniques. In order to be able to implement the resulting amplitudes in a numerical
context, the cancellation between virtual divergences (from
Eq.~(\ref{alpha})) and those arising from real emission (when any $p_{i\perp}\to0$) 
must be made explicit.  

We will consider the cancellation of divergences order by order in
\as. For simplicity and without loss of generality, let us consider the
divergence related to the middle gluon in $gg\to gghg$ going soft; again,
without loss of generality we will take the rapidity of the middle gluon to
be larger than that of the Higgs boson. In this case, $j=n=1$ in
Eq.~(\ref{FKL}), and in the limit $p_{1\perp}\to 0$ with all other outgoing
momenta fixed we get:
\begin{align}
  \label{eq:kisoftlimit}
  \left|\mathcal{M}_{\mathrm{HE}}^{p_a p_b\to p_0 p_1 p_h
    p_2}\right|^2\ \stackrel{{\bf p}^2_{1}\to 0}{\longrightarrow}\
\left(\frac{4\gs^2\ca}{{\bf p}_{1}^2}\right)
  \left|\mathcal{M}_{\mathrm{HE}}^{p_a p_b\to p_0 p_h
    p_2}\right|^2
\end{align}
The structure of divergences and their cancellation will turn out to be very
similar to the one arising in the BFKL equation, and can be regularised 
using a phase space slicing method, which has been successful at regularising
the iterative approach to solving the BFKL equation at LL
\cite{Kwiecinski:1996fm,Schmidt:1997fg,Orr:1997im} and NLL
\cite{Andersen:2003an,Andersen:2003wy}. By integrating over the soft
part ${\bf p}_i^2<\lambda^2$ of phase space in $D=4+2\varepsilon$ dimensions, we find
\begin{align}
  \begin{split}
    \label{eq:realdiv1}
    \int_0^\lambda& \frac{\mathrm{d}^{2+2\varepsilon}{\bf p}\
      \mathrm{d}y_1}{(2\pi)^{2+2\varepsilon}\ 4\pi} \left(\frac{4 \gs^2
        \ca}{{\bf p}^2}\right)\mu^{-2\varepsilon}\\
    &=\frac{4\gs^2\ca}{(2\pi)^{2+2\varepsilon}4\pi}\Delta y_{0h}
    \frac{\pi^{1+\varepsilon}}{\Gamma(1+\varepsilon)} \frac 1 \varepsilon (\lambda^2/\mu^2)^\varepsilon
  \end{split}
\end{align}
The divergence as $\varepsilon\to 0$ is cancelled by the virtual
corrections from the matrix element on the right hand side of
Eq.~(\ref{eq:kisoftlimit}), arising from the Reggeised $t$-channel propagator
between parton 0 and the Higgs boson. Indeed, one finds for $\hat
\alpha(t_i)$ (see e.g.~\cite{Fadin:1998sh})
\begin{align}
  \label{eq:ahatdimreg}
  \hat
  \alpha(t)=-\frac{\gs^2\ca\Gamma(1-\varepsilon)}{(4\pi)^{2+\varepsilon}}\frac
  2 \varepsilon\left({\bf q}^2/\mu^2\right)^\varepsilon.
\end{align}
The square of the matrix element on the left hand side of
Eq.~(\ref{eq:kisoftlimit}) contains the exponential $\exp(2\alpha(t_1)\Delta
y_{0h})$. By expanding the exponential to first order in \as and in
$\varepsilon$, the resulting pole in $\varepsilon$ does indeed cancel
that of Eq.~(\ref{eq:realdiv1}), and one is left with a
contribution 
\begin{align}
  \label{eq:virtualexponent}
  \Delta y_{0h}\frac{\as\Nc}{\pi} \ln\left(\frac{\lambda^2}{\mu^2}\right),
\end{align}
which is the regularised form of the exponent describing the virtual (and
soft) emission in the FKL factorised amplitude of Eq.~(\ref{FKL}). It is
clear that the nested rapidity integrals of additional soft, factorising
radiation in multi-parton amplitudes will build up the exponential needed to
cancel the poles from the virtual corrections to all orders in
\as. The divergence arising from a given real emission is
therefore cancelled by that arising from the virtual corrections in the
Reggeised $t$-channel propagator of the matrix element without the real emission.

\subsubsection{Performing the Explicit Resummation}
\label{sec:adding-nll-running}
The regularisation discussed in the previous section allows one to construct
fully inclusive event samples, where each event contains a Higgs boson and
$n\geq 2$ partons in the final state. The framework which emerges is
therefore similar in application to the one suggested in
Ref.\cite{Andersen:2006sp} for solving the BFKL equation. Corrections beyond
those entering the NLL BFKL kernel can be taken into account by applying
directly the effective Feynman rules for Reggeised particles as discussed in
Ref.\cite{Bogdan:2006af}. This would automatically solve the problems
associated with energy and momentum conservation in the NLL corrections to
the BFKL kernel,
as discussed in Ref.\cite{Andersen:2006kp}.

The regularised FKL amplitudes for incoming gluon states (quark states 
change only the colour factor)
and the production of a Higgs boson between the extremal partons
take the form (using the same notation as in Eq.~(\ref{FKL}))
\begin{eqnarray}
\left|{\cal M}_{\mathrm{HE, r}}^{ab\rightarrow p_0\ldots p_jhp_{j+1}\ldots p_{n+1}}\right|^2&=&4\hat s^2
\left(\frac{\gs^2 \ca}{t_i} \right)\nonumber\\
&\cdot&
\prod_{i=1}^j \left(\exp[\omega(q_i)(y_{i-1}-y_i)]\left( \gs^2 \ca\right)\frac{-C_{\mu_i}(q_i,q_{i+1})C^{\mu_i}(q_i,q_{i+1})}{t_it_{i+1}}\right)\nonumber\\
&\cdot&\left(\exp[\omega(q_i)(y_{j}-y_h)]\frac{C_{H}(q_{j+1},q_{h})C_{H}(q_{j+1},q_{h})}{t_ht_{j+1}}\right)\label{FKLreg}
\\
&\cdot&\prod_{i=j+1}^n
\left(\exp[\omega(q_i)(y'_{i-1}-y'_i)]\left(\gs^2\ca\right)\frac{-C_{\mu_i}(q_i,q_{i+1})C^{\mu_i}(q_i,q_{i+1})}{t_i
  t_{i+1}}\right)\nonumber\\
&\cdot&
\exp[\omega(q_{n+1})(y'_{n}-y'_{n+1})]\left(\frac{\gs^2\ca}{t_{n+1}}
\right)\prod_{i=1}^n\Theta({\bf p_i}^2-\lambda^2)\nonumber
\end{eqnarray}
with 
\begin{align}
  \label{eq:ahatr}
  \omega(q)=-\frac{\as\Nc}\pi\ln\left(\frac{|{\bf
      q}|^2}{\lambda^2}\right).
\end{align}
In order to ensure the correct cancellation between real and virtual
corrections, the matrix elements in Eq.~(\ref{FKLreg}) should always be used
in fully inclusive samples. That is, in our studies we study the ensemble of
partonic processes
\begin{align}
  \label{eq:ensemble}
  \sigma^{ab\to hX} \sim \sum_{j,n=0}^\infty\ \prod_{h,i=0\ldots j+n}\int
  \mathrm{d}\mathcal{P}_i \left|{\cal M}_{\mathrm{HE, r}}^{ab\rightarrow p_0\ldots p_jhp_{j+1}p_n}\right|^2,
\end{align}
where the parton densities and flux factor are omitted on the right-hand side for brevity.
In this form, the cross section takes a form which is extremely similar to
the equations for the iterative solution to the BFKL equation derived in
Ref.\cite{Andersen:2005jr,Andersen:2006sp}. Only the integrand has changed
in the description. This means that the phase space sampler
for integrals of this sort developed in Ref.\cite{Andersen:2006sp} can be
applied in the calculation of the fully inclusive sample of
Eq.~(\ref{eq:ensemble}). We choose to evaluate the fixed coupling in
Eq.~(\ref{eq:ensemble}) at scale $m_H$. In principle the effects of the running of the coupling
could have been modelled according to Ref.\cite{Orr:1997im}. However, we
choose not to, anticipating instead to incorporate the full NLL corrections
at a later stage.

The only unregulated divergence in this ensemble arises in the first and last
bracket in Eq.~(\ref{FKLreg}) when $k_{0\perp},k_{n+1\perp}\to 0$. This
divergence we have to regularise by restricting the transverse momentum at
the impact factors. This ensures that our chosen framework remains
appropriate. The divergence could also have been regularised by replacing the
use of impact factors with unintegrated pdfs, thus avoiding a strict
cut-off. However, a comparison with fixed order results would then be less
straightforward. 

\FIGURE[tb]{\epsfig{width=9cm,file=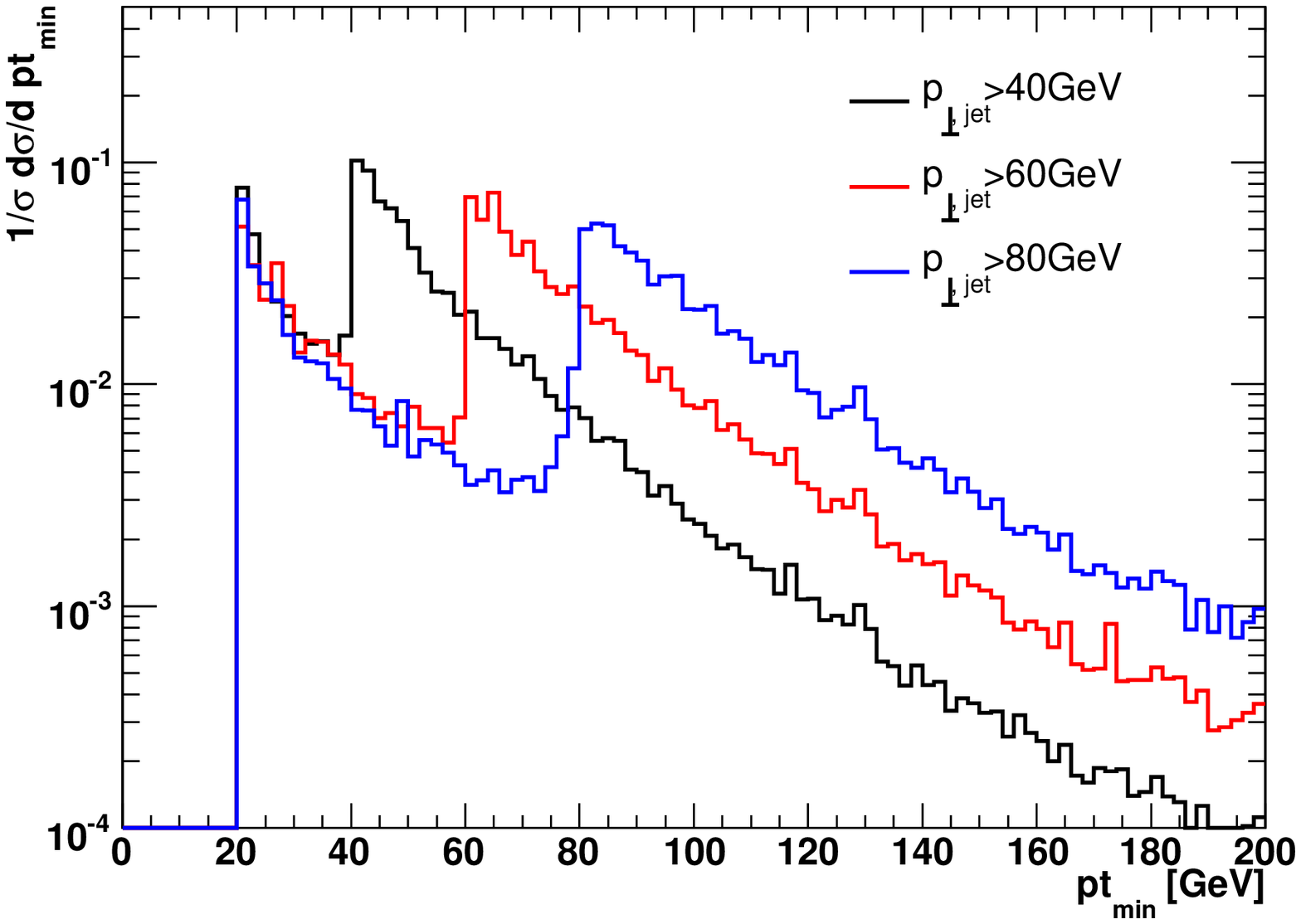}
  \caption{The normalised cross section for the inclusive cuts with three
    different values of the required jet transverse momentum, as a function
    of the minimum transverse momentum of the partons extremal in rapidity. \label{fig:ptIF}}
} On Fig.~\ref{fig:ptIF} we plot the resummed inclusive $H+\!\ge\!2$-jet
cross section for three choices of the transverse momentum cut-off for jets,
as a function of the minimum transverse momentum allowed for the parton
arising from the impact factors. All three results have a similar shape. The
cross section decreases with a universal slope, as the minimum transverse
momentum for partons from the impact factors $p_{\perp\mathrm{min}}$ is
increased above the minimum jet transverse momentum
$p_{j\perp\mathrm{min}}$. There is a `heel' in the spectrum for
$p_{\perp\mathrm{min}}$ just below $p_{j\perp\mathrm{min}}$; this is caused
by a parton from the impact factors being just short of making it as a single
jet, but combined with an additional parton emitted from the $t$-channel
evolution there is sufficient transverse momentum to qualify as a jet. As
$p_{\perp\mathrm{min}}$ is lowered further, a plateau 
in the dependence is
reached, followed by the universal divergence as
$p_{\perp\mathrm{min}}\to0$. The separation between these regions is
increasingly clear as the required hardness of the jets is increased. As
discussed above, we want to remove as much dependence on the singularity at
$p_{\perp\mathrm{min}}\to0$, since this behaviour would be regularised by
effects not included in the current description. We do want to include the
possibility of the partons from the impact factors (i.e.~the partons extremal
in
rapidity) forming a jet with a parton from the evolution, giving
rise to the heel in the distribution. We therefore impose a cut in the
transverse momentum of the parton from the impact factor somewhere in the
plateau region. In our studies we require the observed jets to have a transverse
momentum of at least 40~GeV, and decide to allow for extremal partons down to
30~GeV. In this way, the partons emitted from the impact factors do not
necessarily enter any jet - in other words, the ``evolution'' in rapidity is
at least as long as the largest difference in rapidity between the observed
jets.

\section{Matching to Tree Level Matrix Elements}
\label{matching}
Given that the full tree level matrix elements for two and three parton states are
computationally quick to calculate, they can be implemented alongside the 
modified FKL amplitudes and used to improve the description
of Higgs boson production at low orders in \as. It is then necessary to define a suitable
prescription for matching the perturbative expansion of the two and three jet rates
obtained from the ensemble in Eq.~(\ref{eq:ensemble}) to the results obtained
in fixed order perturbation theory point by point in phase space. In this way
one corrects for both the contribution from non-FKL configurations, and for the
difference between the approximation and the full result for the 2 and 3 jet
rates in FKL configurations. If the virtual corrections to the 2-jet rate
were known in a compact form, one could also match the ensemble to full
$\as^5$-accuracy. 

In practice this works as follows. Our Monte Carlo implementation will start by 
generating a random point in the
full $(n+1)$--particle phase space. It will then sum over each possible partonic
channel, checking whether it corresponds to a FKL configuration. If so, it
evaluates the appropriate scattering amplitude of Eq.~(\ref{FKLreg}) (with
the colour factor $C_A$ in the impact factors replaced with $C_F$ for
incoming and outgoing quarks). If the phase space configuration corresponds
to a $n$-parton final state, $n$ hard jet configuration for $n=2,3$, it will then
evaluate also the appropriate full tree-level matrix element, and apply a suitable matching 
correction to avoid any double counting of radiation.

We consider two distinct methods for matching, inspired by the so-called 
$R$ and $\ln R$-matching in Ref.\cite{Catani:1992ua}. We will describe
the matching procedure for matrix elements with a two parton final state.
The generalisation to the three parton final state is straightforward.

Let ${\cal M}_{\mathrm{HE,\hat\alpha=0}}^{ab\rightarrow p_0 h p_1}$ be the LO 
$hjj$ matrix element coming from the modified FKL approach i.e. this arises
from Eq.~(\ref{FKL}) with the value of $\hat{\alpha}$ in the virtual
corrections set to zero. 
Furthermore, let ${\cal M}_{\mathrm{HE,r}}^{ab\rightarrow p_0 h p_1}$ be the
regularised (i.e.~with virtual corrections 
included), modified
FKL amplitude for $hjj$ (where the 2 partons form 2 jets), and ${\cal M}^{ab\rightarrow p_0 h p_1}$ the full tree-level result for 
$hjj$. One may consider matching these quantities as follows:
\begin{equation}
{\cal M}_{\mathrm{HE,r}}^{ab\rightarrow p_0 h p_1}\rightarrow {\cal M}_{\mathrm{HE,r}}^{ab\rightarrow p_0 h p_1}
+\left({\cal M}^{ab\rightarrow p_0 h p_1}-{\cal M}_{\mathrm{HE,\hat\alpha=0}}^{ab\rightarrow p_0 h p_1}\right),
\label{match1}
\end{equation}
i.e. instead of using the usual FKL amplitude for such a configuration, one
modifies it to include the full tree level matrix element, and subtracts the
LO part of the FKL amplitude. We call this {\it $R$-matching} by analogy with
Ref.\cite{Catani:1992ua}.  

Since the resummed virtual and unresolved
corrections suppress a given $n$-parton final state such that $\left| {\cal
    M}_{\mathrm{HE,r}}^{ab\rightarrow p_0 h p_1} \right|<\left| {\cal
    M}_{\mathrm{HE,\hat\alpha=0}}^{ab\rightarrow p_0 h p_1}\right|$, it seems
reasonable to also suppress the relevant matching corrections.
This can be achieved by instead reweighting those events with a $n$ parton
final state, n hard jet
configuration according to the prescription:
\begin{align}
  \label{eq:matchreweight}
  \left|{\cal M}_{\mathrm{HE,
    r}}^{ab\rightarrow p_0 h p_1}\right|^2 \longrightarrow
  \left|{\cal M}_{\mathrm{HE,
    r}}^{ab\rightarrow p_0 h p_1}\right|^2
  \left(\frac{\left|{\cal M}^{ab\rightarrow p_0 h p_1}\right|^2 }
  {\left|{\cal M}_{\mathrm{HE,
    \hat\alpha=0}}^{ab\rightarrow p_0 h p_1}\right|^2}\right).
\end{align}
This is formally equivalent to Eq.~(\ref{match1}) up to the required order of 
the perturbation expansion. We call this {\it $\ln R$-matching}, since the
correction is additive in the logarithm of the scattering amplitude.
Effectively, this means that the exponentiated virtual corrections are also applied 
to the matching corrections. 

In both $R$- and $\ln R$-matching, the matching correction vanishes in the MRK 
limit, as it must do. Clearly, one can only apply $\ln R$-matching if the
resummed matrix element is non-zero. For all other processes,
one must use $R$-matching. That is, $R$-matching is used for those matrix element in which either
(i) the partonic configuration is not FKL-like; (ii) the partonic species are FKL-like, but they do
not satisfy the required rapidity ordering.

In principle the matching prescription described above can be extended
to any number of hard jets (and indeed any order in the fixed order perturbation 
expansion). However, the computational evaluation of tree level matrix elements for the process
considered here becomes very CPU intensive for more than 3 jets. So far, no estimates
have been reported on the leading-order $H + 4$--jet cross section.

\section{Results}
\label{results}
In this section we present a few physics analyses arising from a Monte Carlo
event generator based on the resummation discussed in the previous sections,
supplemented with matching point by point in phase space to the full tree
level matrix elements for Higgs boson production with two and three jets. The
Multi-Jet EVent generator can be obtained at
\verb!http://andersen.web.cern.ch/andersen/MJEV!.

\subsection{Relations Between Rapidity Span and the Number of Hard Jets}
\label{sec:jet-activity}
The resummation leads to an increase in average jet activity with increasing
rapidity length of the event. This effect is inherent to all $t$-channel
colour octet exchanges admitting this resummation, and independent of the
specific process. In order to study this evolution over a longer rapidity
range, we relax the cut on the rapidity separation between two jets. It is
therefore relevant to start by asking how well the modified FKL amplitudes
approximate the full fixed order results, if the cut is relaxed to a minimum
separation of, say, two units of rapidity. In Figure~\ref{fig:23jetrelaxedcuts}
we plot the equivalent of Figure~\ref{xsecfig} with the cut in
rapidity difference relaxed from 4.2 to 2 units of rapidity.
\FIGURE[!b]{
  \epsfig{width=.49\textwidth,file=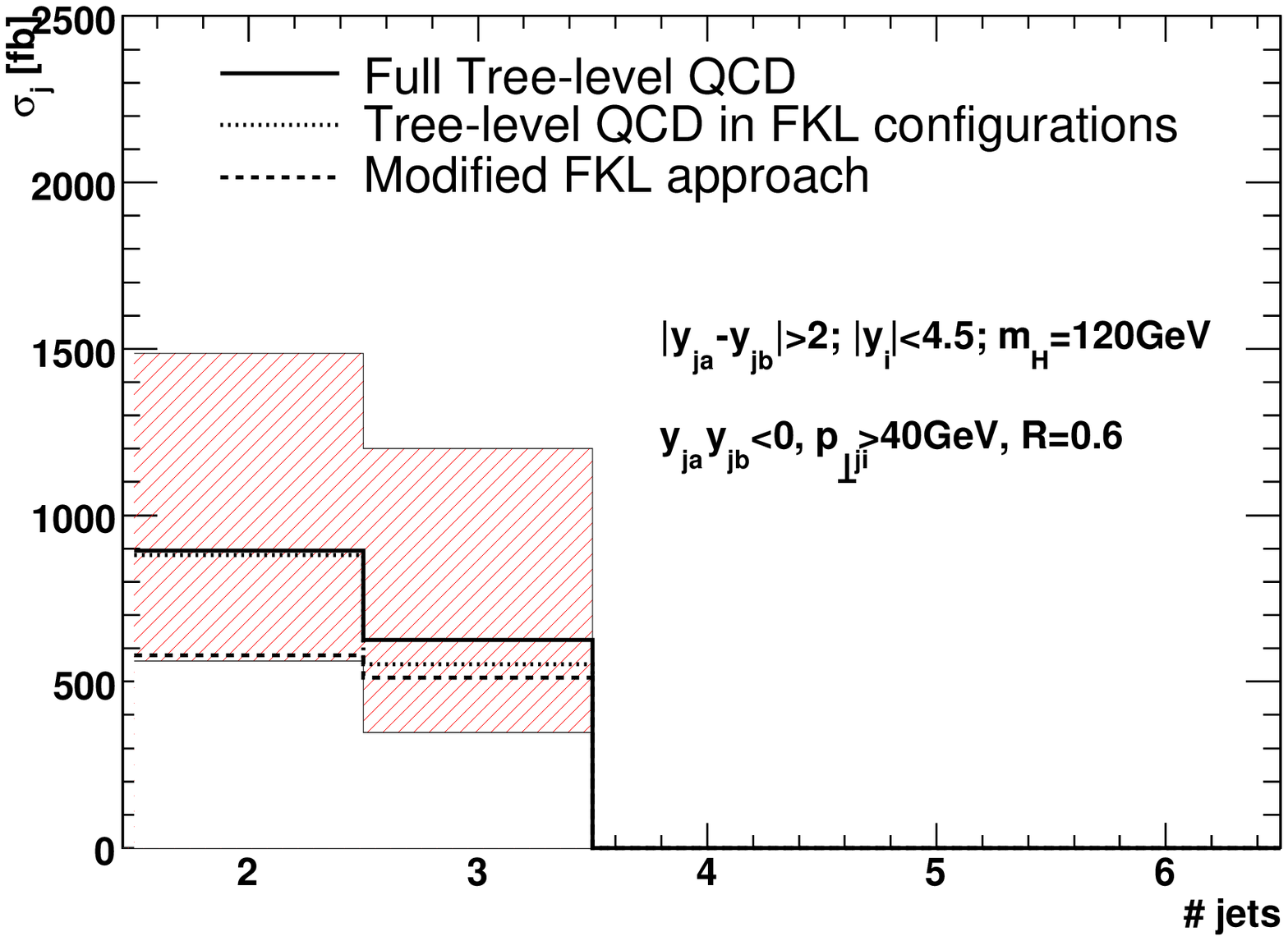}
  \epsfig{width=.49\textwidth,file=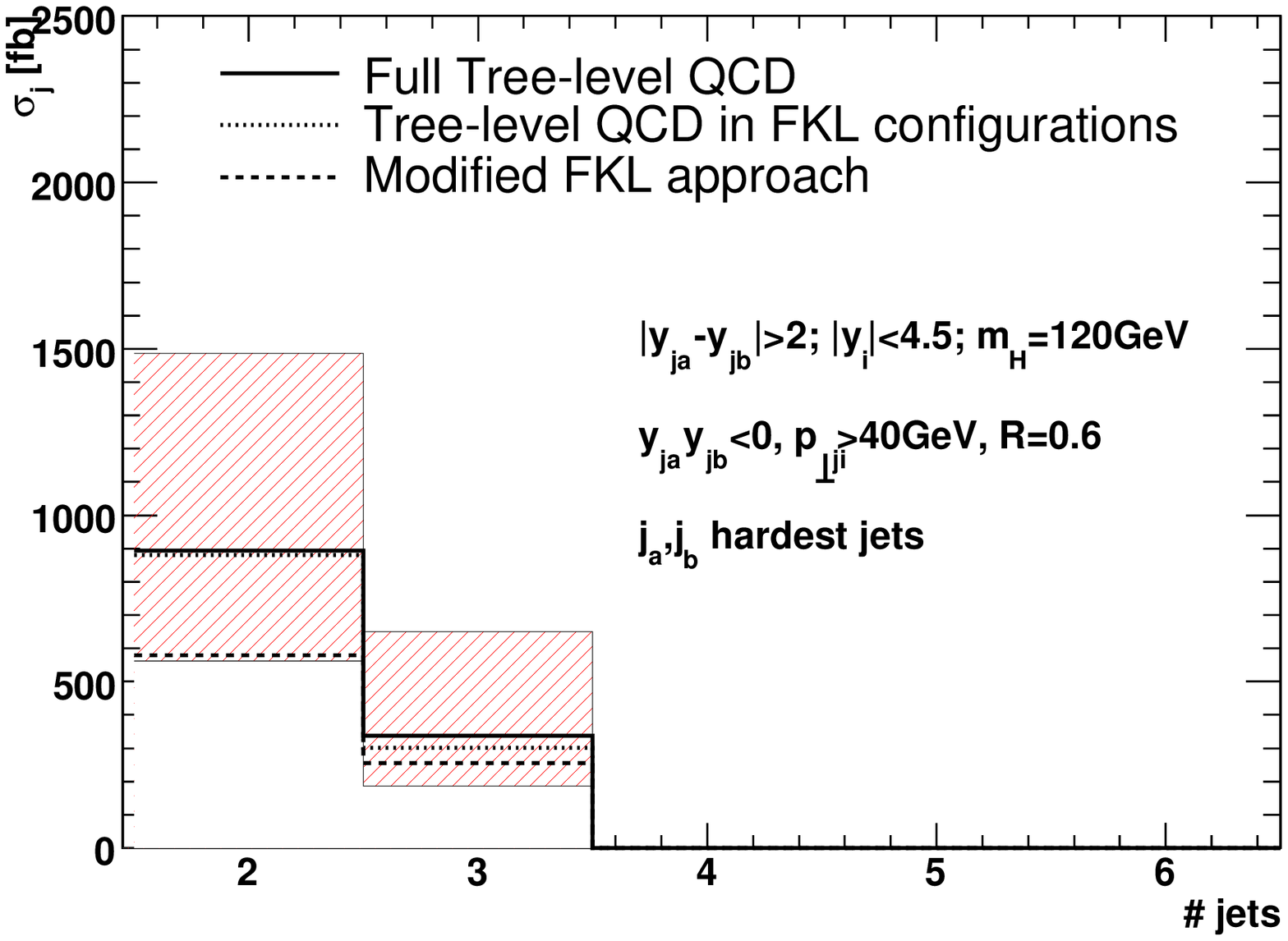}
  \caption{The 2 and 3 jet rates compared between the leading order results
    obtained within full, fixed order QCD and within the modified FKL
    formalism. The rapidity cut has been significantly relaxed compared to
    the results reported in
    Figure~\ref{xsecfig}.\label{fig:23jetrelaxedcuts}}
  }
We see that even with such a small overall rapidity span required, the
factorised formalism still describes the two and three-jet result of the full
tree-level calculation in LL FKL configurations to better than 35\% and
 8\% respectively for the standard cuts, and 35\% and 16\% for the hard cuts. This is
quite remarkable, since the maximum rapidity distance between all jets at
$y_{ja}-y_{jb}=2$ is just one unit of rapidity, and furthermore the Higgs
boson is also often produced within the same rapidity interval, so the
maximal invariant mass between each set of particles is not necessarily
particularly large. 

The $t$-channel colour octet exchange has a characteristic radiation pattern,
in terms of a correlation between the length in rapidity of the resummation
(i.e.~$y_b-y_a$ in Eq.~\eqref{FKL}) and the average number of hard jets. This
quantity is studied in Figure~\ref{fig:avgnhardjet} (left) for the inclusive
cuts, with $\Delta y$ defined as the rapidity span between the most forward
and backward hard jet. We see that the average number of
hard jets rises almost linearly from 2.6 to 4.1, as the rapidity length
increases from 2 to 7.  
\FIGURE{
\epsfig{width=.49\textwidth,file=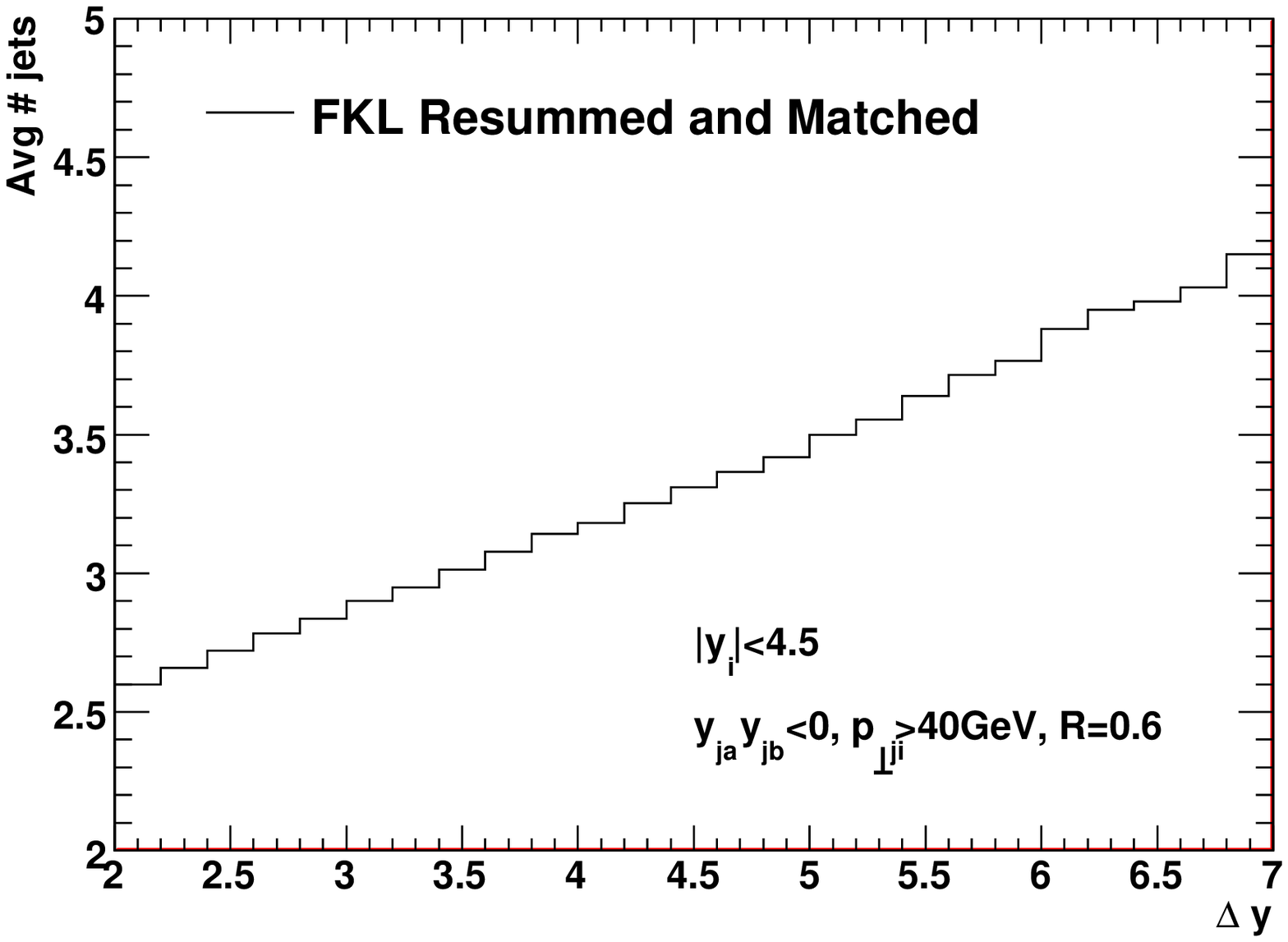}
\epsfig{width=.49\textwidth,file=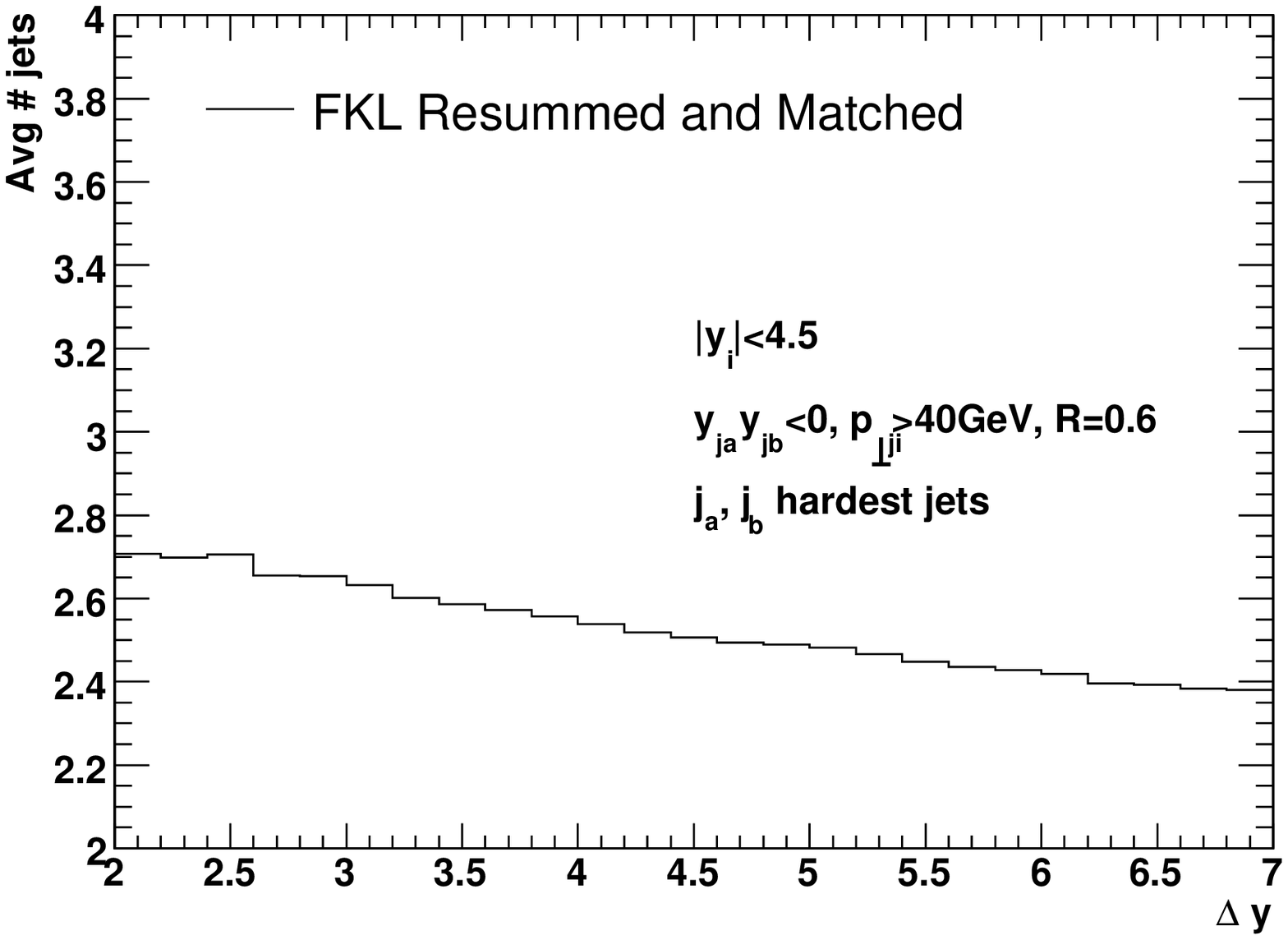}
\caption{The average number of hard jets vs. the rapidity span of event,
  defined as the total rapidity span (left) or the rapidity span between the
  two hardest jets (right).\label{fig:avgnhardjet}}
}

On Figure~\ref{fig:avgnhardjet} (right) we study the average number of hard jets as a
function of the rapidity difference between the two hardest jets, with the
event selection based on the same two hard jets. Obviously, in this case the average jet
count is smaller than with with inclusive cuts, but
we also see that despite the same underlying physics used in the description,
the correlation between average hard jet count and rapidity is partly changed
by the choice of cuts, and partly masked by the definition of the rapidity
variable. One now sees a \emph{decrease} in the average
number of hard jets, as the rapidity difference between the two hardest jets
is increased. This is because effectively the hard jet cuts introduce a veto on
hard radiation in between the two hardest jets, resulting in a large
reduction in the three-jet (and more) cross section. In the time-evolution
language of the parton shower, one could imagine a central jet of say 65GeV transverse momentum
splitting up into a 59GeV and a 6GeV jet. With forward and backwards jets of 60GeV
($>40$GeV) transverse
momentum, the event would be rejected before the splitting, but accepted
after the splitting. This sensitivity to higher order splittings is the
motivation for considering the inclusive cuts, where the acceptance of an
event is insensitive to such splittings of the central partons, although of
course the categorising of the event as a $n$ or $n+1$ hard jet state is
still sensitive. We have checked that the results of Figure~\ref{fig:avgnhardjet} are
relatively insensitive to the scale variations, i.e.~the shape is unchanged
and the average jet count changes by less than 0.2 unit.

\subsection{Relaxed Cuts on Central Jets}
\label{sec:relaxed-cuts-central}
\FIGURE[htb]{
  \epsfig{width=.49\textwidth,file=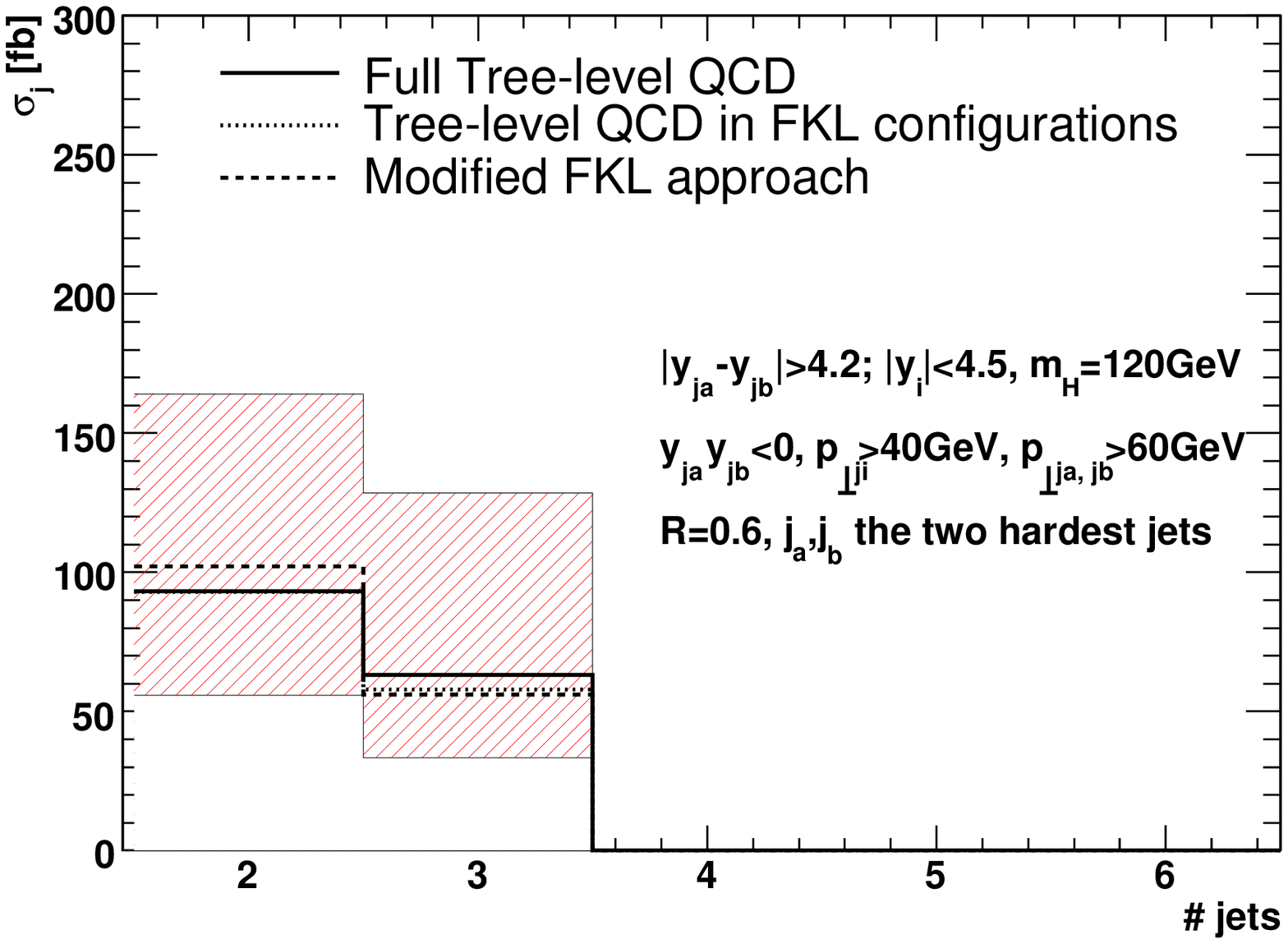}
  \epsfig{width=.49\textwidth,file=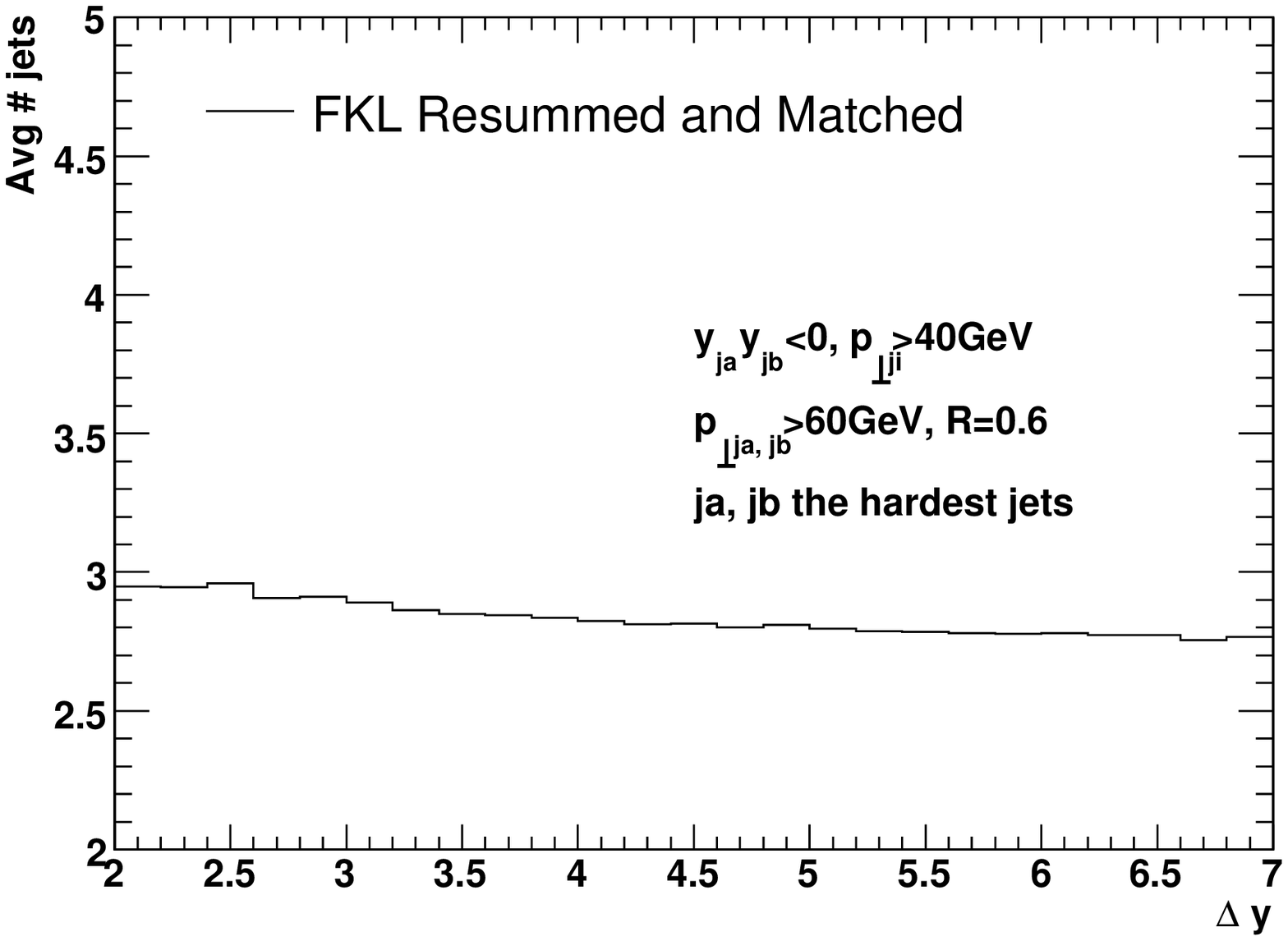}
    \caption{Left: The LO $hjj$ and $hjjj$ cross-sections for full tree-level QCD
      (full line), and those obtained from tree-level QCD in FKL
      momentum-configurations (dotted line), compared with the results from
      the modified FKL approach. Results are for the hard cuts, where the
      tagging jets must satisfy an additional constraint of $p_t>60$GeV. The
      scale uncertainty relates to the tree-level QCD
      results. Right: The average number of hard jets ($p_\perp>40$GeV) vs.~the rapidity
      difference between the two hardest jets ($p_\perp>60$GeV).\label{LOfigs_B}}
  }

In order to better investigate the perturbative activity in-between the two
hardest jets when the hard jet cuts are used, we introduce a variation of the hard jet cuts: The two hardest jets
should have a transverse momentum of at least 60GeV, but the multiplicity is
counted according to the number of jets with a transverse momentum above
40GeV. 

The LO $hjj$ and $hjjj$ cross-sections are shown in figure \ref{LOfigs_B},
together with the results from the modified FKL approach. We see that, just
as for the cuts studied so far, the modified fixed order FKL approach
describes very well the cross sections obtained with the full fixed order
matrix elements. Obviously, only the three-jet cross section depends on the
hardness required by the non-tagging jets. On the same figure we have plotted
the average number of hard jets ($p_\perp>40$GeV) against the rapidity
difference between the two hardest jets ($p_\perp>60$GeV) in the inclusive
sample of the resummed calculation. We see that there
is again a decrease in the average number of jets with increasing rapidity span, albeit milder 
than that of Fig.\ref{fig:avgnhardjet} (right).

In the following sections which investigate the jet activity, we will also report
results using this set of cuts.

\subsection{Cross Sections and Jet Counts within Weak Boson Fusion Cuts}
\label{sec:cross-sections-jet}
We now return to the two set of weak boson fusion cuts, differing only by the
choice of jets which are required to pass the requirement on a separation in
rapidity. To elucidate further the difference appearing between the ``inclusive'' and
``hardest'' choice of jet cuts, we study in Figure~\ref{fig:4jetrapdist} the
rapidity distribution of the most forward and backward jet, and of the
hardest and next-to-hardest jet, all within the inclusive jet cuts. It is worth
recalling that the events passing the cuts on the hardest jets are a sub-set
of the events passing the inclusive cut. The figure demonstrates that
the rapidity distribution of the hardest jet is more central than that of the
next-to-hardest jet. Obviously, cuts based on the two hardest jets
also being far apart in rapidity will reject many of the events with strictly
more than 2 jets, which would
otherwise be accepted by the inclusive cuts.
\FIGURE[tb]{
  \epsfig{width=9cm,file=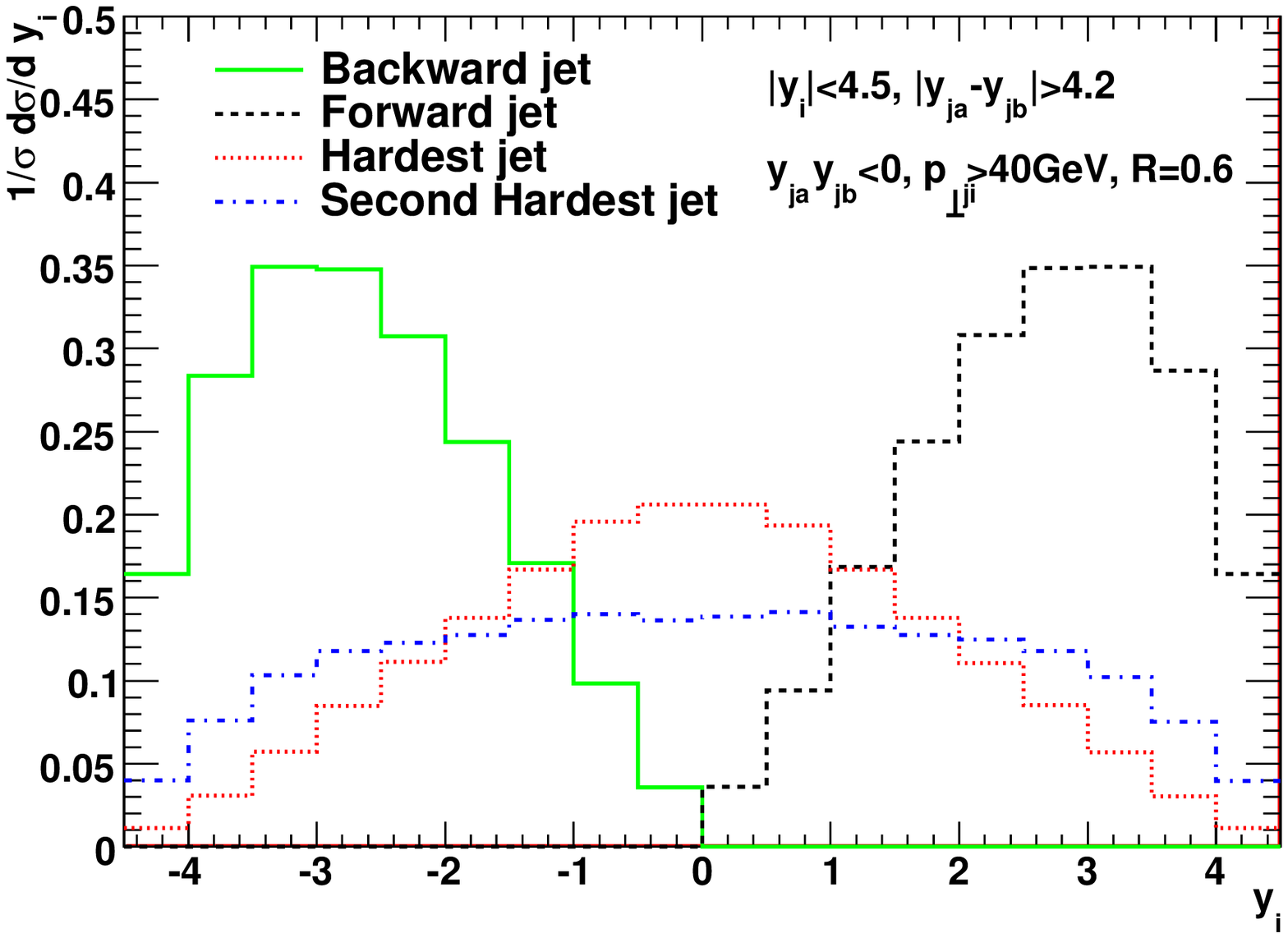}
\caption{The rapidity distribution of the most forward/backward jets, and of
  the two hardest jets within the inclusive cuts.\label{fig:4jetrapdist}}
}

With the inclusive jet cuts and using the resummed and matched calculation,
we find a total cross section of $435^{+439}_{-202}$fb, where the quoted
uncertainty is obtained by variation of the common factorisation and
renormalisation scale by a factor of two (in the resummation, we choose to
evaluate \as at the scale $m_H$, just as in the fixed order calculations). The relative scale uncertainty is
similar to the tree-level results; the scale uncertainty would be reduced if
next-to-leading logarithmic corrections were taken into account. The result
for the resummed cross section corresponds to an increase over the LO rate of
89\%, which is only slightly larger than the K-factor of 1.7-1.8 found at
NLO. The large NLO K-factor arises from the relative large 3-jet rate (only
12\% less than the tree-level 2-jet rate). In
the resummed calculation, the virtual and unresolved real radiation
implemented by the exponentials in Eq.~(\ref{FKLreg}) suppress the cross
section for any fixed number of partons compared to the tree-level
approximation. This suppression is then counterbalanced by the sum over
 further real emissions.

We can apply the jet finding algorithms to the resummed and matched event
sample, and investigate the frequency of multiple hard jets. In
Figure~\ref{fig:njetsresuma} (top left) we show the breakdown of the total
cross section on the number of hard jets in each event. The red uncertainty
bands arise by varying the factorisation and renormalisation scale by a
factor of two. The scale variations change the overall normalisation, but has
a much milder impact on the relative jet counts, as indicated in
Fig.~\ref{fig:hardjetdistr1}. We find that with the inclusive jet cuts and
the chosen parameters for the jet-algorithm, the three-jet rate is larger
than the two-jet rate, and from three to six jets there is an almost linear
decrease in the jet rates. For the cuts on the hardest jets, the jet rates
are uniformly decreasing with the jet count; the two-jet rate is more than
twice as large as the three-jet rate, a pattern which continues for the
higher jet rates. The jet rates are still uniformly decreasing, when
furthermore the hardest jets are required to be harder than 60GeV, but all
jets with a transverse momentum larger than 40GeV are counted, although
obviously the higher jet counts are relatively more important than with a
single jet scale.
\FIGURE[tb]{ \epsfig{width=.49\textwidth,file=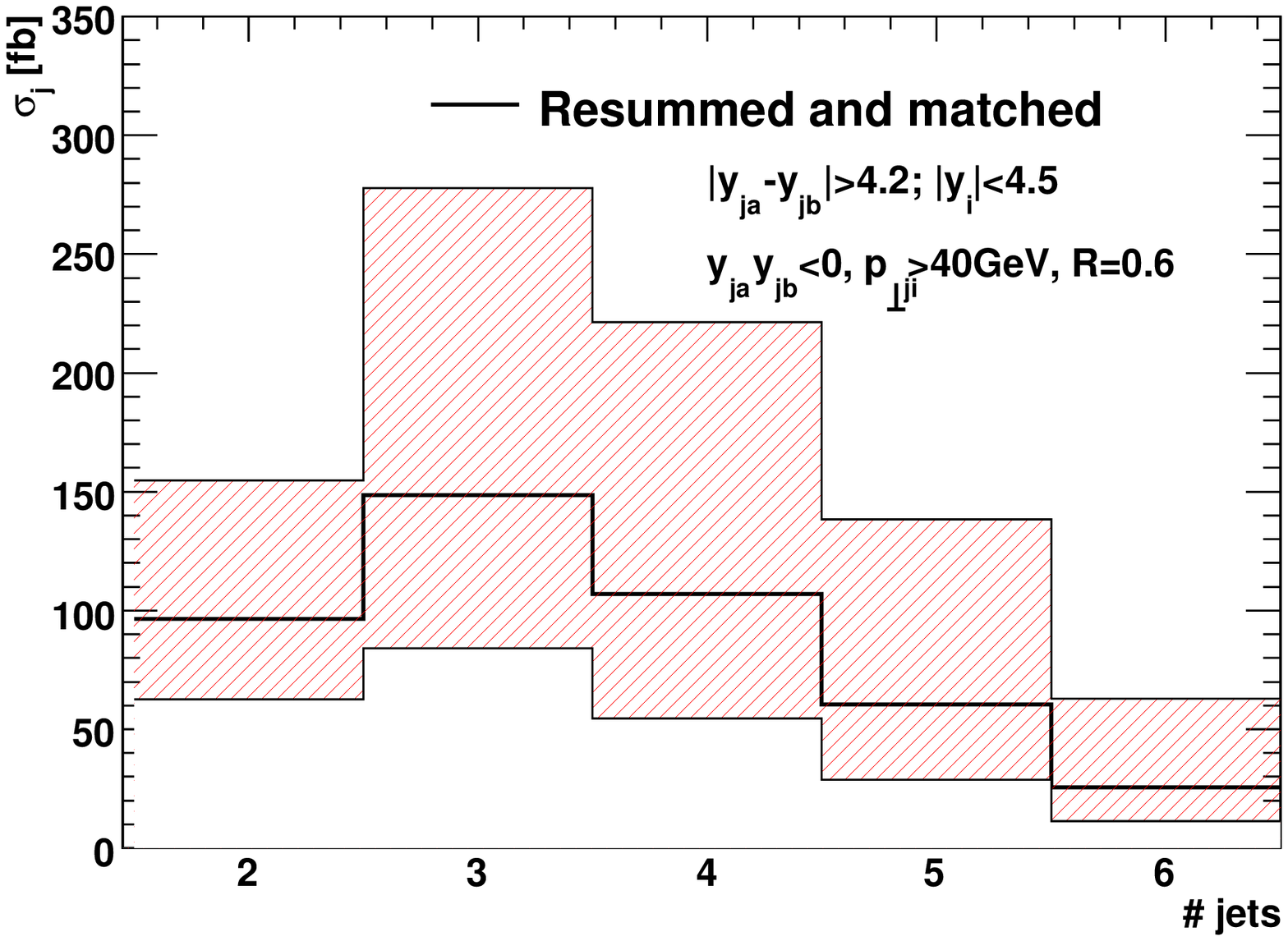}
  \epsfig{width=.49\textwidth,file=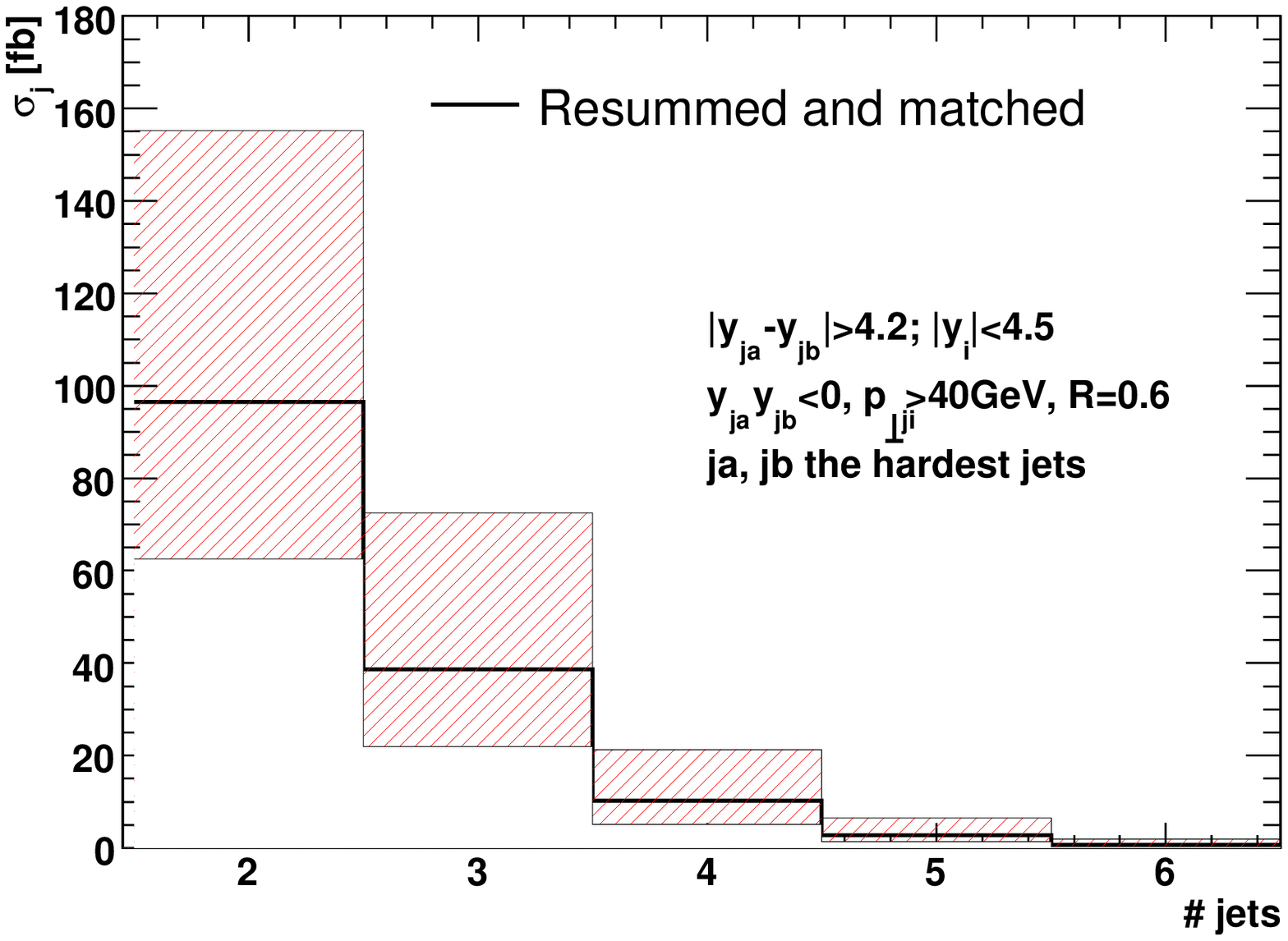}
  \epsfig{width=.49\textwidth,file=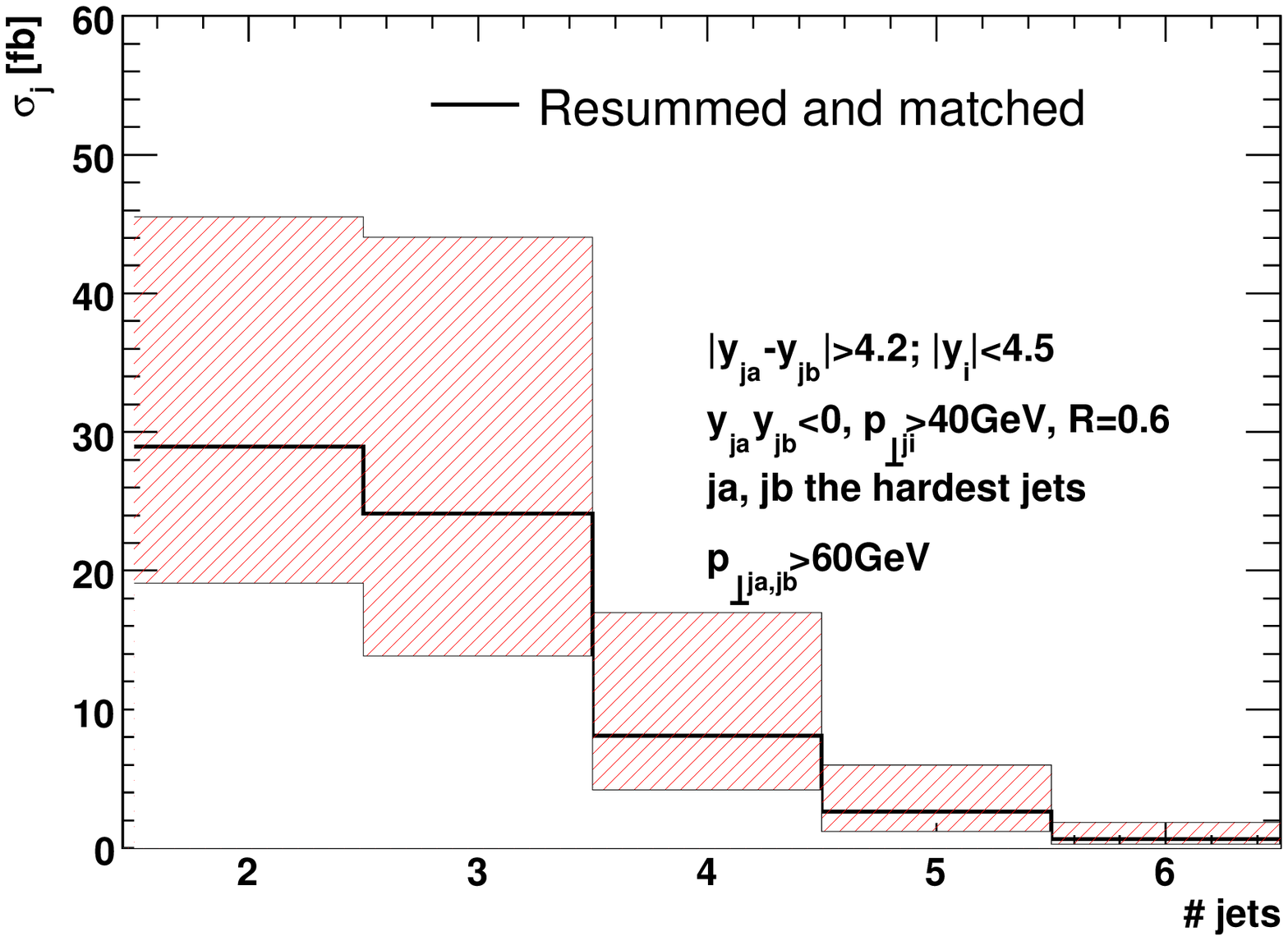}
\caption{The distribution of the number of jets including the uncertainty from scale variation,
         shown for the inclusive cuts (top left) and cuts on the two hardest jets 
         (top right). Also shown is the result when the two hardest jets are
         required to be harder than 60GeV in transverse momentum (bottom) as
         discussed in Section~\ref{sec:relaxed-cuts-central}.\label{fig:njetsresuma}}
} 

When the two hardest jets are required to pass the cuts on rapidity
separation, we find a cross section of $149^{+109}_{-58}$fb, roughly 65\% of
the lowest order tree-level $hjj$ estimate reported in
Section~\ref{sec:central-higgs-+}. We see a \emph{suppression} compared to
the LO estimate, instead of the \emph{increase} of 30-40\%, which is seen in
the NLO calculation with the hardest cuts. As we have already discussed, the
hardest cuts are very sensitive to further hard radiation beyond the 3 jets
which are included in the NLO calculation. When 4 jet configurations are
included, the two hardest jets are likely to be found centrally (see
Fig.~\ref{fig:4jetrapdist}), and all such events will be rejected according
to the hardest cuts. This is not taken into account in the NLO calculation,
but is in the resummed calculation, which therefore results in a smaller
cross section than seen at NLO. The resummed calculation can directly address
the perturbative instability of the hardest cuts at the lowest fixed orders.  The resummed and matched
prediction for the cross section when the cuts are placed on the hardest
jets, which are also required to be harder than 60GeV (see
Sec.~\ref{sec:relaxed-cuts-central}) is $57^{+41}_{-24}$fb, again roughly
65\% of the tree-level $hjj$ cross section found in
Section~\ref{sec:relaxed-cuts-central}.

The predictions for the relative number of hard jets are surprisingly stable
against variations of factorisation and renormalisation scale. This is
illustrated on Figures~\ref{fig:hardjetdistr1} for both the \emph{inclusive}
and \emph{hard} cuts.  \FIGURE[tb]{
  \epsfig{width=.49\textwidth,file=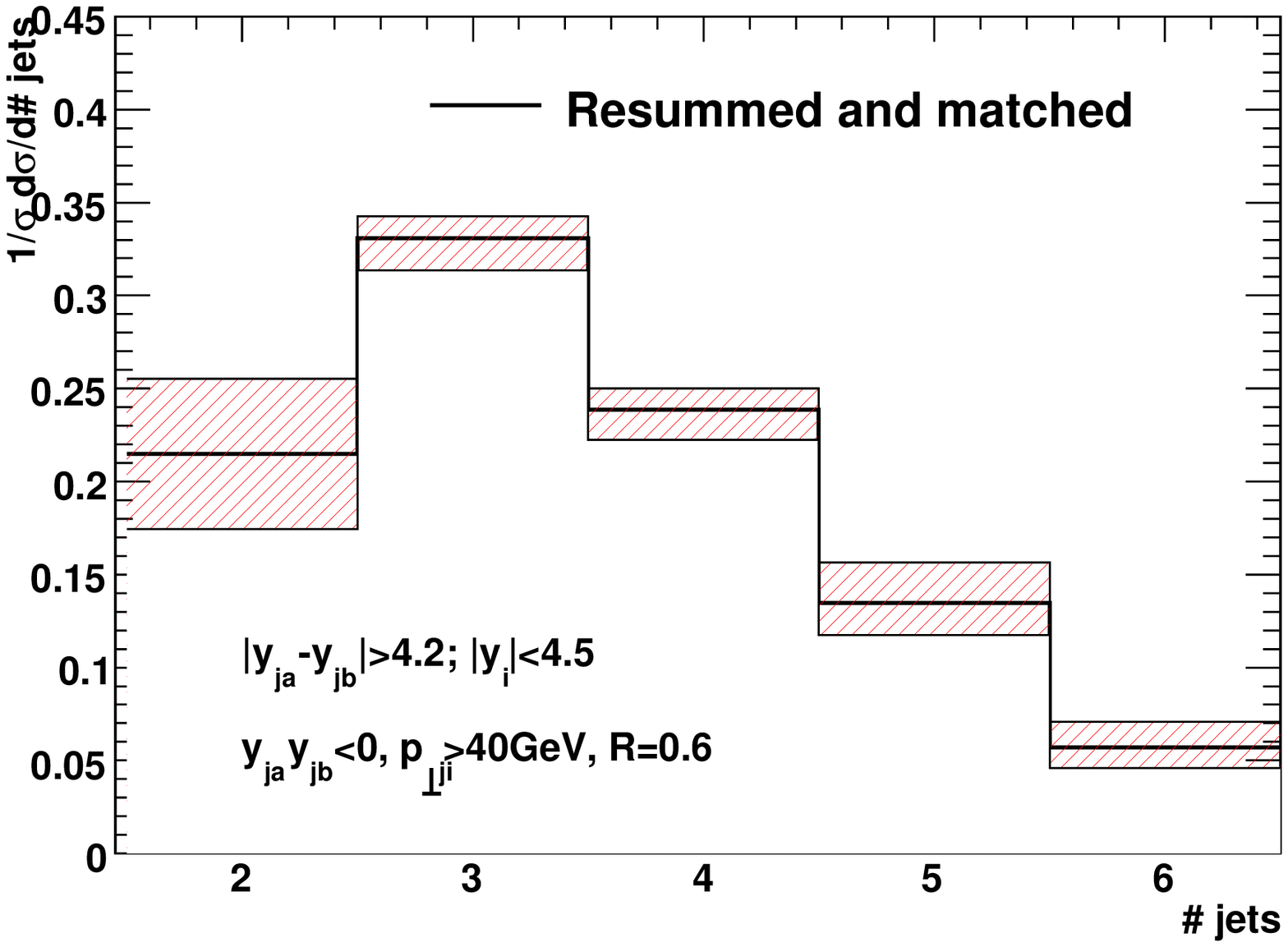}
  \epsfig{width=.49\textwidth,file=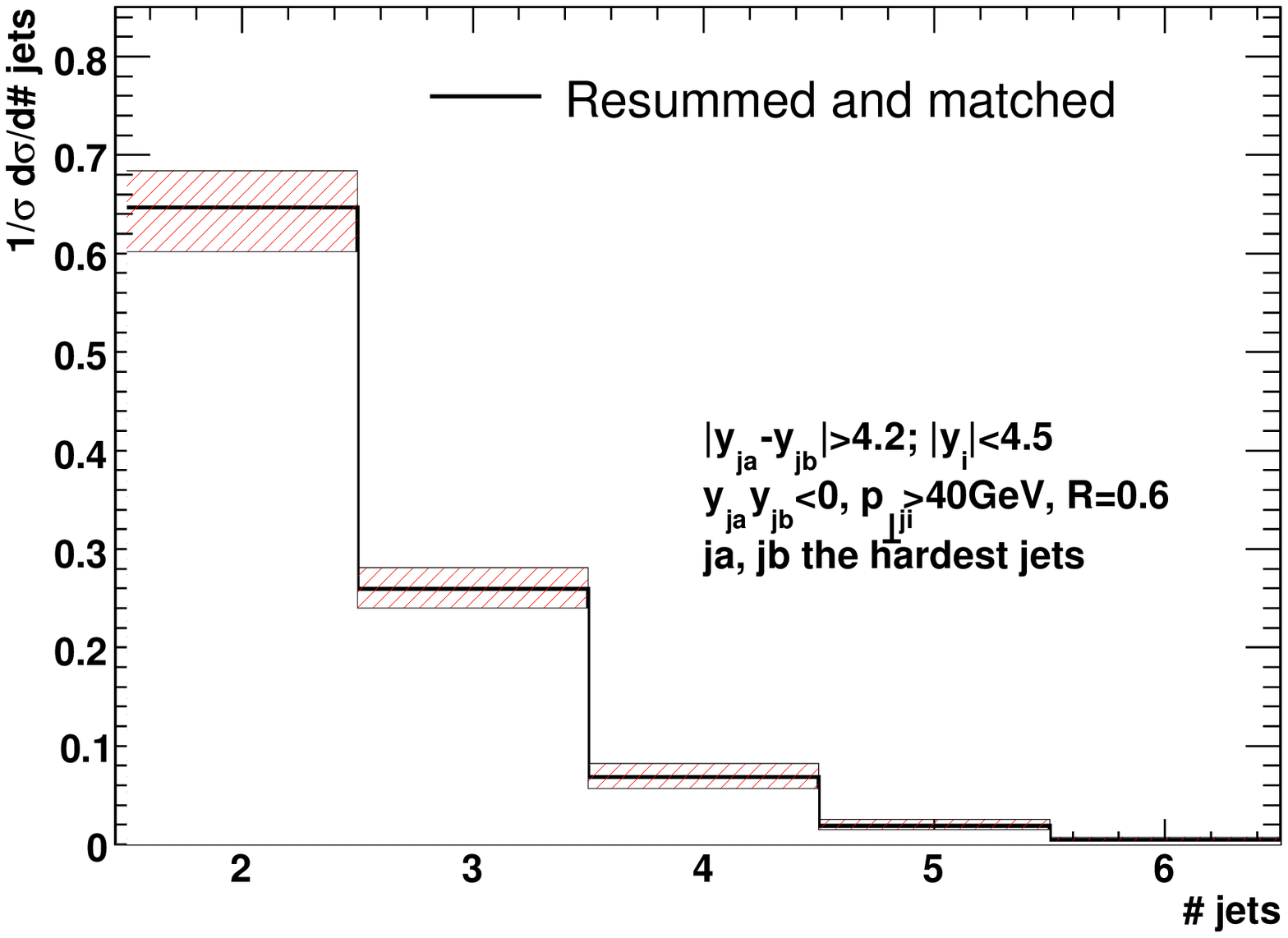}
  \epsfig{width=.49\textwidth,file=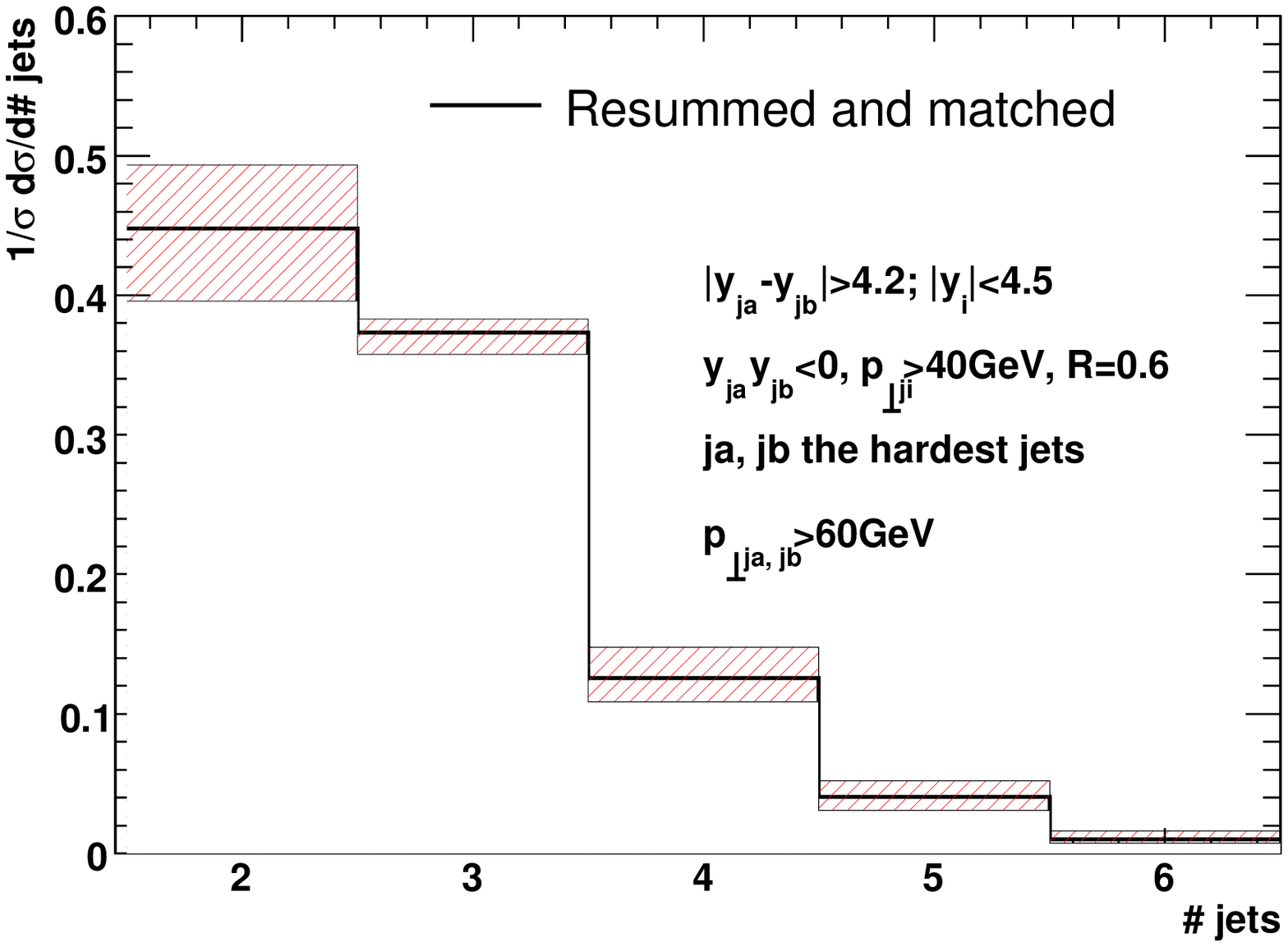}
\caption{The distribution of the number of hard jets including the
  uncertainty from the variation in scale, with the inclusive cuts (top left) and
  cuts on the two hardest jets (top right). Also shown is the result when the two hardest jets are
         required to be harder than 60GeV in transverse momentum (bottom) as
         discussed in Section~\ref{sec:relaxed-cuts-central}\label{fig:hardjetdistr1}}
}

\subsection{The Effects of a Central Rapidity Jet Veto}
\label{sec:effects-centr-rapid}
Since the weak boson fusion process has no kinematically un-suppressed colour
connection between the two jets at the two lowest orders in perturbation
theory, the jet activity between these is expected to be significantly lower
than that of the gluon fusion process\cite{Dokshitzer:1987nc}. It has been
suggested to use this as a further discriminator between the production
mechanisms, and suppress the gluon fusion contribution by vetoing events with
jets in-between the two tagged jets. Obviously, everything will depend on
which two jets are chosen as the tagging ones and many other details, which
necessitates a flexible generator for any study of the effects of central
rapidity jet vetos; in the following we will demonstrate the effect both when
the tagged jets are chosen as the ones furthest apart in rapidity, and when
they are chosen as the hardest jets. We will study the cross section as a
function of vetoing all further hard jets of transverse momentum greater than
$p_{\perp,\mbox{veto}}$ within a distance in rapidity $y_c$ from the centre
of the tagged jets $j_a, j_b$ (always of more than 40GeV transverse momentum)
of rapidity $y_a, y_b$. That is, we require:
\begin{equation}
\forall j\in\{\mbox{jets\ with\ } p_{j\perp}>p_{\perp,\mbox{veto}}\}\setminus\{a,b\}\ :\ \left|y_j-\frac{y_a+y_b}{2}\right|>y_c
\label{eq:yc}
\end{equation}
In the present study we will be interested only in vetoing relatively hard
mini-jets\cite{Barger:1995zq} with transverse momentum of more than 20GeV -
if the transverse momentum scale for the jet veto is significantly lower than
the transverse momentum of the two tagged jets, then a sensitivity is
introduced to potentially large logarithms of soft origin, see
e.g.~Ref.~\cite{Forshaw:2007vb}.

In Figure~\ref{fig:rapidvetoa} (left) we show the cross section as a function
of the variable $y_c$ introduced in Eq.~\eqref{eq:yc}, when the two tagged
jets are those most forward and backward in rapidity (left) and when they are
the hardest (right). We have included the results for
$p_{\perp,\mbox{veto}}=20\mbox{GeV},30\mbox{GeV}$ and $40\mbox{GeV}$. At $y_c=0$ there is no
additional cut, while for $p_{\perp,\mbox{veto}}=40\mbox{GeV}, y_c\to\infty$ the
cross section asymptotes to the two-jet cross section obtained in the
resummed and matched calculation. As $p_{\perp,\mbox{veto}}$ is lowered, more
jets are resolved and more events are vetoed. For the low veto scale
of $p_{\perp,\mbox{veto}}=20\mbox{GeV}$ and as $y_c\to\infty$, the cross section is reduced
to 50fb or roughly a fifth of the tree-level $hjj$-prediction.

Figure~\ref{fig:rapidvetoa}
(right) shows the results when jets $a,b$ are the hardest jets of the
event. This asymptotes to the cross section for 2 hard jets with a
softer third jet (still passing the cut on 40GeV) outside in rapidity.
 \FIGURE[tb]{
  \epsfig{width=.49\textwidth,file=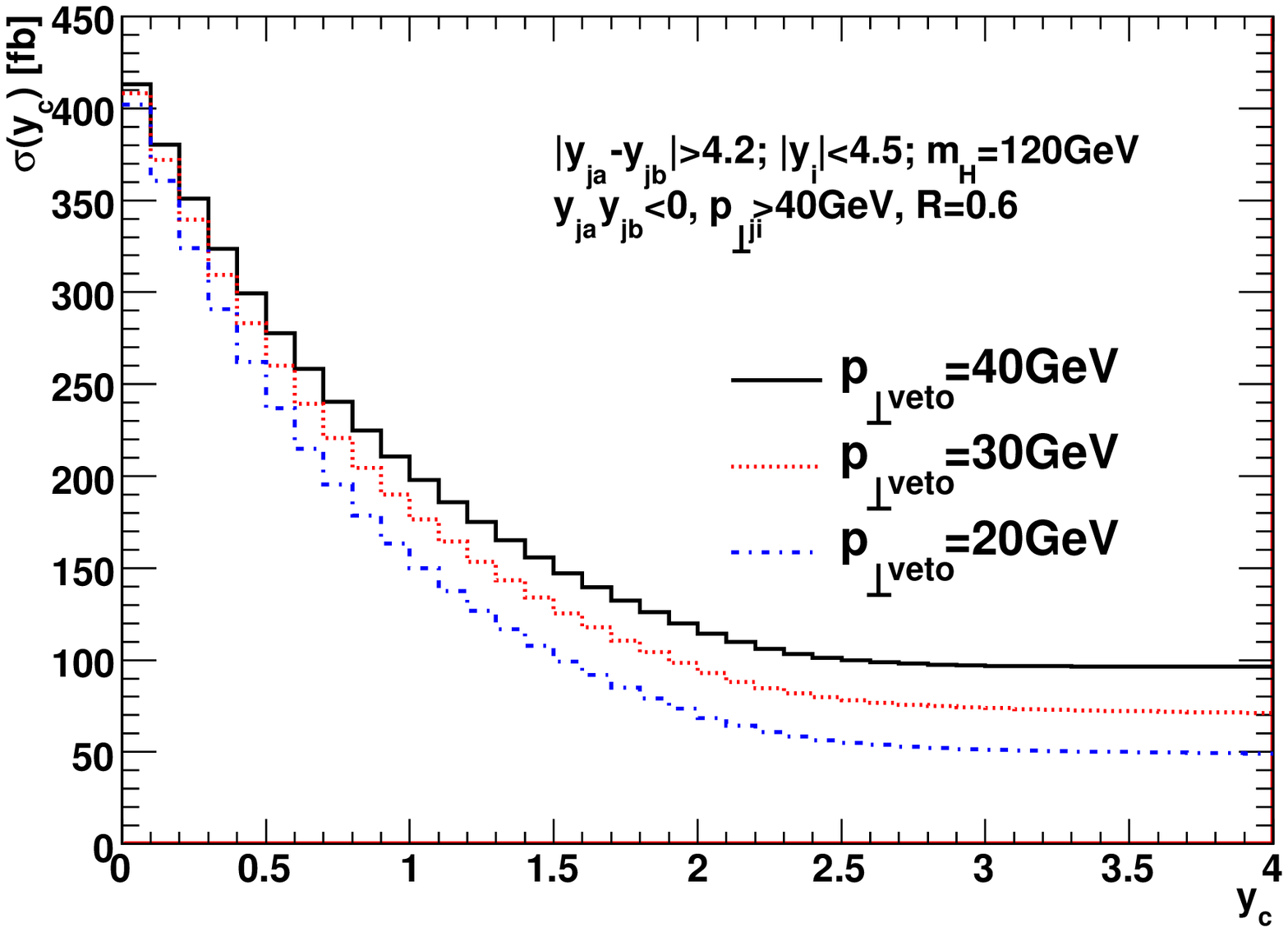}
  \epsfig{width=.49\textwidth,file=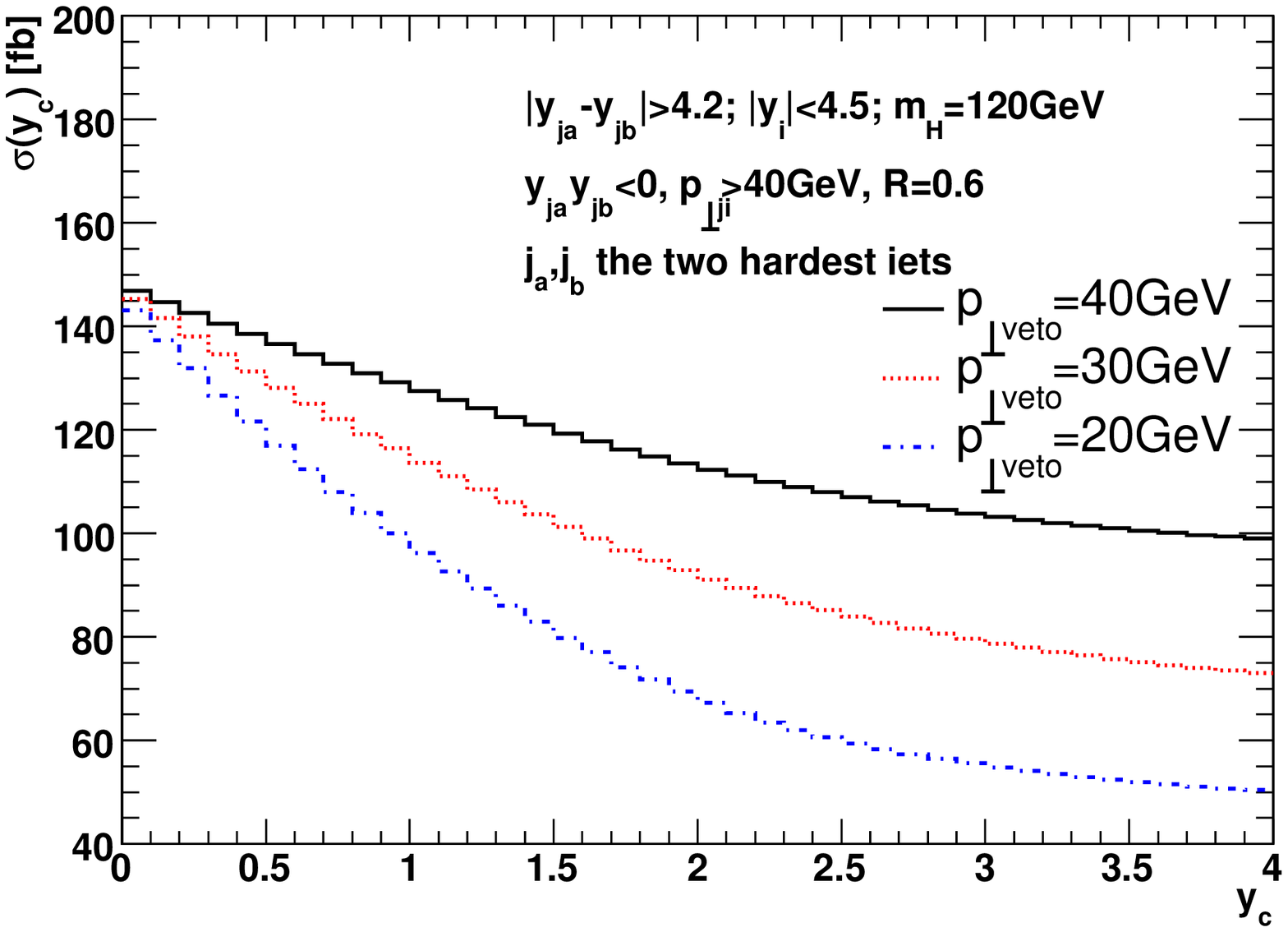}
  \epsfig{width=.49\textwidth,file=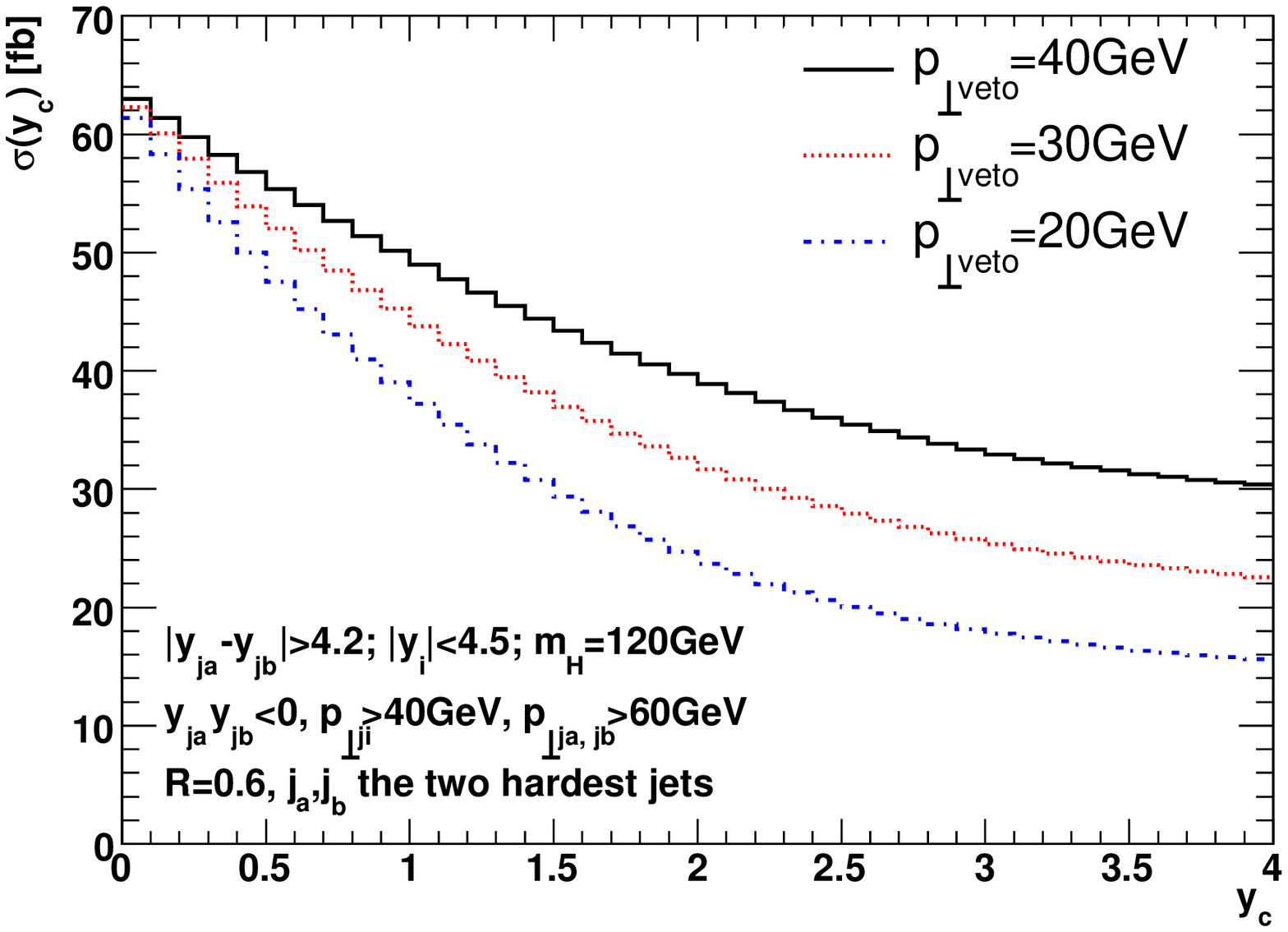}
\caption{The cross section as a function of the central jet veto $y_c$
  defined in Eq.~(\ref{eq:yc}) for three values of the transverse momentum
  parameter of the central rapidity jet veto $p_{\perp,\mbox{veto}}$.\label{fig:rapidvetoa}}
}
On Figure.~\ref{fig:rapidvetoa} we also show the similar distribution obtained
with the modified hard jet cuts discussed in
Section~\ref{sec:relaxed-cuts-central}. Again, a jet scale veto of 20GeV
extended to the whole rapidity span between the two hardest jets reduces the cross
section to about a fifth of the leading order $hjj$ estimate. Reducing the
jet scale to below 20GeV would probably need a study of the interplay with the
underlying event.

\subsection{Azimuthal correlations}
In this subsection we return to the discussion of the azimuthal correlation
between two jets. In Table~\ref{tab:aphitab} we compare the results for $A_\phi$
(defined in Eq.~\eqref{eq:Aphi}) using various calculations, and the two sets
of cuts.
\label{sec:azim-corr}
\begin{table}
\begin{center}
\begin{tabular}{c|c}
Inclusive cuts&$A_\phi$\\
\hline
LO 2-jet&$0.456$\\
Resummed, $=2$-jet & $0.437$\\
LO 3-jet&$0.203$\\
Resummed&$0.133$
\end{tabular}
\begin{tabular}{c|c}
Hardest cuts&$A_\phi$\\
\hline
LO 2-jet&$0.456$\\
Resummed, $=2$-jet & $0.436$\\
LO 3-jet&$0.374$\\
Resummed&$0.372$
\end{tabular}
\caption{Values of the decorrelation parameter $A_\phi$ for both choices of
  cuts. For the resummed results we present the values both for the events
  with two and only two jets (first resummed row), and for all events within
  the respective cuts (second resummed row).\label{tab:aphitab}}
\end{center}
\end{table}
Of particular interest is the difference between the first two lines of
numbers. The first ($A_\phi=0.456$) describes the result obtained in the
two-jet tree-level calculation. The second ($A_\phi=0.437$) is the result
obtained for events of the resummed and matched calculation, classified as
containing only two hard jets, but otherwise completely inclusive. The
difference is slight, and mostly due to the decorrelation caused by the
additional radiation not sufficiently hard to increase the number of hard
jets. These two numbers are obviously the same for the two sets of cuts,
while the azimuthal correlation observed in the tree-level three-jet
calculation and the fully inclusive, resummed approach depends on the choice
of cuts. The value of $A_\phi$ is most stable between the different
calculational procedures for the set of cuts relying on only the hardest jets
in the event. This arises since there are fewer hard jets in the event
samples based on the hardest cuts, which leads to stronger correlation
between the hardest jets. This is illustrated in Fig.~\ref{phiresum}, which
compares the $\phi$ distributions obtained in the resummed calculation, for
both the inclusive and hard cuts.  \FIGURE[tb]{
  \epsfig{width=.49\textwidth,file=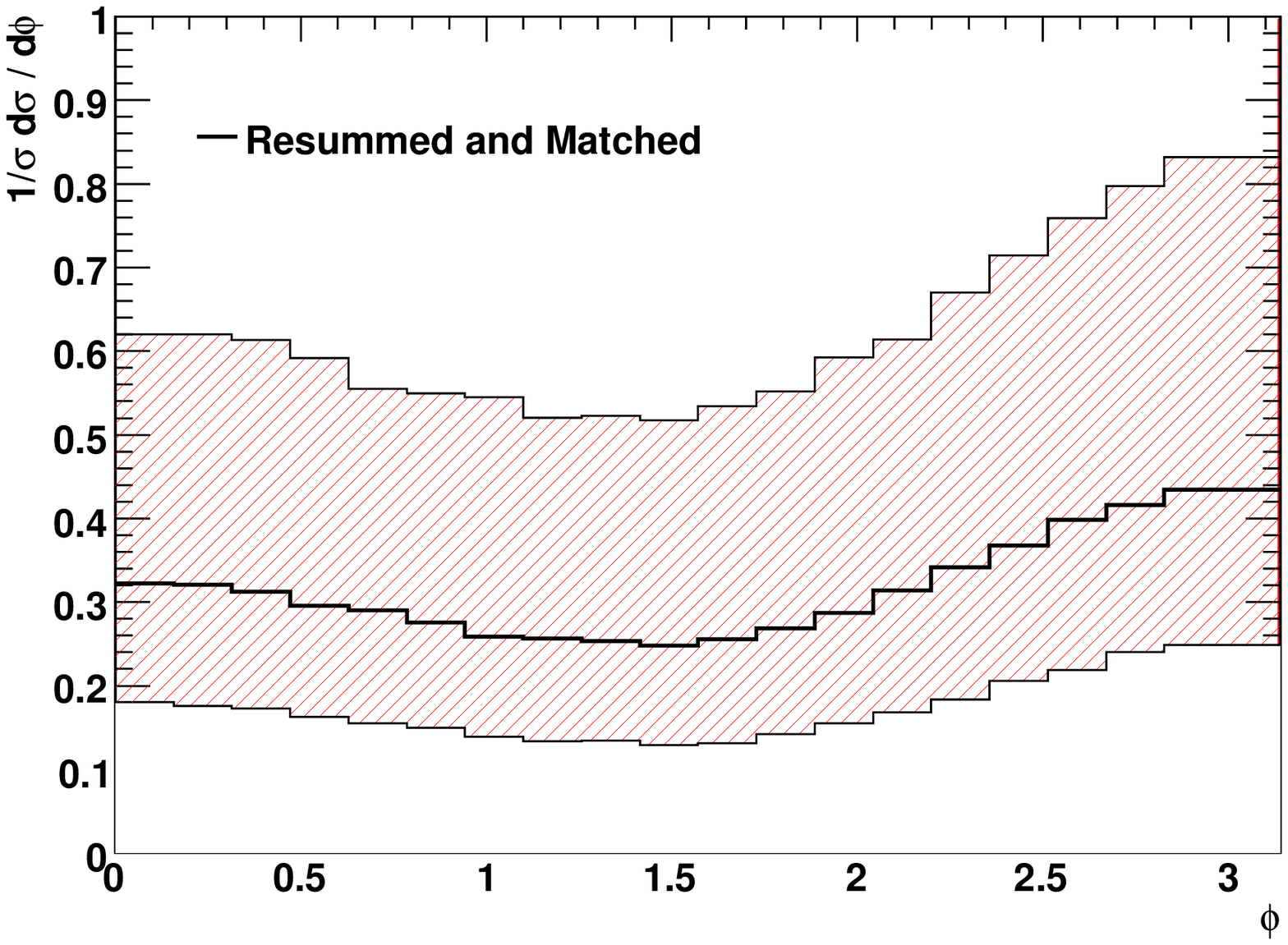}
  \epsfig{width=.49\textwidth,file=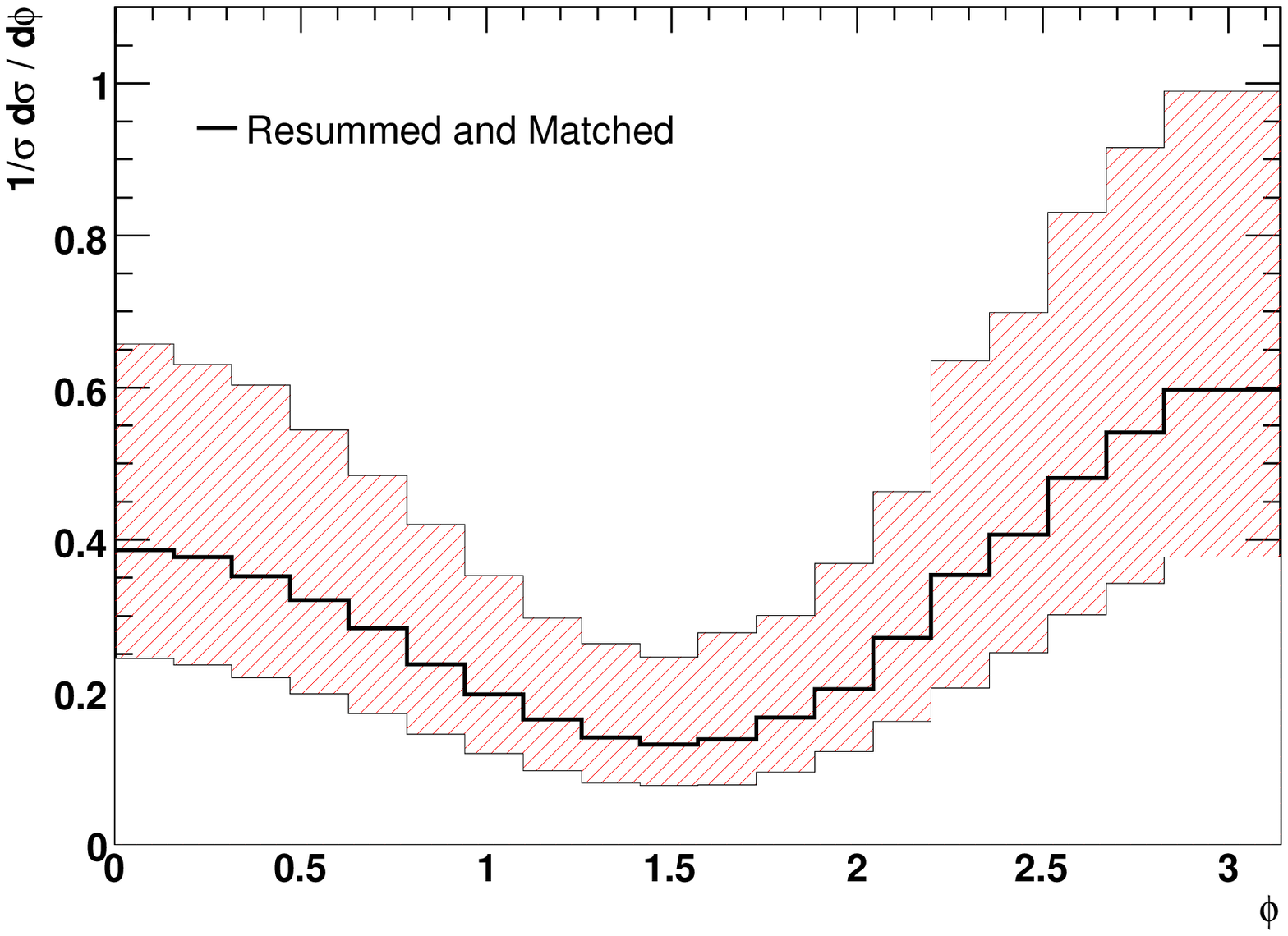}
    \caption{The azimuthal angle between the tagging jets, shown for the inclusive 
cuts of table \ref{tab:cuts} (left) and for the case where these cuts are applied to 
the two hardest jets (right). The uncertainty band corresponds to scale variation by 
a factor of 2.\label{phiresum}}
}

\subsection{Transverse Momentum Spectrum of the Higgs Boson}
\label{sec:transv-moment-spectr-1}
The transverse momentum spectrum of the Higgs boson when produced in
association with at least two hard jets is shown in Fig.~\ref{fig:ptH} for
both the calculation of the tree-level $hjj$, $hjjj$, and completely
inclusive, resummed and matched $hjj$ procedure presented here. The spectrum
is of course completely different to the one for the completely inclusive
Higgs boson production (no jets required), which has previously been
extensively studied in the literature, and can be obtained from the combined
NNLO-NNLL calculation of Higgs boson production through gluon
fusion\cite{Bozzi:2007pn}. The tree level 2 jet result has a bimodal
structure, which arises from the azimuthal correlation between the two jets
combined with the jet cuts. This structure disappears when extra radiation is
added, giving a qualitatively different behaviour. The significant difference
between the fixed-order spectra emphasises the importance of considering 
higher order corrections. The Higgs boson transverse momentum spectrum of the
resummed calculation has a hint of a remnant from the double-hump result
found at tree-level $hjj$ at small ($<100$GeV) transverse momenta, but is far
smoother, while tending to the tree-level $hjjj$ result for large transverse momenta.\FIGURE{ \epsfig{width=9cm,file=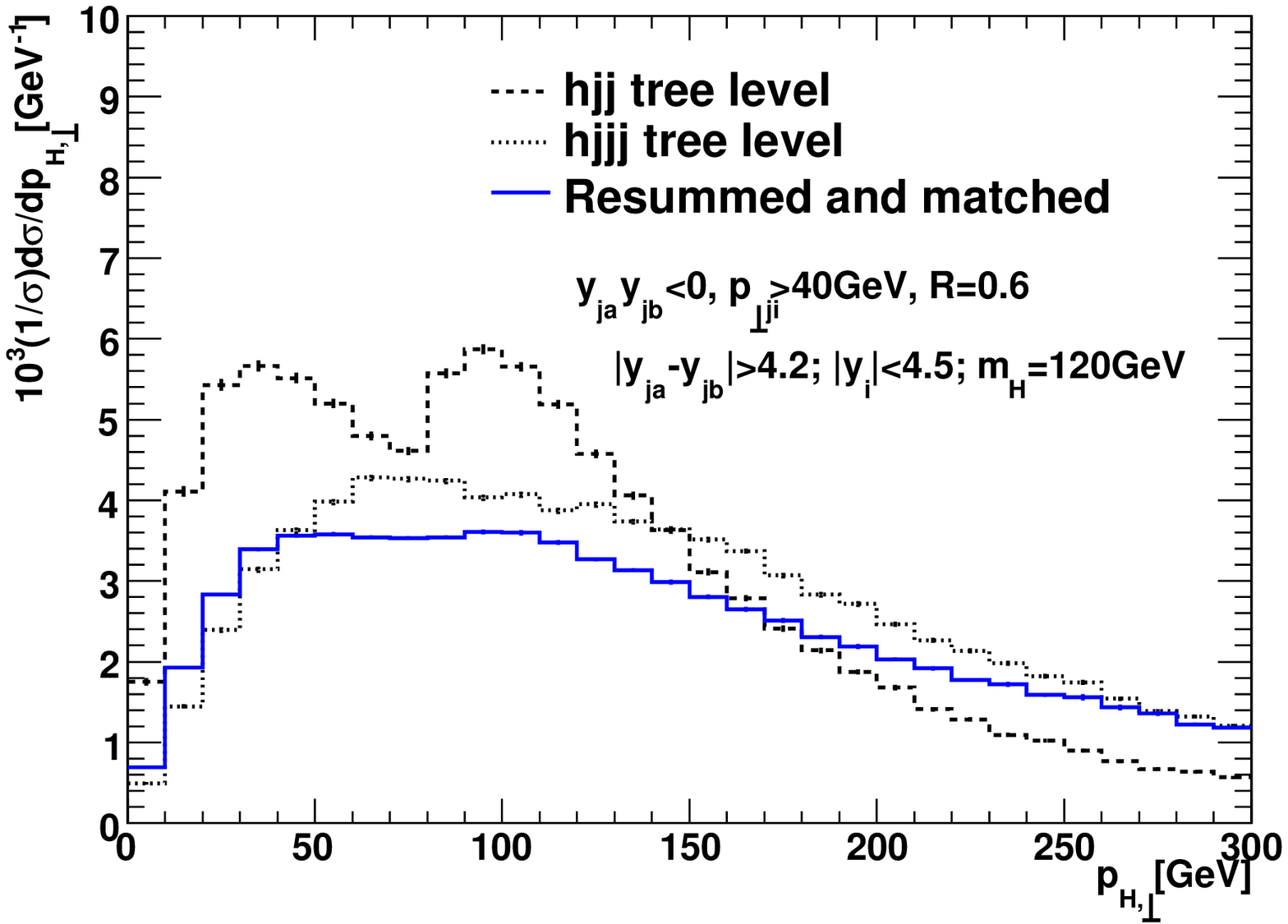}
\caption{The transverse momentum spectrum of the Higgs boson in association with at least two hard jets.\label{fig:ptH}}
}

\section{Discussion}
\label{discussion}
We have outlined a new technique for estimating the higher order QCD
corrections in Higgs boson production via GGF. Our technique takes FKL
factorisation as a starting point, as this implements hard radiation and also
includes some virtual corrections, which controls the normalisation as well
as cancelling singularities associated with soft gluon emissions. This is in
contrast to the inclusion of collinearly enhanced (soft) radiation to hard
matrix elements using a parton shower.

We define our $h$ + multiparton scattering amplitudes covering all of phase space
by the FKL
factorisation formula, implemented with the following requirements: (i) 
Use full 4-momenta for the gluon propagators; (ii) Use the Lipatov vertex as
given by Eq.~(\ref{lip1}). The former implements part of the known singularity
structure of higher order scattering amplitudes, whilst the latter enforces
the Ward identity and positivity of the squared gluon emission vertex over all
of phase space. The
approach has been validated by comparison with known tree level matrix
elements. As demonstrated, the approximations are generally very good, even
for modest rapidity spans, which increases confidence in the reliability of the
resummation procedure.

To further increase the accuracy of the approximation, we include the known
tree level matrix elements for 2 and 3 final state partons using a matching
prescription. In principle the matching could be extended to include higher
order tree level matrix elements, although the evaluation of the tree level
matrix element from standard packages become prohibitively slow. We note that
matching corrections have been included in all cases where the full
tree-level matrix elements have currently been used for predictions of cross
sections.

We have implemented the full framework in a Monte Carlo generator, and
provided example results obtained using the VBF cuts of
table~\ref{tab:cuts}. We have thoroughly discussed differences arising from
the interpretation of cuts in cases with more than 2 jets. When events are
accepted if any two jets satisfy the VBF cuts of table~\ref{tab:cuts}, then
the cross section is stable compared to the result at fixed NLO, but the
impact on the correlation in azimuth between tagged jets is large. If the two
hardest jets are required to also fulfil the VBF cuts, then the cross section
is reduced further compared to the result at NLO, due to the possibility of
additional jets, predominantly at central rapidity. However, the impact on
the azimuthal correlation is much smaller. If therefore the aim is to
suppress the contribution to the inclusive $hjj$ channel from gluon fusion,
while ensuring the azimuthal correlation is modified the least from the LO
estimate, then this is achieved best by tagging on the hardest jets in the
event. This is, however, a result only of the effective suppression of events
with central or many jets when requiring the hardest jets to pass the VBF
cuts, and similar results can be achieved by use of a central rapidity jet
veto.

The modified FKL formalism developed here is not unique to Higgs production
but is more generally applicable to many different scattering processes, such
as production of a $W$ boson with associated jets, and also purely partonic
processes. These can be implemented in the existing
framework, and could also be tested against existing hadron collider data.

\section*{Acknowledgments}
We would like to acknowledge useful discussions with John Campbell, Keith
Ellis and Gavin Salam. We are grateful to the GGI institute, Florence, where part of this work was carried out. CDW is supported by the Dutch Organisation for Fundamental Matter Research (FOM). He is grateful to the Cavendish Laboratory for hospitality, and also to Eric Laenen and Jos Vermaseren for useful discussions.
This work was partly supported by MIUR 
under contract 2006020509$_0$04 and by the EC Marie-Curie Research 
Training Network ``Tools and Precision
Calculations for Physics Discoveries at Colliders''
under contract MRTN-CT-2006-035505 .

\appendix

\bibliographystyle{JHEP}
\bibliography{database}
\end{document}